\PassOptionsToPackage{dvipsnames}{xcolor}
\documentclass[sigconf,nonacm]{acmart}

\usepackage{pvldb}

\newcommand{\para}[1]{\vskip0.5em\noindent\textbf{#1}.}
\newcommand{\apdix}{Appendix\xspace} 

\newif\ifpublish
\publishtrue

\newif\ifvldb
\vldbfalse

\usepackage{algorithm}
\usepackage{algorithmicx}
\usepackage{amsmath}
\makeatletter
\@ifclassloaded{acmart}{}{\usepackage{amssymb}}
\makeatother
\usepackage{amsthm}
\usepackage{array}
\usepackage{arydshln}
\usepackage[noend]{algpseudocode}
\usepackage{boxedminipage}
\usepackage{enumitem}
\usepackage{etoolbox}
\usepackage{float}
\usepackage{graphicx}
\usepackage{hyperref}
\usepackage{latexsym}
\usepackage{listings}
\usepackage{makecell}
\usepackage{multicol}
\usepackage{multirow}
\usepackage{pifont}
\usepackage{setspace}
\usepackage{textcomp}
\usepackage{threeparttable}
\usepackage{url}
\usepackage{subcaption}
\usepackage{tikz,balance}
\usepackage[dvipsnames]{xcolor}
\usepackage{xspace}
\usepackage{cleveref}
\usepackage[normalem]{ulem}

\newcounter{algorithmicinstance}
\BeforeBeginEnvironment{algorithmic}{\stepcounter{algorithmicinstance}}
\makeatletter
\providecommand{\theHALG@line}{}
\renewcommand{\theHALG@line}{\arabic{algorithmicinstance}.\arabic{ALG@line}}
\makeatother

\makeatletter
\@ifclassloaded{acmart}{}{%
\patchcmd{\@makecaption}
{\scshape}
{}
{}
{}
}
\makeatother

\makeatletter
\newcommand{\removelatexerror}{\let\@latex@error\@gobble}

\@ifclassloaded{acmart}{\theoremstyle{acmplain}}{\theoremstyle{plain}}

\newtheorem{lemma}{Lemma}
\newtheorem{corollary}{Corollary}
\newtheorem{proposition}{Proposition}
\algnewcommand\algorithmicforeach{\textbf{for each}}
\algdef{S}[FOR]{ForEach}[1]{\algorithmicforeach\ #1\ \algorithmicdo}

\newenvironment{myproof}[1]{\noindent \emph{Proof{#1}:}}{\hfill$\square$}

\algnewcommand\algorithmicinput{\textbf{INPUT:}}
\algnewcommand\INPUT{\item[\algorithmicinput]}
\algnewcommand\algorithmicoutput{\textbf{OUTPUT:}}
\algnewcommand\OUTPUT{\item[\algorithmicoutput]}
\algnewcommand\algorithmicassume{\textbf{ASSUME:}}
\algnewcommand\ASSUME{\item[\algorithmicassume]}

\newcommand{\mymode}{\mathit{mode}}
\newcommand{\cert}{\mathit{cert}}
\newcommand{\view}{\mathit{view}}
\newcommand{\round}{\mathit{round}}

\newcommand{\vertex}{\mathit{vertex}}
\newcommand{\creator}{\mathit{creator}}

\newcommand{\parents}{\mathit{references}}
\newcommand{\post}{\mathit{SB}}
\newcommand{\aln}{\mathsf{Line}}

\newcommand{\lightbold}[1]{%
 \textcolor{black}{\textmd{\scalebox{1.0}[1.1]{#1}}}
}

\newcommand{\tocommit}{\lightbold{to-commit}}
\newcommand{\toskip}{\lightbold{to-skip}}
\newcommand{\undecided}{\lightbold{undecided}}

\newcommand{\gtext}[1]{\textcolor{gray}{#1}}

\newcommand{\ACSPropose}{\mathit{acs\_propose}}
\newcommand{\ACSDecide}{\mathit{acs\_decide}}

\newcommand{\ignore}[1]{}
\newcommand{\sig}[1]{\langle #1 \rangle}

\algrenewcommand\textproc{}
\algdef{SE}[EVENT]{Event}{EndEvent}[1]{\textbf{upon}\ #1\ \algorithmicdo}{\algorithmicend\ \textbf{event}}%
\algtext*{EndEvent}
\makeatletter
\@ifclassloaded{acmart}{}{%
  \renewcommand{\paragraph}[1]{\medskip\noindent\textbf{#1}}%
}
\makeatother
\newtheorem{claim}{Claim}

\newtheorem{restrict}{Restriction}

\newcommand{\node}[1]{\ensuremath{N_{#1}}\xspace}

\newcommand{\NoVote}{\mathsf{no\text{-}vote}}
\newcommand{\NoVoteCert}{\mathcal{NVC}}

\definecolor{yescolor}{HTML}{026378}

\makeatletter
\renewcommand{\ALG@name}{}
\makeatother

\newenvironment{graytext}{\color{gray}}{\ignorespacesafterend}

\newcommand{\sysname}{Lifefin\xspace}
\newcommand{\membudget}{\ensuremath{M_B}\xspace}

\makeatletter
\newcounter{vldbauthorblock}
\pretocmd{\@mkauthors@iii}{\setcounter{vldbauthorblock}{0}}{}{}
\pretocmd{\@typeset@author@bx}{%
  \stepcounter{vldbauthorblock}%
  \ifcase\value{vldbauthorblock}\relax
  \or\author@bx@wd=.183\textwidth
  \or\author@bx@wd=.171\textwidth
  \or\author@bx@wd=.202\textwidth
  \or\author@bx@wd=.223\textwidth
  \or\author@bx@wd=.181\textwidth
  \fi
}{}{}
\makeatother

\renewcommand\vldbdoi{XX.XX/XXX.XX}
\renewcommand\vldbpages{XXX-XXX}
\renewcommand\vldbavailabilityurl{https://github.com/EtherCS/Lifefin}

\begin{document}

\title{Liveness Attacks on DAG-based BFT and A Generic Solution}

\author{Jianting Zhang}
\affiliation{%
  \institution{Purdue University}
  \country{}
}
\email{zhan4674@purdue.edu}

\author{Sen Yang}
\affiliation{%
  \institution{Yale University}
  \country{}
}
\email{sen.yang@yale.edu}

\author{Alberto Sonnino}
\affiliation{%
  \institution{Mysten Labs/\\ UCL}
  \country{}
}
\email{alberto@mystenlabs.com}

\author{Sebasti\'an Loza}
\affiliation{%
  \institution{UTEC}
  \country{}
}
\email{sebastian.loza@utec.edu.pe}

\author{Aniket Kate}
\affiliation{%
  \institution{Mysten Labs/Purdue University}
  \country{}
}
\email{aniket@purdue.edu}

\begin{abstract}
   Directed Acyclic Graph (DAG)-based Byzantine Fault-Tolerant (BFT) protocols have emerged as promising solutions for high-throughput blockchains. By decoupling data dissemination from transaction ordering and constructing a well-connected DAG in the mempool, these protocols enable zero-message ordering and implicit view changes. However, we identify a liveness vulnerability in a broad class of DAG protocols: their optimistic paths require later-round vertices to commit earlier ones, but these vertices accumulate when commitment stalls, allowing an adversary to trigger \emph{mempool explosions} and prevent transaction commitment, ultimately compromising the protocol's liveness.

    In response, we present \sysname, a generic recovery framework for this class of DAG protocols. \sysname leverages the Agreement on Common Subset (ACS) mechanism, allowing nodes to escape mempool explosions by committing transactions with bounded working storage even in adverse conditions. As a result, \sysname imposes (almost) zero overhead in typical cases while preventing mempool explosions from causing permanent liveness loss under a fixed storage budget.

    To demonstrate the effectiveness of \sysname, we integrate it into two state-of-the-art DAG-based BFT protocols, Sailfish and Mysticeti, resulting in two enhanced variants: Sailfish-\sysname and Mysticeti-\sysname. We implement these variants and compare them with the original Sailfish and Mysticeti systems. Our evaluation demonstrates that \sysname achieves comparable transaction throughput while introducing only minimal additional latency to resist similar attacks.
\end{abstract}

\maketitle

\vldbtopmatter

\section{Introduction}\label{section-introduction}
For several decades, the Byzantine Fault-Tolerant (BFT) state machine replication problem has been a prominent topic of study in the distributed systems literature.
In this problem, a group of nodes executes a BFT protocol to establish a consistent order for transactions continuously submitted by clients~\cite{hotstuff, neiheiser2021kauri, liu2023flexible, sun2023neobft, gupta2023dissecting, suri2021basil, chen2024porygon, autobahn}. This process typically involves a \textit{data dissemination} task where transactions are propagated among nodes and an \textit{ordering} task where nodes decide on an order for the propagated transactions. 

BFT protocols have been widely adopted by several in-production blockchains~\cite{sui, aptos, supra, monad, cosmos, iota, rocket2019scalable} for achieving consensus on transaction ordering. 
Modern deployed BFT protocols~\cite{diembft, jolteon, autobahn, doidge2024moonshot, monadbft} typically operate in the partial synchrony model~\cite{DLT} and as they scale, they decouple data dissemination from ordering: nodes continuously propagate intact transactions in batches or blocks, while the ordering task is performed separately using lightweight metadata (that is, hashes of batches/blocks)~\cite{narwhal}. This decoupling offers two key advantages. First, it improves bandwidth utilization, as nodes can disseminate data even during asynchrony periods, which are known to hinder ordering due to the famous FLP impossibility result~\cite{fischer1985impossibility}. Second, it enhances communication efficiency in the ordering task, as nodes only need to process lightweight metadata instead of full transactions.

Most prior decoupled BFT protocols~\cite{diembft, jolteon, duan2024dashing, autobahn, arete, monadbft, hotstuff, hotstuff-2, hotstuff-1, streamlet, fasthotstuff, librabft} rely on leader-driven ordering, where a designated leader node in each consensus instance proposes a transaction order and \textit{explicitly} coordinates agreement process. Although the decoupled design reduces communication overhead in ordering, the leader-driven approach exposes several practical limitations. First, leader nodes experience uneven communication overhead and become potential bottlenecks, since they need to exclusively disperse order proposals to all nodes. Second, if order proposals fail, these protocols require a costly view-change mechanism to elect new leaders and/or need to manage large backlogs~\cite{autobahn}.

DAG-based BFT protocols~\cite{dag-rider, narwhal, bullshark, shoal, sailfish, bbca-chain, suilutris, cordial-miners, mysticeti, jovanovic2024mahi, danezis2024obelia, frontrundag} introduce an alternative decoupled design that eliminates the limitations of leader-driven approaches. 
Specifically, these protocols operate in rounds: in each round, every node proposes a vertex containing transactions and connections to vertices from the previous round. The proposed vertices and connections are then disseminated to form a directed acyclic graph (DAG), where connections serve as votes for the connected vertices. The ordering task is performed separately: nodes check whether selected leader vertices meet the committing rules (i.e., have enough votes/connections in subsequent rounds), and if they do, nodes order and commit them deterministically. Due to the information encoded in the DAG, once a leader vertex is committed, its entire causal history can also be ordered and committed.

The key innovation of DAG-based BFT lies in the structured \emph{DAG-based memory pool (mempool)}, which nodes maintain to track uncommitted vertices. This design offers two advantages. First, nodes can independently order and commit transactions by interpreting the DAG, where edges represent votes for leader vertices---a property known as \textit{zero-message ordering}. Second, in the presence of faulty leaders, nodes avoid costly view-change procedures; a subsequent valid leader vertex can naturally supersede failed proposals through its causal history, enabling \textit{implicit view-change} (\S~\ref{sec-preliminaries-dag}).

However, these advantages do not come for free.
This paper identifies a resource-bounded liveness vulnerability in a broad class encompassing all modern DAG protocol families surveyed in this paper~\cite{dag-rider,narwhal,bullshark,shoal,bolt,sailfish,bbca-chain,suilutris,cordial-miners,mysticeti,jovanovic2024mahi,danezis2024obelia,frontrundag,Fino,gradeddag,wahoo,lightdag,dai2024remora}. The vulnerability arises from a gap between theory and practice: their liveness guarantees are proved with unbounded working storage, while deployed nodes have fixed storage budgets.
Specifically, nodes can exhaust their working storage by maintaining an ever-growing set of uncommitted vertices in the DAG-based mempool during periods of asynchrony. A finite implementation must then stop generating vertices to remain within its budget, which can leave the latest leader without enough subsequent votes and prevent commitment even after the network returns to synchrony.

More specifically, in these DAG-based BFT protocols, nodes create new vertices without waiting for the commitment of earlier ones. However, before the Global Stabilization Time (GST), a leader vertex is not guaranteed to gather enough votes to be committed. As uncommitted vertices accumulate, nodes risk running out of resources while trying to maintain the mempool. Meanwhile, leader vertices require sufficient votes, manifested as new subsequent vertices, for successful ordering under the zero-message ordering mechanism. As a result, enforcing a storage limit can block nodes from generating new vertices or collecting enough votes, ultimately halting progress.
We refer to this resource exhaustion issue as \emph{mempool explosion} (\S~\ref{sec-liveness-issue}). More importantly, it follows from the continuous-growth optimistic path defined in \S~\ref{sec-liveness-issue}, rather than from a particular memory allocator or implementation bug. Our technical claim is that a storage cap alone cannot preserve progress on that path. A DAG protocol that exhibits this path and lacks an independent bounded-resource committing mechanism is vulnerable to this liveness issue.

To exploit this liveness vulnerability, this paper introduces a class of \emph{inflation attacks} (\S~\ref{sec-inflation-attack}). In the attack, the adversary participates in data dissemination by proposing vertices in every round, but crucially deviates from the ordering process by permanently refraining from voting for leader vertices. As new vertices are continually generated without being committed, correct nodes reach their resource limits while maintaining the DAG-based mempool. 
Crucially, the attack uses no adversarial power beyond the standard $f$ out of $3f+1$ Byzantine nodes and pre-GST message scheduling (see \S~\ref{sec-theory-practice}). 
Such liveness attacks need not be purely griefing: recent work shows that consensus-layer disruption can be monetized through application-layer financial positions, illustrating how attackers may have concrete economic incentives for disruption~\cite{yang2026your}.

In response to the liveness vulnerability and inflation attacks, one might attempt to adopt existing resource-control mechanisms~\cite{beyer2016site, pbft, microsoft2026throttling, amazon-load-shedding, tendermint-timeouts}; however, these naive approaches either introduce new liveness issues or merely slow mempool growth without fundamentally addressing these issues (\S~\ref{sec-attempt}).
To address this root cause, we propose \sysname,
a generic recovery framework for DAG protocols with a continuous-growth optimistic path (\S~\ref{sec:overview}). The key insight is that if nodes can complete the ordering task with finite resource usage, they can ensure enough resources are available for ordering, thereby avoiding mempool explosions.
The core concept behind \sysname is a single-shot agreement that uses a bounded set of recognized DAG messages to commit the backlog in the DAG-based mempool. Specifically, \sysname is built on the Agreement on Common Subset (ACS) paradigm, which allows $n$ nodes with $n$ inputs to agree on a common subset $V$ (where $|V| \geq n-f$, with $f$ being the number of faulty nodes). In \sysname-enhanced DAG-based BFT protocols, nodes continue to efficiently disseminate, order, and commit vertices by following the underlying DAG protocol (on the optimistic path). When nodes suspect an inflation attack or are near resource exhaustion, they collectively activate the ACS mechanism, committing vertices with bounded resources.
\sysname therefore introduces (almost) zero overhead in optimistic cases while providing recovery from inflation attacks.
Moreover, \sysname carefully selects vertices from the ACS output for ordering and incorporates a smooth transition to prevent inconsistencies caused by asynchrony or by switching between the ACS mechanism and the underlying protocol, thereby guaranteeing correctness.

We instantiate \sysname in two representative DAG-based protocols, Sailfish~\cite{sailfish} (\S~\ref{sec:sailfish-sysname}) and Mysticeti~\cite{mysticeti} (\S~\ref{sec:mysticeti-sysname}), and implement \sysname on top of them. We conduct extensive experiments in geo-distributed environments across five regions to evaluate the performance of \sysname and emulate its behavior when encountering inflation attacks. The results demonstrate that \sysname achieves comparable throughput while introducing only minor latency overhead to effectively resist the exploited inflation attacks.

We summarize our contributions as follows:
{
\makeatletter
\def\@listi{\leftmargin10pt \labelwidth\z@ \labelsep5pt}
\makeatother
\begin{itemize}[nosep]
    \item We identify a liveness vulnerability (\S~\ref{sec-liveness-issue}) in a broad class of DAG-based BFT protocols and exploit it with a family of attacks (\S~\ref{sec-inflation-attack}). We show that this vulnerability follows from their ordering design and cannot be eliminated by implementation-level resource control alone (\S~\ref{sec-attempt}).
    \item We present \sysname (\S~\ref{sec:overview}), a generic recovery framework that provides this DAG protocol class with a bounded-resource recovery path and resists the exploited attacks.
    \item We instantiate \sysname in Sailfish (\S~\ref{sec:sailfish-sysname}) and Mysticeti (\S~\ref{sec:mysticeti-sysname}), along with rigorous security analyses. 
    Sailfish and Mysticeti cover certified and uncertified DAGs, demonstrating that the recovery approach is not tied to vertex certification.
    \item We implement \sysname on Sailfish and Mysticeti, resulting in two enhanced variants: Sailfish-\sysname and Mysticeti-\sysname. We evaluate and compare them with the original Sailfish and Mysticeti protocols (\S~\ref{sec-evaluation}), showing that \sysname achieves comparable performance while offering effective recovery from the attacks.
\end{itemize}
}

\section{Preliminaries}\label{sec-preliminaries}
\subsection{Threat, System, and Network Models}
We consider a system with $n = 3f{+}1$ nodes $\{\node{1}, \cdots, \node{n}\}$, of which up to $f$ are Byzantine and can be corrupted by a static adversary. The Byzantine nodes can behave arbitrarily but are computationally bounded. The remaining $2f{+}1$ nodes are correct.
We use a public-key infrastructure and digital signatures for message authentication. We use $\sig{m}_i$ to denote a message $m$ digitally signed by node $\node{i}$ with its private key.
Motivated by finite-memory models~\cite{delporte2008finite, dolev2016possibility}, we consider a bounded working-storage model. Each correct node has a fixed budget \membudget for live consensus state, including admitted but uncommitted data, protocol messages, and synchronization buffers; committed ledger output is not charged to this budget. Vertex and transaction sizes are bounded by protocol parameters. We assume standard admission control: clients retry until admission, offered load is eventually sustainable, and correct nodes fairly include admitted transactions. A resource-aware protocol must keep its live state within \membudget while continuing to execute.
An execution requiring more than \membudget of live consensus state violates the resource bound; it is not modeled as an additional crash or Byzantine fault.

We consider the partial synchrony model of Dwork et al.~\cite{DLT}, where there exists an unknown time called Global Stabilization Time (GST) and a known time bound $\Delta$, such that any message sent by correct nodes after GST is guaranteed to arrive within time $\Delta$.

This paper focuses on the Byzantine Fault-tolerant (BFT) state machine replication problem, where a group of nodes runs a BFT protocol to establish a consistent order for transactions that are continuously submitted by clients. A BFT protocol is designed to guarantee the following properties~\cite{fin}:
{
\makeatletter
\def\@listi{\leftmargin10pt \labelwidth\z@ \labelsep5pt}
\makeatother
\begin{itemize}[nosep]
    \item \textbf{Safety:} If a correct node delivers a transaction $tx$ before $tx'$, then no correct node delivers $tx'$ without first delivering $tx$, i.e., all correct nodes have the same prefix ledger.
    \item \textbf{Liveness:} If a transaction $tx$ is sent to all correct nodes, then all correct nodes will eventually deliver $tx$.
\end{itemize}
}

\subsection{DAG-based BFT}\label{sec-preliminaries-dag}
Directed acyclic graph (DAG)-based BFT protocols have been proposed to enhance performance and adopted by modern blockchains, such as Sui~\cite{sui} and IOTA~\cite{iota}. Based on how they disseminate vertices, they can be categorized into \textit{certified} DAG~\cite{narwhal, bullshark, shoal, bolt, sailfish} and \textit{uncertified} DAG~\cite{cordial-miners, mysticeti, jovanovic2024mahi}. We survey the modern round-based DAG protocol family and summarize the features relevant to our analysis below.

\para{Separating data dissemination and ordering} \label{sec-dag-separation}
DAG-based BFT protocols decouple data dissemination from metadata ordering. Specifically, DAG-based BFT protocols proceed in logical rounds. In each round $r$, every node creates and disseminates a vertex consisting of transactions and a quorum of connections to round $r-1$ vertices. The connected vertices then form a DAG and are maintained in the nodes' mempools until they are later ordered and committed.

The DAG-based mempool grows as nodes move to new rounds, allowing them to perform the ordering task separately by interpreting the DAG. Specifically, a DAG-based BFT protocol designates leader nodes to propose \textit{leader vertices}. In partially synchronous DAG-based BFT protocols, leader vertices are predefined, e.g., using a round-robin approach. By interpreting the DAG, nodes search for leader vertices that are connected (i.e., voted for) by enough vertices in subsequent rounds and then order the qualifying leader vertices and their causal histories using a deterministic algorithm~\cite{frontrundag}.

To achieve a secure separation between data dissemination and ordering, DAG-based BFT defines two core quorums:
{
\makeatletter
\def\@listi{\leftmargin10pt \labelwidth\z@ \labelsep5pt}    
\makeatother
\begin{itemize}[nosep]
       \item \textbf{Dissemination Quorum} $q_d$: The minimum number of distinct nodes whose valid vertices are required for data dissemination to proceed.
\item \textbf{Committing Quorum} $q_c$: The minimum number of distinct nodes whose valid messages are required for the ordering task to proceed.
\end{itemize}
}

Across the surveyed protocols, $q_d$ is set to $2f+1$ to form a threshold logical clock~\cite{threshold-clock}: a node processing round $r$ waits for $q_d$ vertices from round $r$ before advancing to round $r+1$, and a round $r+1$ vertex must connect to at least $q_d=2f+1$ vertices from round $r$. The exact \emph{committing evidence} summarized by $q_c$ varies across protocols.
Certified DAG-based BFT protocols~\cite{dag-rider, narwhal, bullshark, sailfish} rely on reliable broadcast (RBC)~\cite{bracha1985asynchronous} to generate certified vertices for commitment, typically with acknowledgements from $2f+1$ distinct nodes, leading to $q_c=2f+1$.
Meanwhile, a class of uncertified DAG-based protocols~\cite{cordial-miners, mysticeti} forgoes RBC to achieve lower latency but enforces a stricter committing rule.
Briefly, in Mysticeti, a leader vertex in round $r$ is directly committed only when there are at least $2f+1$ vertices in round $r+2$, each of which has $2f+1$ paths reaching it (\S~\ref{sec:mysticeti-protocol-review}); hence, its relevant $q_c$ is $2f+1$.
The dissemination and committing quorums stipulate the formation of a DAG, enabling nodes to maintain a robust DAG-based mempool to securely order transactions with an efficient ordering process.

\para{Two gains with one DAG-based mempool}
Unlike previous decoupled BFT protocols, the DAG-based BFT protocol organizes uncommitted vertices as a DAG in the mempool, enabling two key features for the ordering task:
{
\makeatletter
\def\@listi{\leftmargin10pt \labelwidth\z@ \labelsep5pt}    
\makeatother
\begin{itemize}[nosep]
    \item \textbf{Zero-message ordering:} Nodes can order transactions in the DAG-based mempool without extra communication for ordering.
    \item \textbf{Implicit view-change:} There is no need for explicit view-change mechanisms for handling faulty leaders.
\end{itemize}
}
Once the DAG of vertices is formed, each node can order transactions \textit{locally} by interpreting the DAG in the mempool, with connections serving as votes for leader vertices. This enables zero-message ordering. Additionally, unlike previous leader-driven BFT protocols that require extra communication rounds to handle faulty leaders, DAG-based BFT protocols implement an \textit{implicit} view-change mechanism. Specifically, since the relevant quorums $q_d$ and $q_c$ govern the formation of the DAG, uncommitted leader vertices can be indirectly committed by leader vertices in subsequent rounds. This eliminates the need for an explicit view change to select new leader vertices to replace the faulty ones.

\section{DAG BFT: Theory meets Practice}\label{sec-theory-practice}
\subsection{Observation} \label{sec-liveness-issue}
\noindent\textbf{Mempool explosions}.
We identify an overlooked cost of achieving zero-message ordering and implicit view-change in DAG-based BFT: a node needs to maintain continuously growing data in its DAG-based mempool during periods of asynchrony, potentially causing a \textit{mempool explosion}. Specifically, nodes are allowed to create new vertices even when prior vertices remain uncommitted (cf. implicit view-change). However, due to the nature of asynchrony, there is no guarantee that leader vertices receive enough votes for commitment. Thus, nodes eventually reach their resource limit while maintaining continuously generated uncommitted vertices.

\para{Scope}
Our observation concerns a structural pattern in round-based DAG protocols: (i) data dissemination can advance while commitment is stalled; (ii) committing a pending leader requires a quorum of later vertices; and (iii) uncommitted vertices needed for ordering or availability cannot be discarded before a committed recovery point. We call this a \emph{continuous-growth optimistic path}. All modern DAG protocol families surveyed in this paper exhibit this path and lack an independent bounded-resource committing mechanism~\cite{dag-rider,narwhal,bullshark,shoal,bolt,sailfish,bbca-chain,suilutris,cordial-miners,mysticeti,jovanovic2024mahi,danezis2024obelia,frontrundag,Fino,gradeddag,wahoo,lightdag,dai2024remora}, spanning certified and uncertified DAGs as well as partially synchronous and asynchronous designs.

\para{Liveness issue in practice}
For protocols with a continuous-growth optimistic path, a fixed storage cap cannot preserve both bounded storage and liveness. Before GST, an adversary can allow dissemination to continue while preventing commitment until retained state reaches the storage budget \membudget. Continuing vertex creation exceeds \membudget; stopping it allows Byzantine nodes to withhold their vertices, leaving only $2f<q_d=2f+1$ nodes able to advance rounds. 
This also means that, unlike non-DAG decoupled BFT (such as Hotstuff~\cite{hotstuff} and Autobahn~\cite{autobahn}), existing resource-control mechanisms~\cite{beyer2016site, pbft, microsoft2026throttling, amazon-load-shedding, tendermint-timeouts} cannot resolve the liveness issue in DAG protocols (see further discussions in \S~\ref{sec-attempt} and the full paper~\cite[Appendix~G]{zhang2025lifefin}).
Since all surveyed protocols lack a separate bounded-resource committing path, the vulnerability applies to all of them.

We emphasize that this liveness issue persists even after the network recovers to synchrony, as it arises from the failure of vertex creation rather than message delays. In the following section, 
We introduce a class of attacks, termed \textit{inflation attacks}, that drives a node to its resource limit; enforcing the limit then prevents it from producing protocol messages even after GST. Our estimates show that these inflation attacks can compromise liveness in as little as 11 minutes (\S~\ref{sec-inflation-attack}).



\subsection{Inflation Attack}\label{sec-inflation-attack}
\noindent\textbf{Attack description}.
The goal of an inflation attack is to cause a mempool explosion before GST by exhausting nodes' resources before correct nodes can complete the ordering task. To achieve this, Byzantine nodes actively propagate DAG vertices from correct nodes in every round, enabling the data dissemination task to proceed. However, they simultaneously prevent correct nodes from completing the ordering task. As a result, correct nodes are forced to maintain ever-growing uncommitted DAG vertices in their mempools, eventually leading to resource exhaustion and a mempool explosion.

To make the inflation attack succeed in practice, the adversary can execute a distributed denial-of-service (DDoS) attack on correct nodes, preventing them from receiving and voting for leader vertices before GST. For instance,
the adversary launches a DDoS attack on a correct node to make it temporarily unavailable. Since the $f$ Byzantine nodes do not vote for leader vertices, the remaining $2f$ correct nodes will never collect enough (i.e., $2f\text{<}q_c=2f+1$) votes to commit leader vertices. Moreover, the adversary can also launch DDoS attacks on the leader node every round, preventing it from delivering the leader vertex. In any case, the ordering task keeps failing, causing uncommitted DAG vertices to accumulate and eventually triggering a mempool explosion.


Note that the DDoS attack is temporary and serves solely to trigger the initial mempool explosion. Its effect is modeled by the unbounded message delays permitted before GST in the classic partial synchrony model.
Crucially, after the first correct node reaches its limit and stops creating vertices, the failure to complete ordering tasks becomes self-sustaining. The adversary can either repeatedly drive other correct nodes to their limits without DDoS attacks (since the stopped node cannot create new vertices or certify leader blocks to contribute to the committing quorum $q_c$) or stall progress to new rounds; either outcome could ultimately violate liveness.
We defer the concrete attack description to the full paper~\cite[Appendix~F]{zhang2025lifefin}.

\begin{figure}
    \setlength\abovecaptionskip{0.1\baselineskip}
    \centering
    \begin{subfigure}[t]{0.235\textwidth}
        \setlength\abovecaptionskip{-0.1\baselineskip}
		\centering
        \includegraphics[width=1\linewidth]{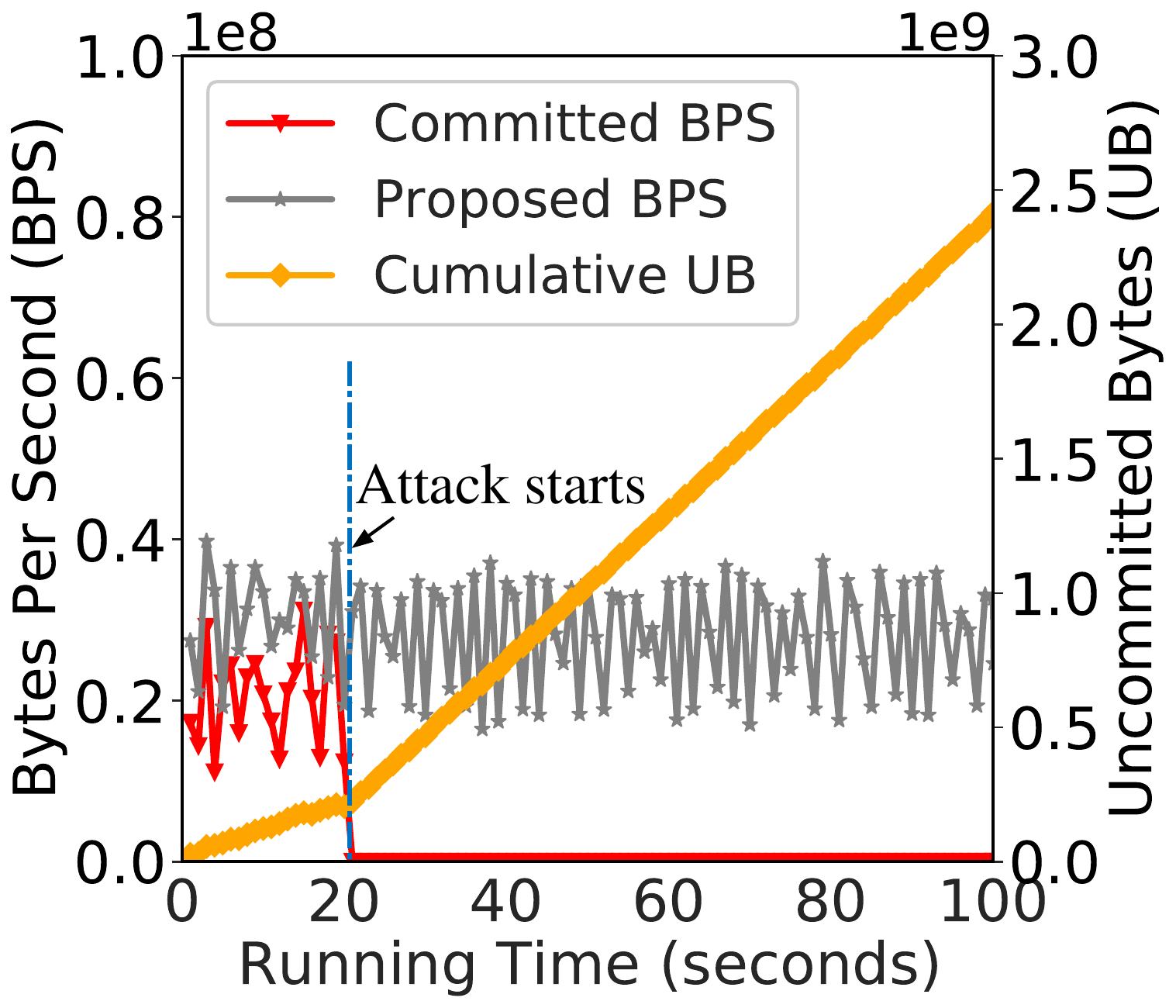}
        \caption{Sailfish under attack}
    \end{subfigure}
    \begin{subfigure}[t]{0.235\textwidth}
        \setlength\abovecaptionskip{-0.1\baselineskip}
        \centering
        \includegraphics[width=1\linewidth]{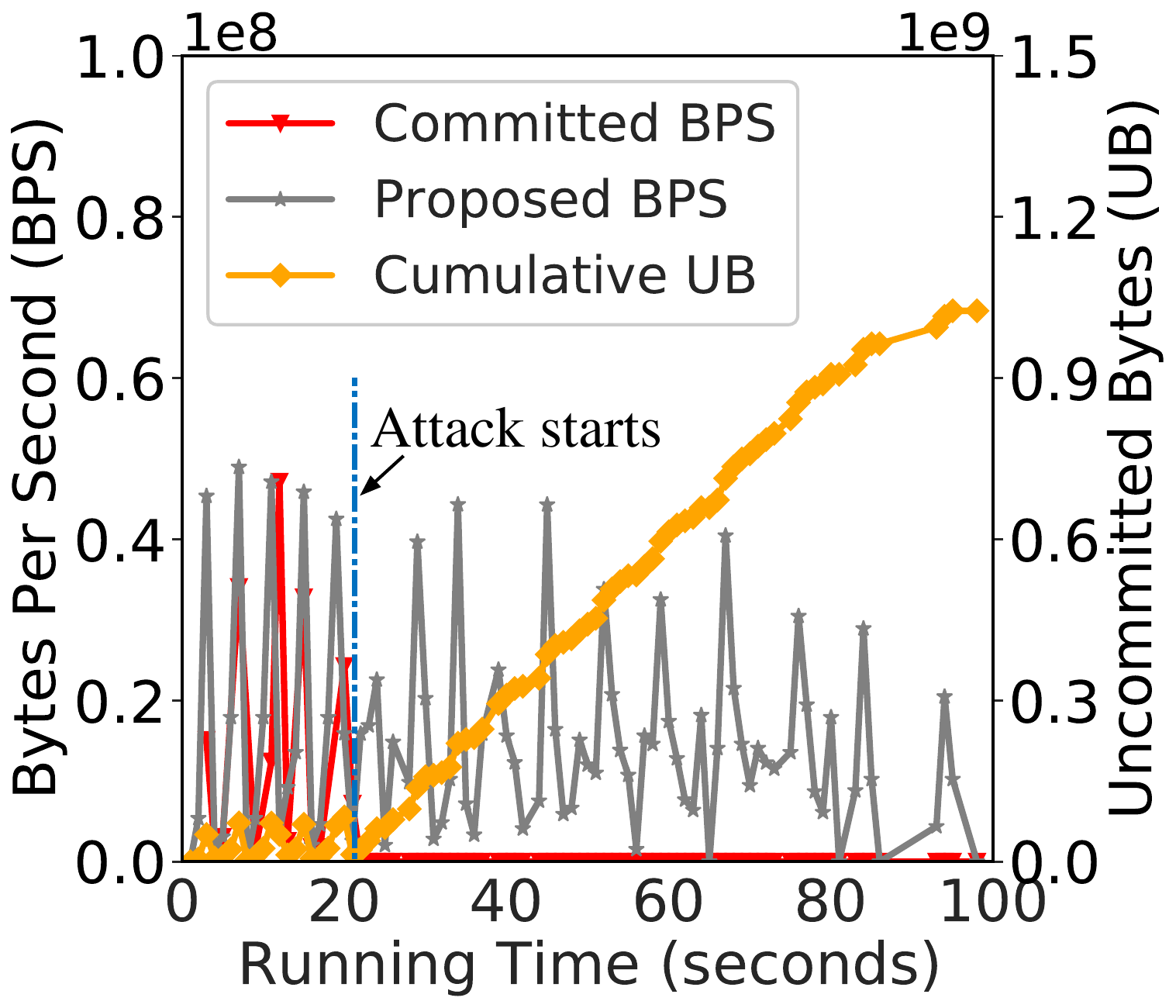}
        \caption{Mysticeti under attack}
    \end{subfigure}
    \caption{Sailfish and Mysticeti under inflation attacks: after the attack starts (at $20$s), the committed BPS drops to $0$ while the proposed BPS continues, causing a sharp increase in uncommitted data maintained in the DAG-based mempool.
    }
    \label{fig-attack-eval}
\end{figure}

\para{Attack evaluation and estimation} We implement the proposed inflation attack on Sailfish and Mysticeti and evaluate its effectiveness. We deploy $n=10$ nodes and clients to submit transactions at a rate of 50,000 tx/s (see \S~\ref{subsec-inflation-tolerance} for more details). We focus on three metrics: (i) committed bytes per second (BPS), which depicts transaction throughput; (ii) proposed BPS, which depicts the proposed vertices per second; and (iii) cumulative uncommitted bytes (UB), which depicts the uncommitted vertices in the mempool. \Cref{fig-attack-eval} illustrates our evaluation results: the attack starts at 20s, after which the throughput (depicted by the committed BPS) drops to 0, while nodes continue proposing new DAG vertices (depicted by the proposed BPS), leading to a sharp increase in uncommitted vertices (depicted by cumulative UB). This demonstrates that it is possible to launch an inflation attack to exhaust nodes' resources and eventually cause mempool explosions.

With the data observed in our later experiments (\S~\ref{sec-evaluation}), we estimate the time required to conduct a successful inflation attack. For example, Mysticeti can achieve over 400 KTps, indicating that a node needs to handle around 0.2GB of data per second. Assuming that a correct node uses 128GB of memory (as suggested by the Sui blockchain~\cite{sui-validator-config}) to maintain its mempool, it would take the adversary
approximately 11 minutes to exhaust a correct node's memory resource.

\section{Towards Escaping Mempool Explosions: Attempts} \label{sec-attempt}
Before presenting \sysname, we discuss several natural attempts, inspired by prior protocols and practical systems, to prevent mempool explosions in DAG-based BFT protocols.

\para{Flow control mechanisms}
One approach is to cap the number of outstanding rounds, analogous to PBFT's watermark mechanism~\cite{pbft}. A node would stop accepting or creating vertices when the number of uncommitted rounds reaches a fixed limit. This policy \emph{bounds the DAG, but it cannot guarantee progress}. PBFT uses watermarks to bound requests outside the active sequence-number window while nodes can still complete agreement for requests already inside the window. In contrast, a DAG protocol represents votes as edges from later vertices. If the DAG reaches the limit before any leader obtains a committing quorum, stopping vertex creation also prevents the protocol from producing the votes needed to commit the existing backlog. The node therefore cannot free space, even after the network becomes synchronous.

\para{Backpressure mechanisms}
Another attempt is to slow mempool growth through backpressure mechanisms~\cite{beyer2016site, pbft, microsoft2026throttling, amazon-load-shedding, tendermint-timeouts}, which reduce the workload when the system becomes overloaded or fails to make progress. For example, nodes may admit fewer client transactions, create smaller vertices, or pace round advancement using longer timeouts, as in classical BFT protocols~\cite{pbft,hotstuff,jolteon}. Such backpressure is valuable operationally: it reduces the growth rate of the uncommitted DAG and gives a transient disruption more time to recover. However, it \emph{does not bound the resources consumed by the uncommitted vertices}. As long as nodes continue creating vertices, an adversary can extend the pre-GST asynchronous period until the backlog exceeds any fixed resource budget. If nodes eventually stop creating vertices, the mechanism reduces to the flow control approach above and may prevent the votes required for commitment.
Moreover, backpressure mechanisms can introduce additional attack surfaces. If backpressure is triggered by ordering failures, malicious leaders may deliberately withhold leader vertices to repeatedly activate backpressure even after GST, allowing them to degrade throughput or increase latency.

\para{Design implication}
Preventing mempool explosions requires more than controlling transaction admission or slowing DAG growth. A successful defense must satisfy two requirements simultaneously: it must bound the resources consumed by uncommitted vertices, and it must preserve the protocol's ability to order and commit those vertices. This observation motivates \sysname's single-shot, ACS-based fallback mechanism.

\section{The \sysname Protocol} \label{sec:overview}

\subsection{Building Blocks} \label{sec-building-block}
\noindent \textbf{Agreement on common subset.}
\sysname uses agreement on common subset (ACS) as a building block. In an ACS, each node $N_i$ invokes $\ACSPropose_i(m)$ to propagate its proposal $m$ and outputs a common set of proposals $V$ via $\ACSDecide_i(V)$. An ACS protocol satisfies the following properties~\cite{fin}:

{
\makeatletter
\def\@listi{\leftmargin10pt \labelwidth\z@ \labelsep5pt}    
\makeatother
\begin{itemize}[nosep]
    \item \textbf{Validity:} If a correct node $\node{i}$ outputs $\ACSDecide_i(V)$, then $|V|\geq n-f$ and $V$ includes proposals from at least $n-2f$ correct nodes who invoke $\ACSPropose$.
    \item \textbf{Agreement:} If a correct node $\node{i}$ outputs $\ACSDecide_i(V)$, then every other correct node $\node{j}$ outputs $\ACSDecide_j(V)$.
    \item \textbf{Termination:} If every correct node $\node{i}$ calls $\ACSPropose_i(m)$, then $\node{i}$ outputs $\ACSDecide_i(V)$.
\end{itemize}
}

\subsection{System Overview}\label{sec:system-overview}
\sysname's key insight is that, unlike DAG-based BFT protocols' optimistic paths, whose retained state can grow continuously while ordering is stalled, a single-shot agreement (such as ACS realized using FIN~\cite{fin}) operates on a fixed set of at most $n$ proposals. It therefore provides a separate path that does not require creating an unbounded sequence of new DAG vertices to commit the backlog.

\begin{figure}[t]
    \setlength\abovecaptionskip{0.1\baselineskip}
    \centering
    \includegraphics[width=0.85\linewidth]{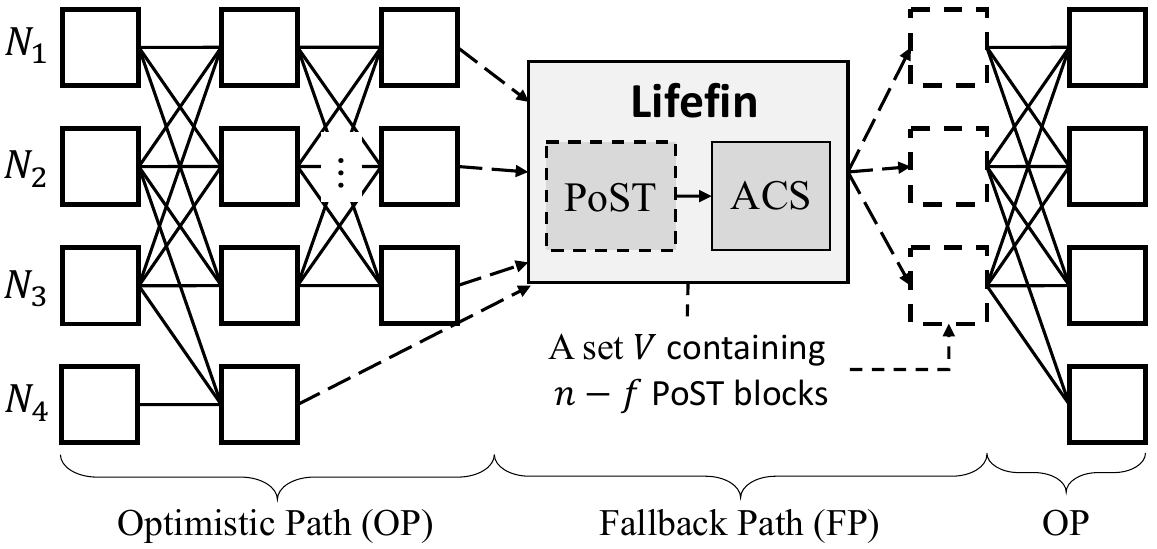}
    \caption{The overview of \sysname: In the optimistic path, nodes disseminate, commit, and order DAG vertices using the underlying DAG protocol. Nodes suspecting inflation attacks or resource exhaustion announce readiness, and after a global activation certificate forms, all correct nodes switch to the fallback path and use ACS to commit and order DAG vertices and their causal histories with bounded resources.}
    \label{fig:overview}
\end{figure}

Based on this observation, \sysname incorporates an ACS protocol into the underlying DAG-based protocol. As the ACS protocol can decide on a set of proposals (i.e., vertices in the DAG), nodes in \sysname can commit these decided vertices with bounded resources. By carefully applying the underlying rules of the original DAG protocols to the committed vertices, \sysname ensures that all correct nodes order historical vertices of the DAG consistently.
Below, we describe the main components of \sysname, \textit{ACS-based fallback} and \textit{Proof-of-STuck (PoST)}, as shown in \Cref{fig:overview}.

\para{ACS-based fallback}
In a \sysname-empowered DAG protocol, nodes primarily run the underlying DAG protocol in the optimistic path but activate an ACS-based fallback mechanism to switch to the fallback path when necessary.

Fallback detection is local, but ACS activation is global. We consider two locally configurable fallback conditions: (i) a node approaches a configured resource limit, or (ii) after receiving a valid $\mathsf{READY}$ message for its current fallback view, the node fails to commit DAG vertices for a long timeout period (e.g., 1 hour). The resource limit is configured to preserve sufficient working storage for fallback execution. These conditions guarantee not only that validators can escape a mempool explosion but also that, after GST, the adversary cannot arbitrarily trigger fallback to degrade performance, since honest validators can commit transactions regularly. A node satisfying either condition broadcasts at most one signed $\mathsf{READY}$ message in that fallback view. It may stop creating new vertices to preserve resources, but it continues processing bounded committing evidence and may return to the optimistic path if it catches up.

Signatures from $2f+1$ distinct nodes on $\mathsf{READY}$ messages for the same fallback view form an \emph{activation certificate}. It contains signatures from at least $f+1$ correct nodes and is disseminated by every correct recipient. Upon receiving the certificate, a correct node irrevocably activates that fallback view, stops creating optimistic-path vertices, and constructs an ACS proposal. 
Separating fallback detection from activation prevents an isolated suspicion from starting an ACS instance that other correct nodes may never join. A node that locally meets a fallback condition during a transient disruption enters only the preparatory phase and can return to the optimistic path if commitment resumes. Only an activation certificate makes the fallback transition irrevocable and coordinates all correct nodes on the same fallback view.

Upon activating a fallback view, a correct node invokes $\ACSPropose$ with its ACS proposal and waits for $\ACSDecide$ to return a set $V$. Since one correct recipient of an activation certificate eventually disseminates it to all correct nodes, every correct node eventually completes the same ACS instance, either by participating or, if the instance has already reached a decision, through ACS agreement.

After deciding $V$, all nodes deterministically select a fallback leader vertex (see \S~\ref{sec:secure-sailfish} and \S~\ref{sec:secure-mysticeti}) and then use $V$ as the one-time transition parent set for their first post-fallback vertex (i.e., the transition vertex). A correct node must create this transition vertex before sending $\mathsf{READY}$ in the next fallback view, ensuring that successive fallback instances select strictly increasing rounds and never redefine the same leader round. The ACS proposals in $V$ may wrap vertices from different rounds, while normal consecutive-round parent rules resume afterward. The nodes then advance the fallback-view counter and return to the optimistic path.

\para{Proof-of-STuck}\label{sec-post-blocks}
The ACS protocol does not inherently ensure the validity of proposals: malicious nodes can $\ACSPropose$ invalid vertices that contain nonexistent causal histories or come from stale rounds.
If an invalid leader vertex is selected by the ACS-based fallback mechanism, nodes may fail to retrieve the uncommitted history or risk committing vertices inconsistently. Briefly, in uncertified DAGs, the adversary can construct an ACS proposal containing phantom histories; in certified DAGs, it can use stale proposals to cause inconsistent commitment (see the full paper~\cite[Appendix~E]{zhang2025lifefin} for more details).

A straightforward solution to these issues is to validate ACS proposals (e.g., synchronize missing ancestors, check their freshness, and validate the corresponding DAG vertices) during the ACS protocol. However, this approach might require substantial modifications to the adopted ACS protocol, depending on the DAG structure.

To address these limitations, \sysname introduces the \emph{Proof-of-STuck (PoST)} block. At a high level, a PoST block encapsulates verifiable information about the DAG state observed by a node, indicating that the node is unable to commit vertices in its latest round.
PoST blocks are constructed using a simple \texttt{Propose} and \texttt{Vote} scheme.
Specifically, after receiving an activation certificate, a node freezes optimistic-path vertex creation, constructs a PoST block containing its fallback view and latest created vertex, and collects signatures from a quorum of $2f+1$ nodes. These signatures certify the validity and availability of that fresh PoST block.
The counter-signed PoST blocks are then used as ACS proposals.

By introducing PoST blocks, \sysname minimizes the coupling between ACS and the underlying DAG structure. Specifically, a PoST block contains a certificate that validates its authenticity. When performing the ACS, nodes simply verify each PoST proposal using its certificate rather than its underlying DAG history. In the following sections, we show how to integrate this framework into two representative protocols: Sailfish~\cite{sailfish} (\S~\ref{sec:sailfish-sysname}) and Mysticeti~\cite{mysticeti} (\S~\ref{sec:mysticeti-sysname}).

\para{Security intuition behind \sysname}
\sysname enables nodes to output a set of PoST blocks $V$ with bounded resources. In our two instantiations, nodes reserve enough resources to perform the ordering task through the ACS-based fallback mechanism, avoiding mempool explosions and preventing the exploited inflation attacks. Regarding liveness, the fallback output contains at least $n-f \geq q_d$ references, allowing nodes to return to the optimistic path. For safety, \sysname carefully selects vertices from the ACS output for commitment and ensures a smooth transition to prevent different nodes from committing inconsistent vertices. In brief, \sysname may select a new leader vertex to replace the originally predefined one, but it always follows the underlying committing rules to order and commit DAG vertices. We present the following proposition, which is used in the proof sketch for safety, and defer its full proof to the full paper~\cite[Appendix~B.1 and Appendix~C.1]{zhang2025lifefin}.

\begin{proposition}\label{proposition-highest-leader-not-commit}
    In each \sysname instantiation considered in this paper, if $r^*$ is the highest round number among the PoST blocks in the output $V$ of a fallback instance, then the predefined round $r^*$ leader vertex must not be committed before the fallback instance terminates.
\end{proposition}

\section{Sailfish-\sysname}\label{sec:sailfish-sysname}
\subsection{The Sailfish Protocol}\label{sec:sailfish-protocol-review}
\begin{figure}
    \setlength\abovecaptionskip{0.1\baselineskip}
    \centering
    \includegraphics[width=0.85\linewidth]{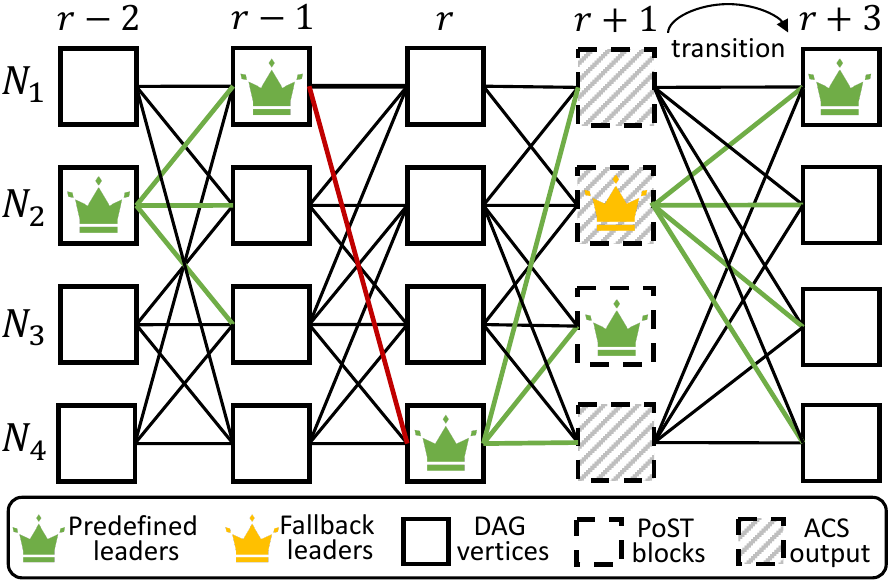}
    \caption{
    Sailfish-\sysname's round-based DAG from a node's local view: a leader vertex connected by green lines (\raisebox{0.5ex}{\textcolor{YellowGreen}{\rule{3mm}{1pt}}}) is committed via the direct committing rule; a leader vertex connected by red lines (\raisebox{0.5ex}{\textcolor{BrickRed}{\rule{3mm}{1pt}}}) is committed via the indirect committing rule. All DAG vertices are disseminated via RBC.
    }
    \label{fig:overview-sailfish}
\end{figure}

Sailfish is a recent \textit{certified} DAG-based protocol designed to optimize latency for early-stage DAG-based protocols~\cite{narwhal, bullshark}. \Cref{fig:overview-sailfish} illustrates its overview. We first review its DAG construction and committing rules.

\para{DAG construction}
The DAG in Sailfish consists of regularly created vertices identified by the round number and the creator. In each round $r$, each node $\node{i}$ contributes to the construction of the DAG using a reliable broadcast (RBC)~\cite{abraham2021good, dasrbc} to propagate a certified vertex $v_i^r$ containing a list of transactions, $2f+1$ references (hash digests) to round $r-1$ vertices, and a quorum of $2f+1$ signatures.

\para{Leader vertices}
Each round $r$ consists of a \textit{predefined} leader vertex $v_L^r$. The predefined leader vertex $v_L^r$ is created by a designated leader $L_r$, which is selected via a method based on the round number $r$ (e.g., round-robin).
In Sailfish, the creation of leader vertices adheres to the following restriction:
\begin{restrict}\label{restrict-sailfish-leader-vertex}
    When creating a new leader vertex $v_L^{r+1}$, $L_{r+1}$ includes either the round $r$ leader vertex $v_L^r$ in its references or a $2f+1$ quorum of messages from other nodes indicating that they do not reference $v_L^r$ in their round $r+1$ vertices.
\end{restrict}

\para{Committing rules} Driven by the leader vertices, each node employs the underlying committing rules to order and deliver DAG vertices (\Cref{fig:sailfish-fallback-commit}). The protocol uses two committing rules: direct commit and indirect commit.

\textit{(1) Direct committing rule:} A leader vertex $v_L^r$ is directly committed by node $\node{i}$ as soon as $\node{i}$ either observes $2f+1$ first messages of the RBC for round $r+1$ vertices with connections to $v_L^r$, or RBC-delivers $2f+1$ such vertices (\Cref{fig:sailfish-fallback-commit}, $\aln$~\ref{step:sailfish-direct-commit-start}-\ref{step:sailfish-direct-commit}). For instance, in \Cref{fig:overview-sailfish}, round $r-2$ leader vertex $v_L^{r-2}$ and round $r$ leader vertex $v_L^r$ are committed via the direct committing rule.

\textit{(2) Indirect committing rule:} Some leader vertices may not receive enough (i.e., $2f+1$) connections from the subsequent round in time and therefore fail to be directly committed. Sailfish deploys an indirect committing rule to commit these failed leader vertices (\Cref{fig:sailfish-fallback-commit}, $\aln$~\ref{step:sailfish-indirect-commit-start}-\ref{step:sailfish-indirect-commit-end}). Specifically, once a round $r$ leader vertex $v_L^r$ is directly committed, $\node{i}$ checks whether there is a path between round $r$ leader vertex and round $r-1$ leader vertex. If so, round $r-1$ leader vertex is indirectly committed, and the mechanism is recursively restarted from round $r-1$ until it reaches a round $r'\text{\textless}r$ in which the round $r'$ leader vertex $v_L^{r'}$ was previously directly committed. In \Cref{fig:overview-sailfish}, round $r-1$ leader vertex $v_L^{r-1}$ is indirectly committed through the directly committed $v_L^r$.

\begin{figure}[t]
    \setlength\abovecaptionskip{0.1\baselineskip}
    \scriptsize
	\begin{boxedminipage}[t]{\linewidth}
		\textbf{Variables:}
            {
                \makeatletter
                \def\@listi{\leftmargin17pt \labelwidth\z@ \labelsep5pt}    
                \makeatother
		\begin{itemize}[nosep]
                \item[] \gtext{$DAG_i[] - $ An array of sets of vertices (indexed by rounds)}
                \item[] $v_{fb} \gets 0$ \Comment{the monotonically increasing fallback-view number}
                \item[] $\mymode \gets OP$ \Comment{$OP$: optimistic; $PP$: preparatory; $FP$: activated fallback}
                \item[] struct PoST block $\post$ {
                    \begin{itemize}[nosep]
                        \item[] $\post.\view$ - the view number of the fallback mechanism
                        \item[] $\post.\vertex$ - the last vertex created by $\node{i}$
                        \item[] $\post.\creator$ - the node that creates $\post$
                        \item[] $\post.\mathit{activation}$ - an activation certificate for $\post.\view$
                        \item[] $\post.\cert$ - a quorum of $2f+1$ signatures for $\post$
                    \end{itemize}
                }
                \vspace{-1mm}
		\end{itemize}
		}
		\begin{algorithmic}[1]
			\Procedure{create\_new\_vertex}{$r$} \label{step:sf-construction-start}
                \If{$\mymode \neq OP$}
                \State \Return $\bot$   \label{step:sf-stop-dag-protocol}
                \EndIf

                \State \gtext{$\cdots$} \Comment{\gtext{return a round $r$ DAG vertex if processing in the optimistic path}}
                \EndProcedure
                \vspace{-0.2em}

                \Event{a local fallback condition after creating the required transition vertex} \label{step:ready-start}
                \If{$\mymode = OP$}
                \State $\mymode \gets PP$ \Comment{preserve resources without invoking ACS}
                \State if not already sent in this view, multicast $\sig{\mathsf{READY}(v_{fb})}_i$
                \EndIf
                \EndEvent \label{step:ready-end}
                \vspace{-0.2em}

                \Event{receiving valid $\sig{\mathsf{READY}(v_{fb})}_j$}
                \State start $T_{st}$ for view $v_{fb}$ if no such timer is running
                \State multicast the latest retained valid first-message commit witness, if any
                \EndEvent
                \vspace{-0.2em}

                \Event{receiving $2f+1$ distinct $\sig{\mathsf{READY}(v_{fb})}_*$} \label{step:activation-cert-start}
                \State $AC_{v_{fb}} \gets$ a certificate formed by the $2f+1$ signatures
                \EndEvent \label{step:activation-cert-end}
                \vspace{-0.2em}

                \Event{forming or receiving a valid $AC_u$} \label{step:switch_fallback}
                \If{$u<v_{fb}$} \State \Return \Comment{ignore a stale view} \EndIf
                \If{$u>v_{fb}$} \State \Call{catch\_up}{$AC_u$}; \Return \Comment{complete earlier views in order, then $AC_u$} \EndIf
                \State multicast $AC_u$ if not already disseminated
                \If{$\mymode \neq FP$}
                \State $\mymode \gets FP$ \Comment{irrevocably activate the fallback view}
                \State $r \gets \max \{ r \mid \exists v \in DAG_i[r] \land v.\creator = \node{i} \}$
                \State \Call{create\_new\_post}{$r, AC_u$} \Comment{snapshot after freezing}
                \EndIf
                \EndEvent
                \vspace{-0.2em}

                \Event{committing transactions while $\mymode=PP$} \label{step:sf_return_op_start}
                \State reset $T_{st}$
                \State $\mymode\gets OP$ 
                \EndEvent   \label{step:sf_return_op_end}
                \vspace{-0.2em}

                \Procedure{create\_new\_post}{$r, AC_{v_{fb}}$} \label{step:create_post_start}  \Comment{PoST \texttt{Propose}}
                \State $\post_{i}.\view \gets v_{fb}$   \Comment{bind the fresh state to the activated view}
                \State $\post_{i}.\vertex \gets$ $v$ s.t $v \in DAG_i[r] \land v.\creator = \node{i}$\strut
                \State $\post_{i}.\creator \gets \node{i}$
                \State $\post_i.\mathit{activation} \gets AC_{v_{fb}}$
                \State multicast $\post_{i}$
                \EndProcedure   \label{step:create_post_end}
                \vspace{-0.2em}

                \Event{receiving $\post_{j}$}   \label{step:vote_post_start}
                \State process $\post_j.\mathit{activation}$ as a received activation certificate
                \If{$\post_j.\view=v_{fb} \land \post_j.\mathit{activation}$ is a valid $AC_{v_{fb}}$}
                    \If{\Call{is\_post\_valid}{$\post_{j}$}} \Comment{creator uniqueness and vertex freshness} \label{step:sf-is-post-valid}
                    \State send $\sig{\post_{j}}_i$ to $\node{j}$ \Comment{sign $\post_j$ and send it back} \label{step:sf_vote-post}
                    \EndIf
                \EndIf
                \EndEvent   \label{step:vote_post_end}
                \vspace{-0.2em}

                \Event{receiving $2f+1$ $\sig{\post_{i}}_*$ for $\post_i$ while $\mymode=FP$ and $\post_i.\view=v_{fb}$} \label{step:acs_propose_start}
                \State $\post_{i}.\cert \gets$ a certificate formed from $2f+1$ $\sig{\post_{i}}_*$
                \State $\ACSPropose_i(v_{fb},\post_{i})$
                \EndEvent   \label{step:acs_propose_end}
                \vspace{-0.2em}
        \algstore{bkbreak1}
		\end{algorithmic}
    \end{boxedminipage}
    \caption{Sailfish-\sysname: vertex creation for node $\node{i}$; the code shown in \gtext{gray} comes from Sailfish~\cite[Figure 2]{sailfish}}
    \label{fig:sailfish-fallback-basic}
\end{figure}

\begin{figure}[t]
    \setlength\abovecaptionskip{0.1\baselineskip}
    \scriptsize
	\begin{boxedminipage}[t]{\linewidth}
		\textbf{Variables:}
		\begin{itemize}[nosep]
			\item[] \gtext{$\mathit{currentRound} \gets 1; \mathit{committedRound} \gets 0; \mathit{leaderStack} \gets \{\} $}
                \item[] \gtext{$\mathit{lastReferences} - $ The parent or transition references for a new block}
                \item[] $\post_L$ - the leader PoST block selected by the ACS instance
		\end{itemize}
		\begin{algorithmic}[1]
                \algrestore{bkbreak1}
                \begin{graytext}
                \Event{receiving $\geq2f+1$ first messages or RBC-delivering $\geq2f+1$ round $r+1$ vertices (set $\mathcal{S}$)} \label{step:sailfish-direct-commit-start}
                \State \Call{try\_commit}{$r, \mathcal{S}$}
                \EndEvent
                \vspace{-0.2em}

			\Procedure{try\_commit}{$r, \mathcal{S}$}
                \State $v \gets$ \Call{get\_leader\_vertex}{$r$}
                \State $\mathit{votes} \gets$ $\{ v' \in \mathcal{S} \mid$ \Call{path}{$v', v$}$\}$
                \If{$\mathit{votes} \geq 2f+1 \land \mathit{committedRound} < r$}
                \State \Call{commit\_leader}{$v$} \Comment{direct commit} \label{step:sailfish-direct-commit}
                \State \textcolor{black}{if $\mathcal{S}$ consists of first messages, retain $\mathit{votes}$ as the latest witness and multicast it if a current-view $\mathsf{READY}$ was received}
                \EndIf
                \EndProcedure
                \vspace{-0.2em}

                \Procedure{commit\_leader}{$v$} \label{step:commit-leader-start}
                \State $\mathit{leaderStack}$.push($v$)
                \State $r \gets v.\round-1$ \label{step:sailfish-indirect-commit-start}
                \State $v' \gets v$
                \While{$r > \mathit{committedRound}$}
                \State $v_s \gets$ \Call{get\_leader\_vertex}{$r$}
                    \If{\Call{path}{$v', v_s$}}
                    \State $\mathit{leaderStack}$.push($v_s$) \Comment{indirect commit} \label{step:sailfish-indirect-commit}
                    \State $v' \gets v_s$
                    \EndIf
                    \State $r \gets r-1$
                \EndWhile   \label{step:sailfish-indirect-commit-end}
                \State $\mathit{committedRound} \gets v.\round$
                \State \Call{order\_vertices}{$ $}  \Comment{any deterministic ordering algorithm}
                \EndProcedure   \label{step:commit-leader-end}
                \vspace{-0.2em}

                \end{graytext}

                \Event{$\ACSDecide_i(v_{fb},V)$} \label{step:acs-decide-start}
                \If{$\forall\post \in V, v_{fb} = \post.\view$}
                \State \Call{finalize\_fallback}{$V$}
                \EndIf
                \EndEvent   \label{step:acs-decide-end}
                \vspace{-0.2em}

                \Procedure{finalize\_fallback}{$V$} \label{step:finalize-fallback-start}
                \State $r_{fb} \gets \max \{\post.\vertex.\round \mid \post \in V\}$ \label{step:set-acs-round}
                \If{$\exists \post' \in V:$ $\post'.\vertex.\creator = L_{r_{fb}}\land \post'.\vertex.\round = r_{fb}$}  \label{step:finalize-fallback-predefine-0}
                \State $\post_L \gets \post'$   \label{step:finalize-fallback-predefine-1}
                \Else   \label{step:finalize-fallback-newdefine-0}
                \State $\post_L \gets$ deterministically select a $\post'\in V$ s.t. $\post'.\vertex.\round = r_{fb}$ \label{step:finalize-fallback-newdefine-deterministic}
                \State $r \gets r_{fb}-1$ \label{step:finalize-fallback-choose-define-leader-0}
                \State $v_L \gets \bot$ \Comment{the last round leader vertex chosen in OP}
                    \If{$r > \mathit{committedRound} \land$
                    $\exists v' \in \cup_{\post\in V}\{\post.\vertex.\parents\}$ s.t. $v'.\creator=L_{r}$
                    }
                    \State $v_L \gets v'$ \label{step:finalize-fallback-choose-define-leader-1}
                    \EndIf
                \If{$v_L \neq \bot$}    \label{step:fallback-commit-predefine-leader}
                \State \Call{commit\_leader}{$v_L$} \label{step:sailfish-fallback-direct-before-post} \Comment{direct commit}
                \EndIf  \label{step:finalize-fallback-newdefine-1}
                \EndIf
                \State $\mathit{L_{r_{fb}}} \gets \post_{L}.\creator$
                \State \Call{commit\_leader}{$\post_L$} \label{step:fallback-commit-new-leader} \Comment{direct commit}
                \State $\mathit{lastReferences \gets V}$ \label{step:fallback-commit-enter-op}
                \State $\mathit{currentRound \gets r_{fb}+2}$ \Comment{graceful transition}  \label{step:fallback-commit-graceful-transition}
                \State clear temporary state and timers for view $v_{fb}$
                \State $v_{fb} \gets v_{fb}+1$ \Comment{complete this fallback view}
                \State $\mymode \gets OP$ \Comment{switch to the optimistic path}
                \EndProcedure   \label{step:finalize-fallback-end}
                \vspace{-0.2em}
		\end{algorithmic}
    \end{boxedminipage}
    \caption{Sailfish-\sysname: vertex commitment for node $\node{i}$; the code shown in \gtext{gray} comes from Sailfish~\cite[Figure 3]{sailfish}}
    \label{fig:sailfish-fallback-commit}
\end{figure}

\subsection{Securing Sailfish with \sysname} \label{sec:secure-sailfish}
\sysname can be integrated into Sailfish to escape mempool explosions. During this process, nodes construct PoST blocks and perform an instance of ACS to commit the backlog.

\para{Global activation and PoST construction (\Cref{fig:sailfish-fallback-basic}, $\aln$~\ref{step:ready-start}-\ref{step:acs_propose_end})}
When a local fallback condition holds, $\node{i}$ enters the preparatory phase and multicasts at most one signed $\mathsf{READY}(v_{fb})$ vote for that view without invoking ACS ($\aln$~\ref{step:ready-start}-\ref{step:ready-end}). The first valid current-view $\mathsf{READY}$ message starts the no-commit timer, which is reset only by actual commit progress. During preparation, $\node{i}$ stops creating vertices to preserve resources ($\aln$~\ref{step:sf-stop-dag-protocol}) but continues processing bounded committing evidence. Upon receiving $\mathsf{READY}(v_{fb})$, a node multicasts its latest retained valid first-message commit witness, if any, and relays any such witness obtained later. A preparatory node verifies the relayed witness using Sailfish's original direct-commit rule and can therefore commit without waiting for the corresponding Byzantine RBC instances to terminate. Each witness contains only $2f+1$ bounded first messages.
It may resume the optimistic path if commitment advances, although its issued $\mathsf{READY}$ signature remains valid ($\aln$~\ref{step:sf_return_op_start}-\ref{step:sf_return_op_end}).

Any node can combine $2f+1$ current-view $\mathsf{READY}$ signatures into an activation certificate $AC_{v_{fb}}$ ($\aln$~\ref{step:activation-cert-start}-\ref{step:activation-cert-end}) and broadcast it. Therefore, once one correct node activates view $v_{fb}$, every correct node eventually receives the same activation certificate, and each node that has not already completed the view enters it.

Upon receiving $AC_{v_{fb}}$, $\node{i}$ atomically stops optimistic-path vertex creation and only then constructs $\post_i$ from its latest created vertex ($\aln$~\ref{step:switch_fallback}-\ref{step:create_post_end}). Hence, $AC_{v_{fb}}$ determines when fallback starts, while the fresh $\post_i$ defines the DAG state proposed to ACS. A PoST carrying a valid current-view activation certificate is treated as receipt of that certificate: the recipient first activates fallback and constructs its own PoST (if not yet), and only then validates and signs the received PoST ($\aln$~\ref{step:vote_post_start}-\ref{step:vote_post_end}). A stale PoST is ignored, while a future-view activation certificate, including one carried by a PoST, invokes \Call{catch\_up}{$AC_u$}, which obtains and finalizes the decided ACS outputs for all unfinished views below $u$ in order and then processes $AC_u$. This is safe because $AC_u$ contains at least $f+1$ correct signers, each of which reaches view $u$ only after completing every earlier view. A correct node signs at most one PoST for each creator in each view. After collecting $2f+1$ signatures, $\node{i}$ initiates an ACS instance using $\post_{i}$ as its proposal ($\aln$~\ref{step:acs_propose_start}-\ref{step:acs_propose_end}).

\para{Commit DAG from ACS (\Cref{fig:sailfish-fallback-commit}, $\aln$~\ref{step:acs-decide-start}-\ref{step:finalize-fallback-end})}
The ACS instance decides a common set of PoST blocks $V$, based on which nodes consistently commit the pending DAG vertices ($\aln$~\ref{step:acs-decide-start}-\ref{step:acs-decide-end}). At a high level, they determine a leader vertex $\post_L$ in the updated round $r_{fb}$ from $V$ and commit its causal history using the underlying committing rules.

Obviously, there should be no conflicting order between $\post_L$ and any previous round $r'\text{\textless}r_{fb}$ leader vertex that may have been committed using the underlying DAG mechanism before the fallback mechanism. To achieve this, $\post_L$ is selected from the PoST blocks in $V$ with the \textit{highest} round number. When $V$ contains a predefined round $r_{fb}$ leader vertex, $\post_L$ is set to that predefined leader vertex ($\aln$~\ref{step:finalize-fallback-predefine-0}-\ref{step:finalize-fallback-predefine-1}). However, there is no guarantee that $V$ contains the predefined round $r_{fb}$ leader vertex, given that $|V| \leq n$ and the leader node $L_{r_{fb}}$ could be malicious or stuck in a round $r'\text{\textless}r_{fb}$ withiout creating the round $r_{fb}$ leader vertex. For example, in \Cref{fig:overview-sailfish}, the ACS output $V$ excludes the predefined round $r+1$ leader vertex $v_3^{r+1}$ (where $r_{fb}=r+1$). In this case, $\post_L$ is chosen using a deterministic algorithm ($\aln$~\ref{step:finalize-fallback-newdefine-deterministic}), e.g., based on its digest.

However, this newly selected leader vertex $\post_L$ might have no path to the previous round $r_{fb}-1$ leader vertex because the creation of $\post_L$ does not follow \Cref{restrict-sailfish-leader-vertex}. This would lead to inconsistent commitments among nodes. For example, in \Cref{fig:overview-sailfish}, the round $r$ leader vertex $v_L^r$ can be directly committed by some nodes that received all $2f+1$ round $r+1$ vertices (i.e., $v_1^{r+1}$, $v_3^{r+1}$, and $v_4^{r+1}$) connecting to it, while the other nodes that have not received $v_3^{r+1}$ skip it through the indirect committing rule applied to $\post_L$. Thus, nodes need to identify the predefined round $r_{fb}-1$ leader vertex $v_L^{r_{fb}-1}$ to which the blocks in $V$ have a path ($\aln$ \ref{step:finalize-fallback-choose-define-leader-0}-\ref{step:finalize-fallback-choose-define-leader-1}).

After selecting $\post_L$, $\node{i}$ commits $v_L^{r_{fb}-1}$ (if it exists) and $\post_L$ via the underlying \Call{commit\_leader}{$ $} function ($\aln$~\ref{step:fallback-commit-predefine-leader}-\ref{step:fallback-commit-new-leader}). It then uses the common output $V$ for its first new vertex, updates states, and switches to the optimistic path ($\aln$~\ref{step:fallback-commit-enter-op}-\ref{step:finalize-fallback-end}).
The graceful transition advances nodes to $r_{fb}+2$ after committing blocks in the fallback path ($\aln$~\ref{step:fallback-commit-graceful-transition}). 
This is because some nodes might have constructed round $r_{fb}+1$ vertices not included in $V$; skipping that round prevents nodes from creating conflicting vertices there and from violating no-equivocation assumptions of DAG protocols. 

\para{Security proof}
We defer the formal proof to the full paper~\cite[Appendix~B]{zhang2025lifefin}. For safety, we only need to consider the commitment of the round $r_{fb}$ leader vertex, as Lifefin may change the leader vertex only in that round. By \Cref{proposition-highest-leader-not-commit}, the predefined round $r_{fb}$ leader vertex cannot be committed before the fallback instance terminates. ACS agreement and deterministic leader selection then ensure that all correct nodes commit the same $\post_L$ in round $r_{fb}$, preventing a conflicting leader from being committed in that round. For liveness, stalled commitment eventually causes an activation certificate to form, which is then received by all correct nodes. Correct nodes therefore either participate in the same terminating ACS instance or obtain its common decision by ACS agreement. Every correct node installs the same decision of at least $n{-}f$ references with which to move to a new round and propose new vertices.

\section{Mysticeti-\sysname}\label{sec:mysticeti-sysname}
\subsection{The Mysticeti Protocol}\label{sec:mysticeti-protocol-review}
\begin{figure}
    \setlength\abovecaptionskip{0.0\baselineskip}
    \centering
        \includegraphics[width=0.85\linewidth]{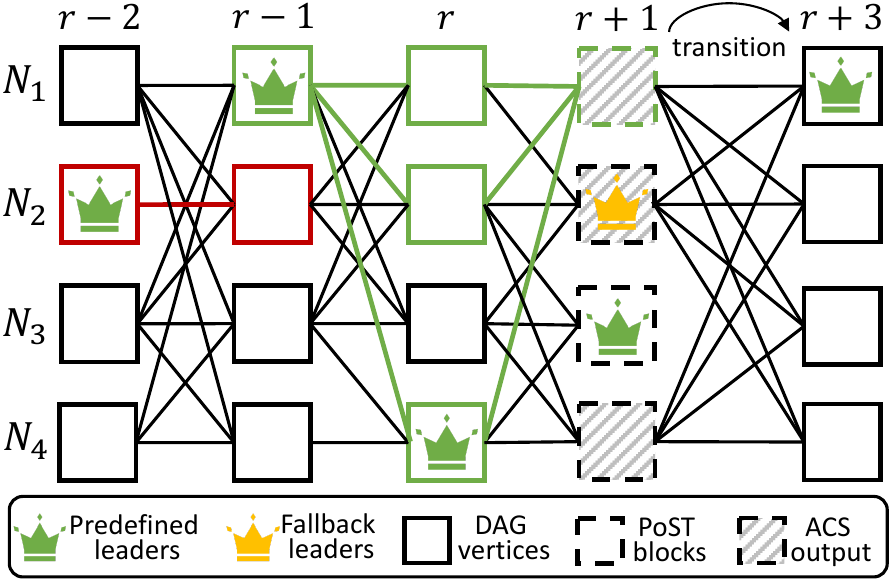}
        \caption{Mysticeti-\sysname's DAG from a node's view: The \textcolor{BrickRed}{red} pattern (\fcolorbox{BrickRed}{white}{\hspace{0.5mm}\vspace{1.5mm}} and \raisebox{0.5ex}{\textcolor{BrickRed}{\rule{3mm}{1pt}}}) represents a skip pattern on $v_2^{r-2}$. The \textcolor{YellowGreen}{green} pattern (\fcolorbox{YellowGreen}{white}{\hspace{0.5mm}\vspace{1.5mm}} and \raisebox{0.5ex}{\textcolor{YellowGreen}{\rule{3mm}{1pt}}}) in round $r$ represents a certificate pattern on $v_1^{r-1}$, and vertex $v_1^{r+1}$ is called a certificate for $v_1^{r-1}$. All DAG vertices are disseminated via best-effort broadcast. }
        \label{fig:overview-mysticeti}
\end{figure}

Mysticeti is an \textit{uncertified} DAG-based protocol that replaces RBC with best-effort broadcast to disseminate DAG vertices, achieving better confirmation latency (\Cref{fig:overview-mysticeti}).
We first review Mysticeti's DAG construction and committing rules.
The pseudocode for Mysticeti-\sysname is presented in the full paper~\cite[Appendix~D, Figures~11 and~12]{zhang2025lifefin}.

\para{DAG construction and leader vertices}
The Mysticeti DAG is composed of vertices that are created in rounds. Specifically, for every round $r$, $\node{i}$ propagates a vertex consisting of a list of transactions and $2f+1$ references to round $r-1$ vertices via best-effort broadcast; each reference contains the parent's round, creator, and digest, and the referenced parent is signed by its creator. Upon receiving $2f+1$ round $r$ vertices, $\node{i}$ moves to round $r+1$ and propagates a new vertex. All leader vertices are predefined and pre-ordered based on their round numbers, with a leader node $L_r$ designated to propose a round $r$ leader vertex $v_L^r$.

\para{DAG patterns}
Vertices in Mysticeti are not explicitly certified, that is, nodes do not need to acknowledge DAG vertices with their signatures during the DAG construction process. To prevent equivocation, Mysticeti implements implicit certification, in which nodes interpret DAG vertices using two patterns:
(i) \textit{skip pattern}, identifying a round $r$ vertex that is not explicitly connected by at least $2f+1$ vertices of round $r+1$. For instance, the leader vertex $v_2^{r-2}$ in \Cref{fig:overview-mysticeti} is identified as a \textit{skip pattern};
(ii) \textit{certificate pattern}, identifying a round $r$ vertex that is connected by at least $2f+1$ vertices from round $r+1$.
A vertex is considered \textit{certified} if a certificate pattern identifies it. Additionally, any subsequent vertex (in round $\text{\textgreater}r+1$) that includes such a pattern in its causal history is called a \textit{certificate} for the vertex. For instance, the vertex $v_1^{r-1}$ in \Cref{fig:overview-mysticeti} is identified by a \textit{certificate pattern} (from round $r$), and vertex $v_1^{r+1}$ is a certificate for $v_1^{r-1}$.

\para{Committing rules}
Mysticeti commits leader vertices by identifying DAG patterns. While running the underlying DAG protocol, nodes check whether a DAG pattern can be applied to each leader vertex, assigning the leader vertex one of two final statuses: $\tocommit$ or $\toskip$.

The full paper~\cite[Figure~12]{zhang2025lifefin} presents Mysticeti's commit rule. Specifically, every time $\node{i}$ receives a new valid vertex, it invokes the \Call{try\_decide}{$ $} function to try to decide previous leader vertices~\cite[Figure~12, Lines~45--57]{zhang2025lifefin}. Similarly, Mysticeti's committing rules consist of a direct decision rule and an indirect decision rule.

\textit{(1) Direct decision rule:} $\node{i}$ first tries to directly decide the round $r$ leader vertex $v_L^r$~\cite[Figure~12, Line~49]{zhang2025lifefin}, which leads to three results: (i) $v_L^r$ is $\tocommit$ if $\node{i}$ observes that $2f+1$ round $r+2$ vertices are certificates for $v_L^r$; that is, each round $r+2$ vertex references $\geq2f+1$ round $r+1$ vertices that connect to $v_L^r$; (ii) $v_L^r$ is $\toskip$ if $\node{i}$ observes a \textit{skip pattern} over the vertex; that is, at least $2f+1$ round $r+1$ vertices do not reference $v_L^r$; and (iii) otherwise, $v_L^r$ is marked as $\undecided$. For instance, in \Cref{fig:overview-mysticeti}, the round $r-1$ leader vertex $v_L^{r-1}$ is directly decided as $\tocommit$ since all round $r+1$ vertices are certificates for $v_L^{r-1}$.

\textit{(2) Indirect decision rule:} For any leader vertex $v_L^r$ marked as $\undecided$ via the direct decision rule, $\node{i}$ applies the indirect decision rule to it~\cite[Figure~12, Line~51]{zhang2025lifefin}. Specifically, $\node{i}$ searches for the first subsequent leader vertex $v_L^{r'}$ (where $r'\text{\textgreater} r+2$) that has been marked as either $\tocommit$ or $\undecided$. Depending on the status of $v_L^{r'}$ and the relevant DAG pattern, $v_L^r$ is assigned one of the following statuses: (i) $\tocommit$ if $v_L^{r'}$ is $\tocommit$ and causally references a \textit{certificate pattern} over $v_L^r$; (ii) $\toskip$ if $v_L^{r'}$ is $\tocommit$ but does not reference a \textit{certificate pattern} over $v_L^r$; and (iii) $\undecided$ if $v_L^{r'}$ is $\undecided$.

Using these two decision rules, all correct nodes will consistently mark the predefined leader vertices as either $\tocommit$ or $\toskip$. By applying any deterministic ordering algorithm to the $\tocommit$ leader vertices, nodes eventually order all DAG vertices consistently.

\subsection{Securing Mysticeti with \sysname} \label{sec:secure-mysticeti}
The integration of \sysname with Mysticeti closely mirrors its integration with Sailfish, involving the construction of PoST blocks and the execution of an ACS instance.

\para{Global activation and PoST construction \cite[Figure~11, Lines~5--44]{zhang2025lifefin}}
Mysticeti-\sysname uses the same local fallback conditions, view discipline, and global activation procedure as Sailfish-\sysname. It further needs to monitor both the amount of uncommitted data and the span between its current and committed rounds. The round-span guard is configured so that synchronizing the missing histories of all eligible PoST blocks, together with the remaining fallback state, fits in the capacity reserved below \membudget. During preparation, $\node{i}$ stops creating vertices but continues processing bounded committing evidence spanning four consecutive rounds. The first three rounds contain the direct-commit pattern, while a correct node that observes this pattern uses its next-round vertex as a carrier referencing the $2f+1$ certificate vertices used for the decision. Upon receiving such a carrier, $\node{i}$ synchronizes its direct-commit pattern and reconstructs the same direct-commit evidence (cf. \S~\ref{sec:mysticeti-protocol-review}).

A locally fallback-ready node multicasts at most one signed $\mathsf{READY}(v_{fb})$ vote but does not invoke ACS. A $2f+1$-signature activation certificate for $v_{fb}$ is disseminated to all correct nodes, and each node that has not already completed the view atomically stops optimistic-path vertex creation and constructs a fresh PoST block from its latest vertex~\cite[Figure~11, Lines~11--32]{zhang2025lifefin}. Thus, even a node that advances after sending $\mathsf{READY}$ proposes its state at activation rather than its earlier state. As in Sailfish-\sysname, receiving a PoST with a valid current-view activation certificate first activates fallback and constructs the recipient's fresh PoST. A future-view activation certificate instead invokes \Call{catch\_up}{$AC_u$}, which obtains and finalizes the decided ACS outputs for all unfinished views below $u$ in order and then processes $AC_u$.

Recall that in Mysticeti, the DAG vertices are not explicitly certified. A malicious node may construct a PoST block with a phantom vertex whose history vertices are never shared with correct nodes. When $\node{i}$ receives a PoST block $\post_{j}$, there is no guarantee that it can retrieve $\post_{j}$'s causal history. To prevent the ACS instance from outputting PoST blocks containing phantom vertices, $\post_j$ carries the creator's signature for every parent digest referenced by $\post_j.\vertex$, binding the parent's round, creator, and digest. A recipient rejects $\post_j$ unless it contains valid matching signatures from $2f+1$ distinct claimed parent creators. Moreover, if $\node{i}$ receives a $\post_{j}$ wrapping a vertex in a round equal to or smaller than $\node{i}$'s next round, it must synchronize the complete causal history of $\post_{j}$ before certifying it and retain that history until it can be safely garbage-collected~\cite[Figure~11, Lines~37--39]{zhang2025lifefin}. Every valid PoST also carries the activation certificate for its fallback view.
Note that the resources used for synchronization are bounded, as $\node{i}$ only needs to synchronize missing vertices within a limited range of round numbers (see the full paper~\cite[Appendix~C.2]{zhang2025lifefin}). This ensures that $\node{i}$ will not exhaust its resources after activating the fallback mechanism, enabling $\node{i}$ to terminate the fallback mechanism with bounded resources. As shown in the full paper~\cite[Appendix~C.1]{zhang2025lifefin}, this design is sufficient to guarantee that the causal history of any PoST block certified by $2f+1$ nodes is retained by at least one correct node, thereby ensuring data availability.

\para{Commit DAG from ACS \cite[Figure~12, Lines~58--74]{zhang2025lifefin}}
Committing vertices after the ACS instance in Mysticeti-\sysname is straightforward. Nodes use the same rule (cf. \S~\ref{sec:secure-sailfish}) to select a leader vertex $\post_L$ from the ACS output $V$, replace the predefined leader vertex from the same round with $\post_L$, and mark it as $\tocommit$. Nodes then run the underlying deterministic ordering rules and use the common output for the one-time transition to round $r_{fb}+2$.
Unlike Sailfish-\sysname, finalizing the fallback mechanism in Mysticeti-\sysname eliminates the need to commit at least one predefined leader vertex after the ACS instance. This is because Mysticeti's underlying DAG protocol does not require a leader vertex to connect to its predecessor from the previous round. Consequently, changing leader vertices does not cause conflicting orders.

The proof for Mysticeti-\sysname is similar to that for Sailfish-\sysname. We defer the formal security proof to~\cite[Appendix~C]{zhang2025lifefin}.

\section{Evaluation}\label{sec-evaluation}
In \sysname, nodes continue to efficiently disseminate, order, and commit vertices following the underlying DAG protocol in typical cases, thereby maintaining the original protocol's performance. The fallback mechanism is only triggered under adverse conditions.
Thus, our evaluation focuses solely on scenarios when fallback occurs and aims to answer the following questions:

{
\makeatletter
\def\@listi{\leftmargin10pt \labelwidth\z@ \labelsep5pt}    
\makeatother
\begin{itemize}[nosep]
    \item \textbf{Performance}: How does \sysname perform in terms of throughput and latency once fallback occurs? (\S~\ref{subsec-performance})
    \item \textbf{Scalability}: How well does \sysname scale with an increasing number of nodes $n$? (\S~\ref{subsec-scalability})
    \item \textbf{Inflation Resistance}: Can \sysname effectively resist inflation attacks? (\S~\ref{subsec-inflation-tolerance})
\end{itemize}
}

\para{Implementation details} We implement \sysname in Rust, integrating it into Sailfish~\cite{sailfish-code} and Mysticeti~\cite{mysticeti-code} to derive Sailfish-\sysname and Mysticeti-\sysname, respectively.
We adopt FIN~\cite{fin} as our ACS protocol and use the RBC protocol adopted by Narwhal~\cite{narwhal} because of its practicality. We use Tokio~\cite{tokio} for networking and ed25519-dalek~\cite{ed25519} for signatures.

\para{Baselines} We use unmodified Sailfish and Mysticeti as paired baselines to isolate the overhead of integrating \sysname. Their performance represents \sysname's optimistic path, whereas Sailfish-\sysname and Mysticeti-\sysname are measured when fallback is activated. Although Bullshark~\cite{bullshark} is conceptually related, both its optimistic and randomized fallback paths require continued DAG growth and thus do not resolve the resource-bounded issue studied here (\S~\ref{sec:related-work}). We thus do not treat Bullshark as a baseline for either path.

\para{Optimization} During our initial evaluation, we observed that Mysticeti-\sysname exhibited latency on the order of several seconds. We identified the cause as Mysticeti's original synchronization mechanism, where a node \emph{randomly} fetches missing vertices, introducing unbounded latency. In our implementation, we eliminate this overhead by replacing it with a deterministic synchronization mechanism that fetches missing vertices from all nodes. This ensures that correct nodes can receive missing vertices within one round-trip. Although this approach may introduce redundant messages, we emphasize that it does not meaningfully affect efficiency because fallbacks are rare in practice.

\para{Experimental setup} We evaluate all systems on AWS, using c5a.4xlarge EC2 instances spread across 5 regions (us-east-1, us-east-2, us-west-1, eu-west-1, and eu-west-2). Each instance provides 16 vCPUs, 32GB of RAM, and up to 10 Gbps of bandwidth and runs Ubuntu Server 20.04.

We deploy one client per node to send raw transactions, each of which consists of 512 random bytes.
To evaluate the performance of \sysname under adverse conditions, we configure nodes in the following experiments to trigger the fallback mechanism and commit DAG vertices in a specific round. We focus on several performance metrics, including \textit{throughput}---the number of transactions committed per second (Tps)---and \textit{end-to-end (e2e) latency}, measured from when clients send transactions to when nodes commit them. We use the default leader timeout set by the baselines, i.e., 5s in Sailfish and 1s in Mysticeti. We run each experiment for 240 seconds. In the following evaluation, we stop the transaction-workload experiments as soon as we observe a notable latency increase or verify that \sysname introduces minimal overhead. We also allow clients to begin issuing transactions after a short setup period (40s), which may add queuing delay to the e2e latency because transactions cannot be immediately packaged into vertices. We limit our experiments due to cost considerations.

\begin{figure*}[htbp]
    \setlength\abovecaptionskip{0.1\baselineskip}
    \centering
    \begin{subfigure}[t]{0.32\textwidth}
        \setlength\abovecaptionskip{-0.1\baselineskip}
		\centering
		\includegraphics[width=0.9\linewidth]{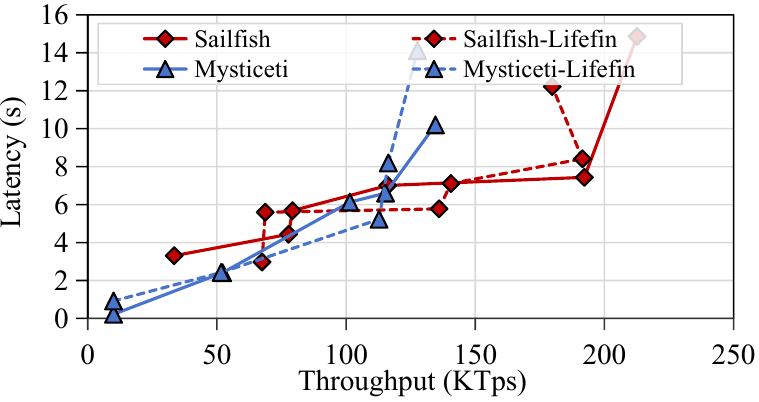}
		\caption{$n=4$}
		\label{fig-tps-latency-4}
	\end{subfigure}
	\begin{subfigure}[t]{0.32\textwidth}
            \setlength\abovecaptionskip{-0.1\baselineskip}
		\centering
		\includegraphics[width=0.9\linewidth]{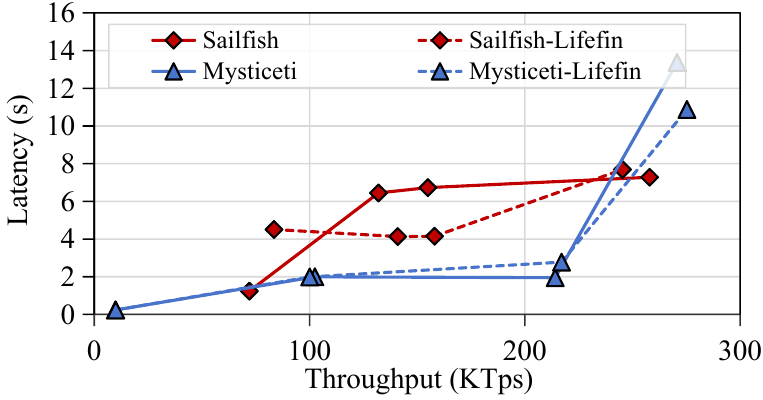}
		\caption{$n=10$}
		\label{fig-tps-latency-10}
	\end{subfigure}
	\begin{subfigure}[t]{0.32\textwidth}
            \setlength\abovecaptionskip{-0.1\baselineskip}
		\centering
		\includegraphics[width=0.9\linewidth]{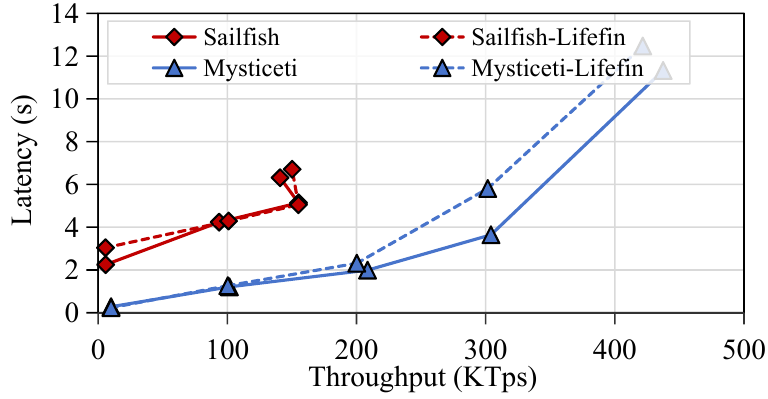}
		\caption{$n=20$}
		\label{fig-tps-latency-20}
	\end{subfigure}
    \caption{Throughput vs. end-to-end latency under various network sizes (no crash failures), where Sailfish-\sysname and Mysticeti-\sysname operate in adverse cases with fallback activated, while Sailfish and Mysticeti operate in good cases \label{fig-tps-latency}}
\end{figure*}

\begin{figure}[ht]
    \setlength\abovecaptionskip{0.1\baselineskip}
    \centering
    \begin{subfigure}[t]{0.235\textwidth}
        \setlength\abovecaptionskip{-0.1\baselineskip}
		\centering
        \includegraphics[width=\linewidth]{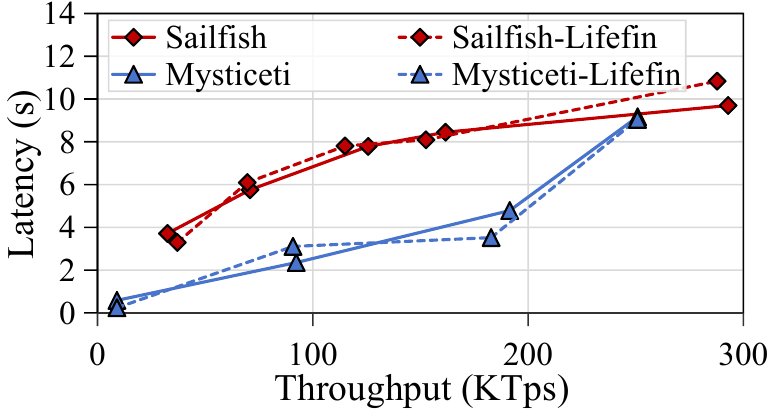}
        \caption{$n=10$, $f=1$}
    \end{subfigure}
    \begin{subfigure}[t]{0.235\textwidth}
        \setlength\abovecaptionskip{-0.1\baselineskip}
        \centering
        \includegraphics[width=\linewidth]{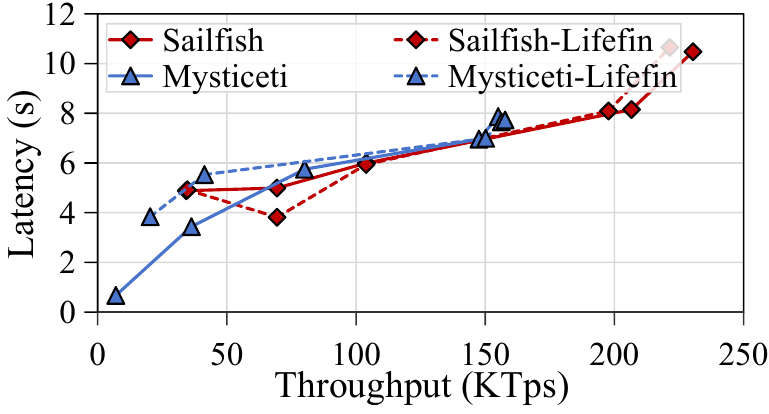}
        \caption{$n=10$, $f=3$}
    \end{subfigure}
    \caption{Throughput vs. end-to-end latency under varying crash failures, where \sysname operates in adverse cases}
    \label{fig-tps-latency-crash}
\end{figure}

\subsection{Performance}\label{subsec-performance}
We first evaluate the performance under different transaction workloads by running $n=10$ nodes.

\para{Fault-free performance} \Cref{fig-tps-latency-10} shows the throughput and e2e latency of all systems under different transaction workloads without crash faults. Sailfish-\sysname and Mysticeti-\sysname achieve throughput and latency comparable to those of vanilla Sailfish and Mysticeti, respectively, showing that \sysname maintains decent performance. For instance, Sailfish-\sysname achieves 245KTps with a 7.6s e2e latency, while Sailfish achieves 258KTps with a 7.3s e2e latency.
Moreover, in some cases, Sailfish-\sysname achieves even lower latency than Sailfish. This is because the ACS-based fallback mechanism eliminates the timeout (i.e., leader timeout) used to commit predefined leader vertices, enabling nodes to commit leader vertices within realistic message delays.

\para{Performance under crash faults} \Cref{fig-tps-latency-crash} illustrates a performance comparison under varying numbers of crash faults, $f=1$ and $f=3$. All systems experience performance degradation as the number of crash faults increases. Sailfish-\sysname and Mysticeti-\sysname exhibit performance-degradation trends similar to those of Sailfish and Mysticeti, respectively, showing comparable resilience to crash faults.

\begin{figure}[t]
    \setlength\abovecaptionskip{0.1\baselineskip}
    \centering
    \begin{subfigure}[t]{0.235\textwidth}
        \setlength\abovecaptionskip{-0.1\baselineskip}
		\centering
        \includegraphics[width=\linewidth]{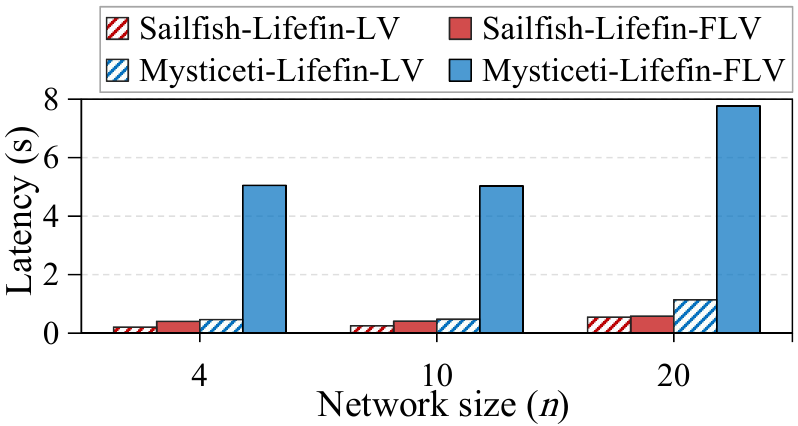}
        \caption{Various $n$ with $f=0$}
        \label{fig-network-leader-latency}
    \end{subfigure}
    \begin{subfigure}[t]{0.235\textwidth}
        \setlength\abovecaptionskip{-0.1\baselineskip}
        \centering
        \includegraphics[width=\linewidth]{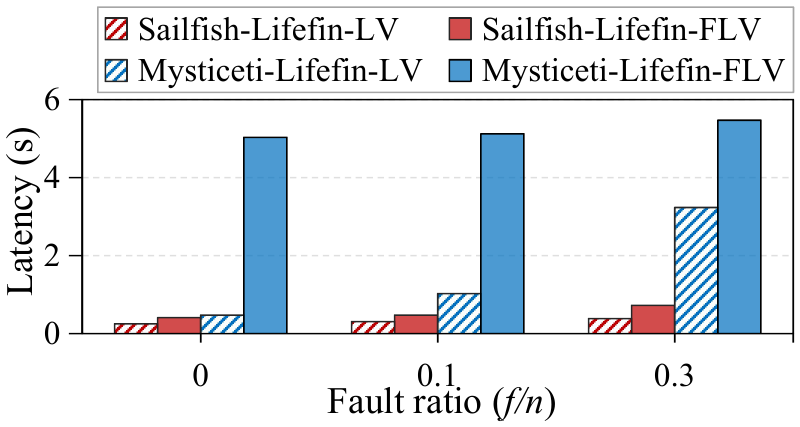}
        \caption{Various $f$ with $n=10$}
        \label{fig-fault-leader-latency}
    \end{subfigure}
    \caption{Leader vertex latencies when fallback occurs: LV = predefined leader vertex, FLV = fallback leader vertex}
    \label{fig-leader-latency}
\end{figure}

\para{Fallback performance} To evaluate the performance of the ACS-based fallback mechanism of \sysname, we conduct experiments to measure the commit latency of the fallback leaders (\Cref{fig-leader-latency}).
Thanks to the certified DAG structure used by Sailfish, Sailfish-\sysname commits fallback leader vertices quickly, with slightly higher latency than for predefined leader vertices. In contrast, in Mysticeti-\sysname, committing fallback leader vertices introduces more latency than committing predefined leader vertices. This overhead arises because its uncertified DAG structure requires nodes to synchronize missing vertices before they can finalize the fallback and propose new DAG vertices. Nevertheless, the overhead is acceptable and can be significantly amortized in practice because nodes proceed in the optimistic path most of the time, while fallback occurs rarely. \Cref{fig-tps-latency} and \Cref{fig-tps-latency-crash} demonstrate this minimal overhead after the fallback-path latency is amortized by the optimistic-path latency.

\begin{figure*}[htbp]
    \setlength\abovecaptionskip{0.0\baselineskip}
    \centering
    \begin{subfigure}[t]{0.325\textwidth}
        \setlength\abovecaptionskip{-0.1\baselineskip}
		\centering
		\includegraphics[width=0.9\linewidth]{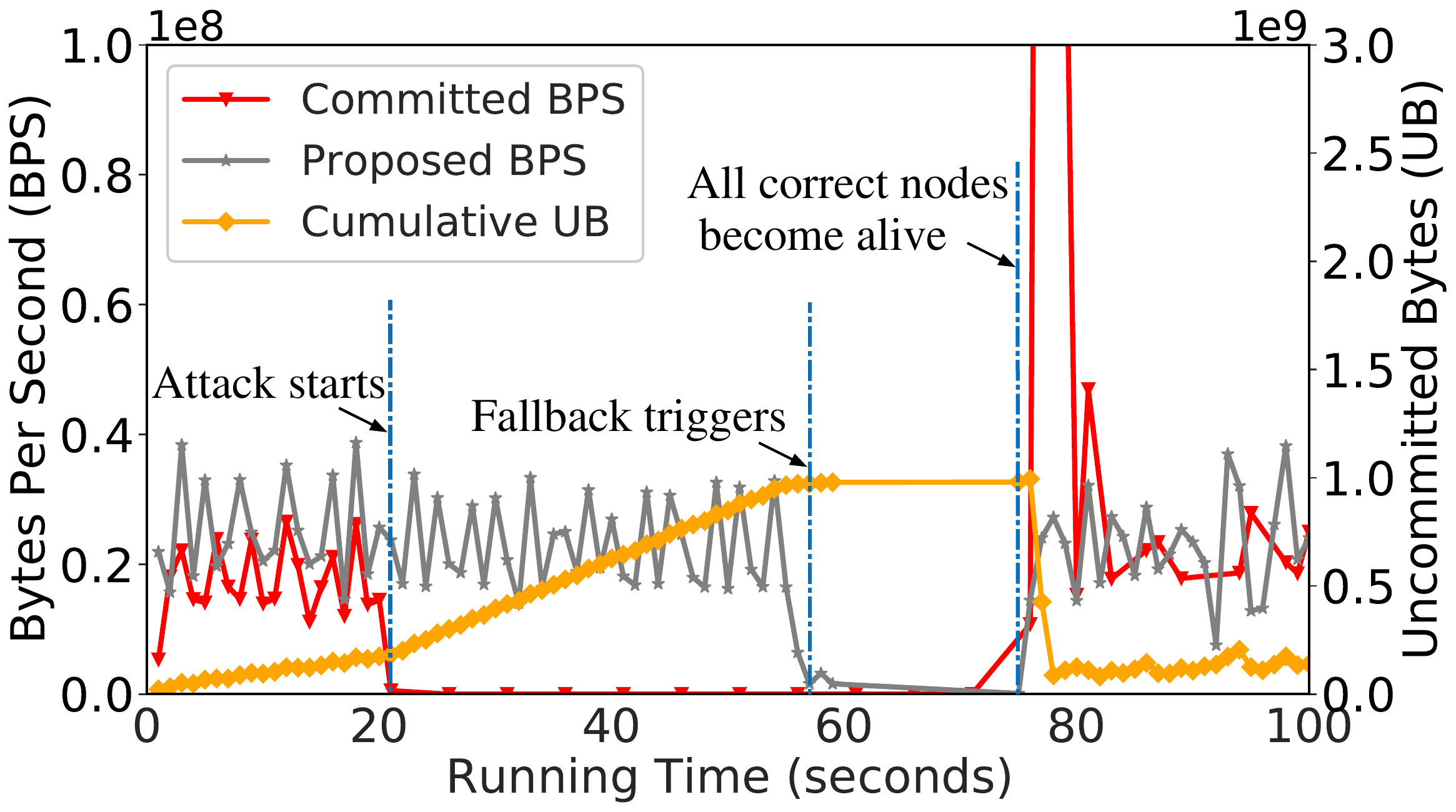}
			\caption{Sailfish-\sysname under attack, $n=10$}
		\label{fig-saifish-resist}
	\end{subfigure}
	\begin{subfigure}[t]{0.325\textwidth}
            \setlength\abovecaptionskip{-0.1\baselineskip}
		\centering
		\includegraphics[width=0.9\linewidth]{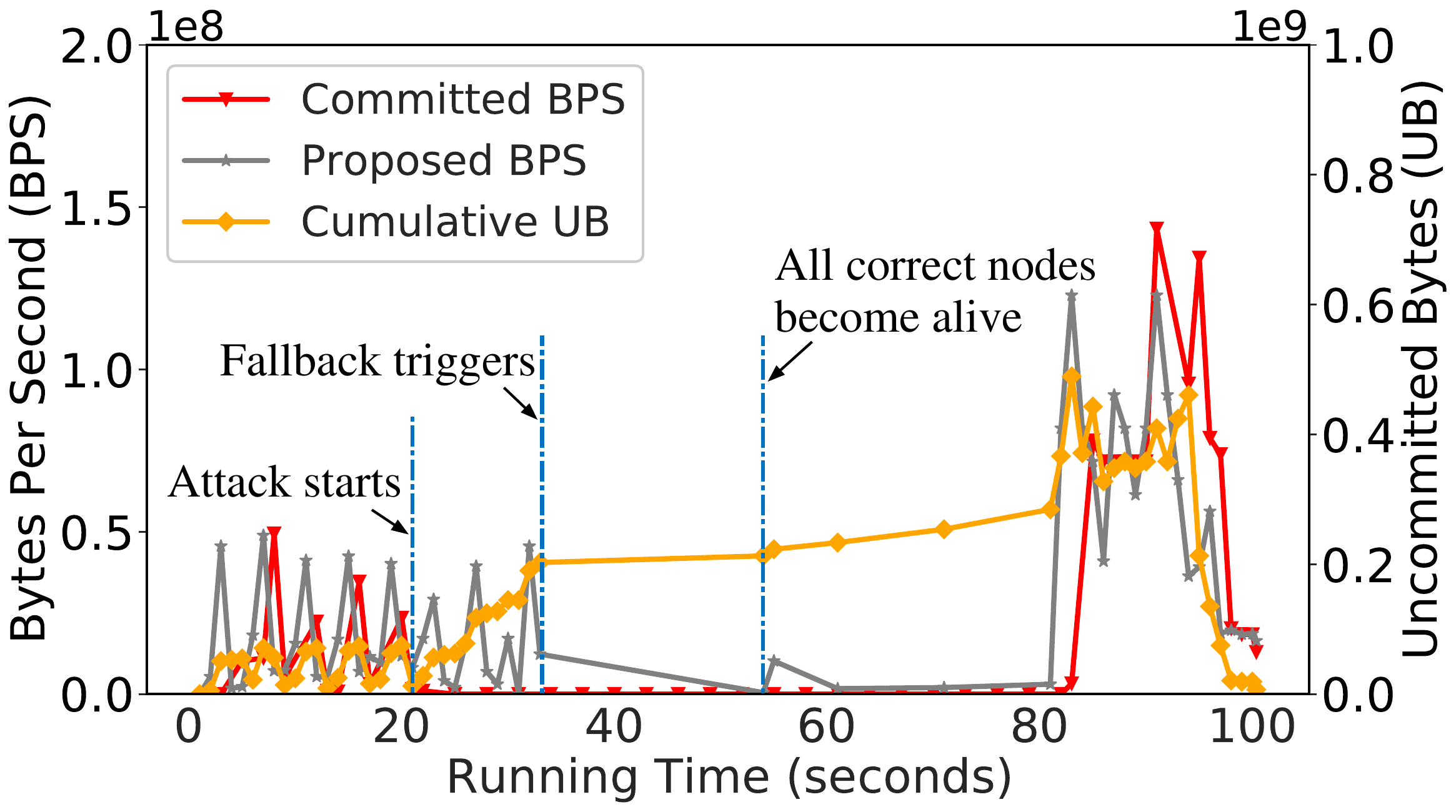}
			\caption{Mysticeti-\sysname under attack, $n=10$}
		\label{fig-mysticeti-resist}
	\end{subfigure}
	\begin{subfigure}[t]{0.31\textwidth}
            \setlength\abovecaptionskip{-0.1\baselineskip}
		\centering
		\includegraphics[width=0.9\linewidth]{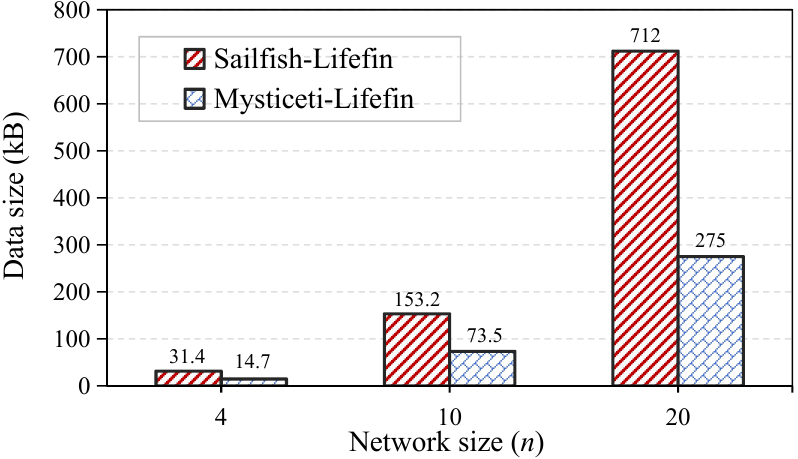}
			\caption{Fallback data size}
		\label{fig-data-size}
	\end{subfigure}
    \caption{The resistance evaluation of \sysname under inflation attacks. The evaluations without \sysname are given in \Cref{fig-attack-eval}}
    \label{fig-attack-resist}
\end{figure*}

\subsection{Scalability}\label{subsec-scalability}
We then evaluate \sysname's scalability by running varying numbers of nodes, $n=4, 10$, and $20$ (\Cref{fig-tps-latency}), following the same settings as Autobahn~\cite{autobahn}. For both Sailfish and Mysticeti, Sailfish-\sysname and Mysticeti-\sysname match the throughput and latency of their corresponding baselines under the same workload and $n$. For instance, under $n=20$, both Sailfish and Sailfish-\sysname achieve 155KTps with a 5.1s e2e latency; similarly, Mysticeti maintains 304KTps with a 3.6s e2e latency, while Mysticeti-\sysname achieves 302KTps with a 5.8s e2e latency. Moreover, Mysticeti-\sysname (and Mysticeti) can achieve better throughput than Sailfish-\sysname (and Sailfish) as the network scales because it eliminates DAG certification, including RBC and certificate verification for DAG vertices. \Cref{fig-tps-latency} shows that incorporating \sysname preserves the scalability of the underlying DAG-based protocol.

\subsection{Inflation Resistance}\label{subsec-inflation-tolerance}
We finally evaluate whether \sysname can efficiently resist inflation attacks. Due to the potential risks of attacking AWS infrastructure, we chose to conduct these experiments on a local machine in a controlled environment. The experimental settings are the same as those of \Cref{fig-attack-eval}. Specifically, we perform the experiments on an Ubuntu server with 48 CPU cores and 128GB of RAM and set a relatively low transaction input rate ($50,000$ tx/s). We begin the attack by making one correct node unresponsive. Then, at round 80, we simultaneously activate the nodes to emulate delivery of an activation certificate and to trigger the fallback mechanism. We use a small round number to activate the fallback quickly and avoid crashing our server during the experiments; in practice, the timeout-based fallback condition uses a long timeout (i.e., 1 hour), as discussed in \S~\ref{sec:system-overview}.
Finally, we let the unresponsive correct node recover and become responsive 20 seconds after the fallback mechanism is triggered.

We run the protocols for approximately 100 seconds. \Cref{fig-saifish-resist} and \Cref{fig-mysticeti-resist} illustrate the performance of Sailfish-\sysname and Mysticeti-\sysname, respectively, under inflation attacks. We focus on three metrics: committed BPS, proposed BPS, and cumulative UB. Specifically, committed BPS (the left y-axis) represents the committed data size (in committed vertices) per second, i.e., byte throughput. Proposed BPS (the left y-axis) indicates the per-second size of vertices proposed by nodes, including committed and uncommitted vertices. Cumulative UB (the right y-axis) represents the cumulative bytes that have been proposed but not committed by nodes. The experiments show that after the attack starts at around 20 seconds, the throughput (i.e., committed BPS) drops to 0 while nodes still propose new vertices (depicted by proposed BPS). During this period, the cumulative UB increases sharply because no vertices are committed.
However, after the protocol triggers the fallback mechanism, nodes stop creating new DAG vertices, causing the cumulative UB to remain unchanged (i.e., during the 56s--76s interval in \Cref{fig-saifish-resist} and the 34s--54s interval in \Cref{fig-mysticeti-resist}). This shows that our fallback mechanism can prevent the adversary from exhausting nodes' resources.
After all correct nodes recover and become responsive, the fallback mechanism can terminate and the uncommitted data can be handled. Compared to the original DAG-based protocols (cf. \Cref{fig-attack-eval}), \sysname-empowered DAG protocols effectively resist inflation attacks by preventing the adversary from exploiting mempool explosions.

\Cref{fig-data-size} illustrates the sizes of fallback data for varying $n$. The fallback data mainly includes PoST blocks, auxiliary data used to verify PoST blocks, and messages used by the FIN protocol. This experiment demonstrates that \sysname can commit DAG vertices using a fixed and relatively small amount of resources, enabling nodes to escape mempool explosions. For instance, in Sailfish-\sysname, running $n=20$ nodes requires $\text{\textless}1$MB of memory. Even with 100 nodes, given the quadratic storage complexity of the ACS mechanism, the overhead is expected to grow to a modest $18$MB.

\section{Related Work} \label{sec:related-work}
All modern DAG protocol families surveyed in this paper exhibit the continuous-growth optimistic path and lack an independent bounded-resource committing mechanism~\cite{dag-rider,narwhal,bullshark,shoal,bolt,sailfish,bbca-chain,suilutris,cordial-miners,mysticeti,jovanovic2024mahi,danezis2024obelia,frontrundag,Fino,gradeddag,wahoo,lightdag,dai2024remora}. Thus, our resource-bounded liveness observation applies to all of them. Our Sailfish and Mysticeti instantiations show that \sysname's recovery approach is not tied to vertex certification.
We discuss the closest related work below. Additional work on bounded-memory consensus and prefix-based protocols is summarized in the full paper~\cite[Appendix~A]{zhang2025lifefin}.

\para{Asynchronous DAG-based BFT}
Our concrete inflation attacks target partially synchronous DAG protocols, which predetermine leaders. Asynchronous DAG protocols such as DAG-Rider~\cite{dag-rider}, Tusk~\cite{narwhal}, and Mahi-mahi~\cite{jovanovic2024mahi} select leaders randomly, so our exact leader-targeting schedules do not directly apply. Nevertheless, their probabilistic termination does not impose a hard bound on retained state before a decision: they exhibit the same continuous-growth dependency and therefore fall within our resource-bounded liveness observation.
Moreover, asynchronous DAG protocols cannot directly serve as a bounded recovery mode for an existing partially synchronous protocol. In contrast, \sysname is a generic framework integrated into these partially synchronous protocols without replacing their certified or uncertified optimistic DAG. Asynchronous protocols also introduce randomization using a global perfect coin~\cite{blum2020asynchronous,cachin2000random,dag-rider,loss2018combining} for every leader vertex. This approach significantly increases latency. For instance, DAG-Rider requires at least 12 message delays per block commitment; in contrast, state-of-the-art partially synchronous protocols commit blocks in just 3 message delays~\cite{mysticeti}.
\sysname minimizes this overhead by operating within a partially synchronous protocol and resorting to ACS protocols only in rare, adversarial scenarios.

\para{Dual-mode protocols}
The family of protocols most conceptually similar to \sysname comprises dual-mode consensus. These protocols, exemplified by Ditto~\cite{jolteon}, Bullshark~\cite{bullshark}, Flexico~\cite{flexico}, and Bolt-Dumbo~\cite{boltdumbo}, combine a partially synchronous path with a path designed for asynchronous conditions. Bullshark is the closest DAG-based design, but its two paths address network asynchrony rather than finite-resource recovery: steady-state and randomized fallback leaders are both committed by interpreting later vertices in the same continuously growing certified DAG. Consequently, Bullshark does not rule out arbitrarily long no-commit prefixes, and neither path can progress if a fixed working-storage budget prevents further vertex creation. Its garbage collection reclaims state following later leader commitments but does not provide an independent committing mechanism that runs within reserved storage. In contrast, \sysname globally stops optimistic DAG growth and uses a bounded-resource ACS instance to commit the retained state; it also supports uncertified DAGs.

\para{Liveness issues in DAG-based BFT}
Prior work has identified several liveness pitfalls in DAG-based BFT. Sailfish~\cite{sailfish} identifies a fault in Mysticeti's liveness proof~\cite{mysticeti} and suggests a larger leader timeout, while concurrent work by Qiu et al.~\cite{qiu2026mechanized} shows that round jumping in Mysticeti's implementation can compromise liveness. These issues concern Mysticeti's timeout assumptions or implementation mechanics and admit local repairs, such as increasing the timeout or disabling round jumping. More broadly, Wong et al.~\cite{wong2024beyond} document storage and garbage-collection mistakes in real BFT implementations, including DAG-based systems, that can cause memory exhaustion or make committed data unavailable. These vulnerabilities arise when implementations deviate from, or underspecify, the intended protocol. A separate line of work exposes a memory--validity tradeoff in fully asynchronous DAG protocols: preserving strong fairness in DAG-Rider~\cite{dag-rider} can require retaining an unbounded DAG, so Bullshark~\cite{bullshark} enables garbage collection by weakening its fairness guarantee before GST. Duan et al.~\cite{duan2024dashing} similarly identify unbounded-memory liveness dependencies in DAG-Rider and Dumbo-NG~\cite{dumbo-ng}. In contrast, our work studies the ordering liveness of DAG protocols whose commit rules depend on continued vertex creation. We show how an adversary can exploit this dependency to exhaust finite resources before GST and leave the protocol unable to recover even after synchrony returns. Rather than repairing a protocol-specific timeout or weakening validity, \sysname provides a generic, globally activated, bounded-resource recovery path for both certified and uncertified DAG protocols.

\section{Conclusion}\label{section-conclusion}
We identify an exploitable liveness vulnerability in DAG-based BFT and propose \sysname, a generic recovery protocol. Our analysis and evaluation show that \sysname restores progress with bounded resources and modest performance overhead.
\ifvldb
\else
\begin{acks}
The authors would like to thank Nibesh Shrestha for his insightful discussion on the Sailfish implementation. This work is supported in part by a research gift by Supra Labs.
\end{acks}
\fi

\bibliographystyle{ACM-Reference-Format}
\bibliography{cite}

@article{librabft,
  title={State machine replication in the libra blockchain},
  author={Baudet, Mathieu and Ching, Avery and Chursin, Andrey and Danezis, George and Garillot, Fran{\c{c}}ois and Li, Zekun and Malkhi, Dahlia and Naor, Oded and Perelman, Dmitri and Sonnino, Alberto},
  journal={The Libra Assn., Tech. Rep},
  volume={7},
  year={2019}
}

@article{DLT,
    author  = {Dwork, Cynthia and Lynch, Nancy and Stockmeyer, Larry},
    title   = {Consensus in the Presence of Partial Synchrony},
    year    = {1988},
    volume  = {35},
    number  = {2},
    issn    = {0004-5411},
    journal = {J. ACM},
    month   = {apr},
    pages   = {288–323}
}

@inproceedings{pbft,
    title     = {Practical Byzantine fault tolerance},
    author    = {Castro, Miguel and Liskov, Barbara},
    booktitle = {OSDI},
    volume    = {99},
    pages     = {173--186},
    year      = {1999}
}

@inproceedings{hotstuff,
    title     = {Hotstuff: Bft consensus with linearity and responsiveness},
    author    = {Yin, Maofan and Malkhi, Dahlia and Reiter, Michael K and Gueta, Guy Golan and Abraham, Ittai},
    booktitle = {ACM PODC},
    pages     = {347--356},
    year      = {2019}
}

@inproceedings{narwhal,
    title     = {Narwhal and Tusk: a DAG-based mempool and efficient BFT consensus},
    author    = {Danezis, George and Kokoris-Kogias, Lefteris and Sonnino, Alberto and Spiegelman, Alexander},
    booktitle = {ACM EuroSys},
    pages     = {34--50},
    year      = {2022}
}

@inproceedings{bullshark,
    title     = {Bullshark: Dag bft protocols made practical},
    author    = {Spiegelman, Alexander and Giridharan, Neil and Sonnino, Alberto and Kokoris-Kogias, Lefteris},
    booktitle = {ACM CCS},
    pages     = {2705--2718},
    year      = {2022}
}

@inproceedings{shoal,
    title   = {Shoal: Improving DAG-BFT Latency And Robustness},
    author  = {Spiegelman, Alexander and Aurn, Balaji and Gelashvili, Rati and Li, Zekun},
    booktitle = {International Conference on Financial Cryptography and Data Security},
    pages = {92–109},
    year = {2024},
    organization = {Springer}
}

@article{Fino,
    title   = {Maximal Extractable Value (MEV) Protection on a DAG},
    author  = {Malkhi, Dahlia and Szalachowski, Pawel},
    journal = {arXiv preprint arXiv:2208.00940},
    year    = {2022}
}

@inproceedings{mysticeti,
    title   = {Mysticeti: Reaching the Limits of Latency with Uncertified DAGs},
    author  = {Babel, Kushal and Chursin, Andrey and Danezis, George and Kichidis, Anastasios and Kokoris-Kogias, Lefteris and Koshy, Arun and Sonnino, Alberto and Tian, Mingwei},
    booktitle={Network and Distributed Systems Security Symposium (NDSS)},
    year={2025}
}

@inproceedings{sailfish,
    title     = {Sailfish: Towards Improving Latency of DAG-based BFT},
    author    = {Shrestha, Nibesh and Kate, Aniket and Nayak, Kartik},
    booktitle = {IEEE S\&P},
    year      = {2025}
}

@inproceedings{suilutris,
    title     = {Sui lutris: A blockchain combining broadcast and consensus},
author = {Blackshear, Sam and Chursin, Andrey and Danezis, George and Kichidis, Anastasios and Kokoris-Kogias, Lefteris and Li, Xun and Logan, Mark and Menon, Ashok and Nowacki, Todd and Sonnino, Alberto and Williams, Brandon and Zhang, Lu},
    booktitle = {ACM CCS},
    year      = {2024}
}

@misc{sui,
    author       = {The Sui Team},
    year         = {2026},
    title        = {The sui blockchain},
    howpublished = {\url{https://sui.io/}},
    note         = {Accessed: 2026}
}

@misc{monad,
    author       = {The Monad Team},
    title        = {The Monad blockchain},
    howpublished = {\url{https://www.monad.xyz/}},
    year         = {2025},
    note         = {Accessed: 2025}
}

@misc{supra,
    author       = {The Supra Research},
    title        = {The Supra blockchain},
    howpublished = {\url{https://supra.com/}},
    year         = {2025},
    note         = {Accessed: 2025}
}

@misc{cosmos,
    author       = {The Cosmos Team},
    title        = {Cosmos},
    howpublished = {\url{https://cosmos.network/}},
    year         = {2025},
    note         = {Accessed: 2025}
}

@inproceedings{jolteon,
    title        = {Jolteon and ditto: Network-adaptive efficient consensus with asynchronous fallback},
    author       = {Gelashvili, Rati and Kokoris-Kogias, Lefteris and Sonnino, Alberto and Spiegelman, Alexander and Xiang, Zhuolun},
    booktitle    = {FC},
    pages        = {296--315},
    year         = {2022},
    organization = {Springer}
}

@misc{aptos,
    author       = {The Aptos Team},
    year         = {2026},
    title        = {Aptos Networks},
    howpublished = {\url{https://aptosfoundation.org/}},
    note         = {Accessed: 2026}
}

@inproceedings{duan2024dashing,
    title     = {Dashing and Star: Byzantine fault tolerance with weak certificates},
    author    = {Duan, Sisi and Zhang, Haibin and Sui, Xiao and Huang, Baohan and Mu, Changchun and Di, Gang and Wang, Xiaoyun},
    booktitle = {ACM EuroSys},
    pages     = {250--264},
    year      = {2024}
}

@article{fischer1985impossibility,
    title   = {Impossibility of distributed consensus with one faulty process},
    author  = {Fischer, Michael J and Lynch, Nancy A and Paterson, Michael S},
    journal = {JACM},
    pages   = {374--382},
    year    = {1985}
}

@article{autobahn,
    title   = {Autobahn: Seamless high speed BFT},
    author  = {Giridharan, Neil and Suri-Payer, Florian and Abraham, Ittai and Alvisi, Lorenzo and Crooks, Natacha},
    journal = {ACM SOSP},
    year    = {2024}
}

@inproceedings{bolt,
    title   = {Shoal++: High Throughput DAG BFT Can Be Fast and Robust!},
    author  = {Arun, Balaji and Li, Zekun and Suri-Payer, Florian and Das, Sourav and Spiegelman, Alexander},
    booktitle = {Usenix NSDI},
    year    = {2025}
}

@article{arete,
    title   = {Optimal Sharding for Scalable Blockchains with Deconstructed SMR},
    author  = {Zhang, Jianting and Luo, Zhongtang and Ramesh, Raghavendra and Kate, Aniket},
    journal = {Proceedings of the VLDB Endowment},
    year    = {2025},
    publisher={VLDB Endowment},
}

@inproceedings{fin,
    title     = {FIN: practical signature-free asynchronous common subset in constant time},
    author    = {Duan, Sisi and Wang, Xin and Zhang, Haibin},
    booktitle = {ACM CCS},
    pages     = {815--829},
    year      = {2023}
}

@inproceedings{frontrundag,
  title={No fish is too big for flash boys! frontrunning on DAG-based blockchains},
  author={Zhang, Jianting and Kate, Aniket},
  booktitle={2025 IEEE Annual Computer Security Applications Conference (ACSAC)},
  pages={1065--1080},
  year={2025},
  organization={IEEE}
}

@article{jovanovic2024mahi,
    title   = {Mahi-Mahi: Low-Latency Asynchronous BFT DAG-Based Consensus},
    author  = {Jovanovic, Philipp and Kogias, Lefteris Kokoris and Kumara, Bryan and Sonnino, Alberto and Tennage, Pasindu and Zablotchi, Igor},
    journal = {arXiv preprint arXiv:2410.08670},
    year    = {2024}
}

@article{danezis2024obelia,
    title   = {Obelia: Scaling DAG-Based Blockchains to Hundreds of Validators},
    author  = {Danezis, George and Kokoris-Kogias, Lefteris and Sonnino, Alberto and Tian, Mingwei},
    journal = {arXiv preprint arXiv:2410.08701},
    year    = {2024}
}

@article{bracha1985asynchronous,
    title     = {Asynchronous consensus and broadcast protocols},
    author    = {Bracha, Gabriel and Toueg, Sam},
    journal   = {Journal of the ACM (JACM)},
    volume    = {32},
    number    = {4},
    pages     = {824--840},
    year      = {1985},
    publisher = {ACM New York, NY, USA}
}

@inproceedings{dag-rider,
    author    = {Idit Keidar and Eleftherios Kokoris-Kogias and Oded Naor and Alexander Spiegelman},
    title     = {\href{https://dl.acm.org/doi/10.1145/3465084.3467905}{All You Need is DAG}},
    booktitle = {PODC'21: Proceedings of the 2021 ACM Symposium on Principles of Distributed Computing},
    year      = {2021}
}

@inproceedings{dumbo-ng,
    title     = {\href{https://dl.acm.org/doi/abs/10.1145/3548606.3559379}{Dumbo-ng: Fast asynchronous bft consensus with throughput-oblivious latency}},
    author    = {Gao, Yingzi and Lu, Yuan and Lu, Zhenliang and Tang, Qiang and Xu, Jing and Zhang, Zhenfeng},
    booktitle = {Proceedings of the 2022 ACM SIGSAC Conference on Computer and Communications Security},
    pages     = {1187--1201},
    year      = {2022}
}

@inproceedings{gradeddag,
    title        = {Gradeddag: An asynchronous dag-based bft consensus with lower latency},
    author       = {Dai, Xiaohai and Zhang, Zhaonan and Xiao, Jiang and Yue, Jingtao and Xie, Xia and Jin, Hai},
    booktitle    = {2023 42nd International Symposium on Reliable Distributed Systems (SRDS)},
    pages        = {107--117},
    year         = {2023},
    organization = {IEEE}
}

@article{wahoo,
    title     = {Wahoo: A DAG-based BFT Consensus with Low Latency and Low Communication Overhead},
    author    = {Dai, Xiaohai and Zhang, Zhaonan and Guo, Zhengxuan and Ding, Chaozheng and Xiao, Jiang and Xie, Xia and Hao, Rui and Jin, Hai},
    journal   = {IEEE Transactions on Information Forensics and Security},
    year      = {2024},
    publisher = {IEEE}
}

@inproceedings{cordial-miners,
    author    = {Idit Keidar and Oded Naor and Ouri Poupko and Ehud Shapiro},
    title     = {\href{https://drops.dagstuhl.de/entities/document/10.4230/LIPIcs.DISC.2023.26}{Cordial Miners: Fast and Efficient Consensus for Every Eventuality}},
    booktitle = {37th International Symposium on Distributed Computing (DISC 2023)},
    year      = {2023}
}

@article{dai2024remora,
    title     = {Remora: A Low-latency DAG-based BFT through Optimistic Paths},
    author    = {Dai, Xiaohai and Li, Wei and Wang, Guanxiong and Xiao, Jiang and Chen, Haoyang and Li, Shufei and Zomaya, Albert Y and Jin, Hai},
    journal   = {IEEE Transactions on Computers},
    year      = {2024},
    publisher = {IEEE}
}

@inproceedings{lightdag,
  author={Dai, Xiaohai and Wang, Guanxiong and Xiao, Jiang and Guo, Zhengxuan and Hao, Rui and Xie, Xia and Jin, Hai},
  booktitle={2024 IEEE International Parallel and Distributed Processing Symposium (IPDPS)}, 
  title={LightDAG: A Low-latency DAG-based BFT Consensus through Lightweight Broadcast}, 
  year={2024},
  volume={},
  number={},
  pages={998-1008},
  organization={IEEE}
}

@article{bbca-chain,
    title   = {BBCA-CHAIN: One-Message, Low Latency BFT Consensus on a DAG},
    author  = {Malkhi, Dahlia and Stathakopoulou, Chrysoula and Yin, Maofan},
    journal = {arXiv preprint arXiv:2310.06335},
    year    = {2023}
}

@article{threshold-clock,
    author  = {Bryan Ford},
    title   = {\href{https://arxiv.org/abs/1907.07010}{Threshold Logical Clocks for Asynchronous Distributed Coordination and Consensus}},
    journal = {CoRR},
    volume  = {abs/1907.07010},
    year    = {2019}
}

@inproceedings{blum2020asynchronous,
    title        = {\href{https://link.springer.com/chapter/10.1007/978-3-030-64375-1_13}{Asynchronous byzantine agreement with subquadratic communication}},
    author       = {Blum, Erica and Katz, Jonathan and Liu-Zhang, Chen-Da and Loss, Julian},
    booktitle    = {Theory of Cryptography: 18th International Conference, TCC 2020, Durham, NC, USA, November 16--19, 2020, Proceedings, Part I 18},
    pages        = {353--380},
    year         = {2020},
    organization = {Springer}
}

@misc{loss2018combining,
    title         = {\href{https://eprint.iacr.org/2018/235}{Combining asynchronous and synchronous byzantine agreement: The best of both worlds}},
    author        = {Loss, Julian and Moran, Tal},
    year          = {2018},
    eprint        = {2018/235},
    archiveprefix = {IACR}
}

@inproceedings{cachin2000random,
    title     = {\href{https://dl.acm.org/doi/10.1145/343477.343531}{Random oracles in constantipole: practical asynchronous byzantine agreement using cryptography}},
    author    = {Cachin, Christian and Kursawe, Klaus and Shoup, Victor},
    booktitle = {Proceedings of the nineteenth annual ACM symposium on Principles of distributed computing},
    pages     = {123--132},
    year      = {2000}
}

@article{flexico,
    title   = {\href{https://journals.plos.org/plosone/article?id=10.1371/journal.pone.0277092}{Flexico: An efficient dual-mode consensus protocol for blockchain networks}},
    author  = {Shuyang Ren and Choonhwa Lee and Eunsam Kim and Sumi Helal},
    journal = {PLoS ONE},
    year    = {2022}
}

@inproceedings{boltdumbo,
    author    = {Yuan Lu and Zhenliang Lu and Qiang Tang},
    title     = {\href{https://dl.acm.org/doi/abs/10.1145/3548606.3559346}{Bolt-Dumbo Transformer: Asynchronous Consensus As Fast As the Pipelined BFT}},
    booktitle = {CCS '22: Proceedings of the 2022 ACM SIGSAC Conference on Computer and Communications Security},
    year      = {2022}
}

@inproceedings{abraham2021good,
  title={Good-case latency of byzantine broadcast: A complete categorization},
  author={Abraham, Ittai and Nayak, Kartik and Ren, Ling and Xiang, Zhuolun},
  booktitle={Proceedings of the 2021 ACM Symposium on Principles of Distributed Computing},
  pages={331--341},
  year={2021}
}

@inproceedings{dasrbc,
  title={Asynchronous data dissemination and its applications},
  author={Das, Sourav and Xiang, Zhuolun and Ren, Ling},
  booktitle={Proceedings of the 2021 ACM SIGSAC Conference on Computer and Communications Security},
  pages={2705--2721},
  year={2021}
}

@misc{tokio,
    author       = {Tokio},
    title        = {Tokio - An asynchronous Rust runtime},
    year         = {2025},
    howpublished = {\url{https://tokio.rs/}},
    note         = {Accessed: 2025}
}

@misc{ed25519,
    author       = {Dalek Cryptography},
    title        = {Dalek elliptic curve cryptography},
    howpublished = {\url{https://github.com/dalek-cryptography/ed25519-dalek}},
    year         = {2025},
    note         = {Accessed: 2025}
}

@inproceedings{doidge2024moonshot,
  title={Moonshot: Optimizing Block Period and Commit Latency in Chain-Based Rotating Leader BFT},
  author={Doidge, Isaac and Ramesh, Raghavendra and Shrestha, Nibesh and Tobkin, Joshua},
  booktitle={2024 54th Annual IEEE/IFIP International Conference on Dependable Systems and Networks (DSN)},
  pages={470--482},
  year={2024},
  organization={IEEE}
}

@misc{jolteon-code,
title = {Jolteon},
author = {{Alberto Sonnino}},
year = 2025,
howpublished = {\url{https://github.com/asonnino/hotstuff}}
}

@misc{hotstuff-code,
title = {HotStuff},
author = {{Alberto Sonnino}},
year = 2025,
howpublished = {\url{https://github.com/asonnino/hotstuff/tree/3-chain}}
}

@misc{diem-code,
title = {Diem},
author = {{The Diem Team}},
year = 2025,
howpublished = {\url{https://github.com/diem/diem}}
}

@misc{sailfish-code,
title = {Sailfish Codebase},
author = {{Nibesh Shrestha}},
year = 2025,
howpublished = {\url{https://github.com/nibeshrestha/sailfish}}
}

@misc{mysticeti-code,
title = {Mysticeti Codebase},
author = {{Alberto Sonnino}},
year = 2025,
howpublished = {\url{github.com/asonnino/mysticeti}}
}

@misc{autobahn-code,
title = {Autobahn Codebase},
author = {{Neil Giridharan}},
year = 2025,
howpublished = {\url{https://github.com/neilgiri/autobahn-artifact}}
}

@misc{arete-code,
title = {Arete Codebase},
author = {{Jianting Zhang}},
year = 2025,
howpublished = {\url{https://github.com/EtherCS/arete}}
}

@misc{sui-validator-config,
title = {Sui Validator Node Configuration},
author = {{Sui}},
year = 2025,
howpublished = {\url{https://docs.sui.io/guides/operator/validator-config}}
}

@article{fasthotstuff,
  title={Fast-hotstuff: A fast and resilient hotstuff protocol},
  author={Jalalzai, Mohammad M and Niu, Jianyu and Feng, Chen and Gai, Fangyu},
  journal={arXiv preprint arXiv:2010.11454},
  year={2020}
}

@article{hotstuff-1,
  title={HotStuff-1: Linear Consensus with One-Phase Speculation},
  author={Kang, Dakai and Gupta, Suyash and Malkhi, Dahlia and Sadoghi, Mohammad},
  journal={arXiv preprint arXiv:2408.04728},
  year={2024}
}

@article{hotstuff-2,
  title={Hotstuff-2: Optimal two-phase responsive bft},
  author={Malkhi, Dahlia and Nayak, Kartik},
  journal={Cryptology ePrint Archive},
  year={2023}
}

@inproceedings{streamlet,
  title={Streamlet: Textbook streamlined blockchains},
  author={Chan, Benjamin Y and Shi, Elaine},
  booktitle={Proceedings of the 2nd ACM Conference on Advances in Financial Technologies},
  pages={1--11},
  year={2020}
}

@article{monadbft,
  title={MonadBFT: Fast, Responsive, Fork-Resistant Streamlined Consensus},
  author={Jalalzai, Mohammad Mussadiq and Babel, Kushal},
  journal={arXiv preprint arXiv:2502.20692},
  year={2025}
}

@misc{diembft,
    author       = {The LibraBFT Team},
    title        = {State machine replication in the libra blockchain},
    howpublished = {\url{https://developers.diem.com/docs/technical-papers/state-machine-replication-paper/}},
    year         = {2020},
}

@misc{iota,
    author = {The IOTA team},
    title = {IOTA is a decentralized blockchain infrastructure to build and secure our digital world},
    year = {2025},
    howpublished = {\url{https://www.iota.org}}
}

@inproceedings{liu2023flexible,
  title={Flexible advancement in asynchronous bft consensus},
  author={Liu, Shengyun and Xu, Wenbo and Shan, Chen and Yan, Xiaofeng and Xu, Tianjing and Wang, Bo and Fan, Lei and Deng, Fuxi and Yan, Ying and Zhang, Hui},
  booktitle={Proceedings of the 29th Symposium on Operating Systems Principles},
  pages={264--280},
  year={2023}
}

@inproceedings{sun2023neobft,
  title={Neobft: Accelerating byzantine fault tolerance using authenticated in-network ordering},
  author={Sun, Guangda and Jiang, Mingliang and Khooi, Xin Zhe and Li, Yunfan and Li, Jialin},
  booktitle={Proceedings of the ACM SIGCOMM 2023 Conference},
  pages={239--254},
  year={2023}
}

@inproceedings{gupta2023dissecting,
  title={Dissecting bft consensus: In trusted components we trust!},
  author={Gupta, Suyash and Rahnama, Sajjad and Pandey, Shubham and Crooks, Natacha and Sadoghi, Mohammad},
  booktitle={Proceedings of the Eighteenth European Conference on Computer Systems},
  pages={521--539},
  year={2023}
}

@inproceedings{suri2021basil,
  title={Basil: Breaking up BFT with ACID (transactions)},
  author={Suri-Payer, Florian and Burke, Matthew and Wang, Zheng and Zhang, Yunhao and Alvisi, Lorenzo and Crooks, Natacha},
  booktitle={Proceedings of the ACM SIGOPS 28th Symposium on Operating Systems Principles},
  pages={1--17},
  year={2021}
}

@article{rocket2019scalable,
  title={Scalable and probabilistic leaderless BFT consensus through metastability},
  author={Rocket, Team and Yin, Maofan and Sekniqi, Kevin and van Renesse, Robbert and Sirer, Emin G{\"u}n},
  journal={arXiv preprint arXiv:1906.08936},
  year={2019}
}

@inproceedings{neiheiser2021kauri,
  title={Kauri: Scalable bft consensus with pipelined tree-based dissemination and aggregation},
  author={Neiheiser, Ray and Matos, Miguel and Rodrigues, Lu{\'\i}s},
  booktitle={Proceedings of the ACM SIGOPS 28th Symposium on Operating Systems Principles},
  pages={35--48},
  year={2021}
}

@article{chandra1996unreliable,
  title={Unreliable failure detectors for reliable distributed systems},
  author={Chandra, Tushar Deepak and Toueg, Sam},
  journal={Journal of the ACM (JACM)},
  volume={43},
  number={2},
  pages={225--267},
  year={1996},
  publisher={ACM New York, NY, USA}
}

@inproceedings{ricciardi1996impossibility,
  title={Impossibility of (repeated) reliable broadcast},
  author={Ricciardi, Aleta},
  booktitle={Proceedings of the fifteenth annual ACM symposium on Principles of distributed computing},
  pages={342},
  year={1996}
}

@inproceedings{delporte2008finite,
  title={With finite memory consensus is easier than reliable broadcast},
  author={Delporte-Gallet, Carole and Devismes, St{\'e}phane and Fauconnier, Hugues and Petit, Franck and Toueg, Sam},
  booktitle={International Conference On Principles Of Distributed Systems},
  pages={41--57},
  year={2008},
  organization={Springer}
}

@article{dolev2010consensus,
  title={When consensus meets self-stabilization},
  author={Dolev, Shlomi and Kat, Ronen I and Schiller, Elad M},
  journal={Journal of Computer and System Sciences},
  volume={76},
  number={8},
  pages={884--900},
  year={2010},
  publisher={Elsevier}
}

@article{dolev2016possibility,
  title={Possibility and impossibility of reliable broadcast in the bounded model},
  author={Dolev, Danny and Spielrien, Meir},
  journal={arXiv preprint arXiv:1611.05161},
  year={2016}
}

@inproceedings{chen2024porygon,
  title={Porygon: Scaling blockchain via 3d parallelism},
  author={Chen, Wuhui and Xia, Ding and Cai, Zhongteng and Dai, Hong-Ning and Zhang, Jianting and Hong, Zicong and Liang, Junyuan and Zheng, Zibin},
  booktitle={2024 IEEE 40th International Conference on Data Engineering (ICDE)},
  pages={1944--1957},
  year={2024},
  organization={IEEE}
}

@inproceedings{qiu2026mechanized,
  title={Mechanized safety and liveness proofs for the Mysticeti consensus protocol under the LiDO-DAG framework},
  author={Qiu, Longfei and Xiao, Jingqi and Shao, Zhong},
  booktitle={Proceedings of the 2026 IEEE Symposium on Security and Privacy (SP)},
  year={2026}
}

@article{wong2024beyond,
  title={Beyond the Whitepaper: Where BFT Consensus Protocols Meet Reality},
  author={Wong, David and Kolegov, Denis and Mikushin, Ivan},
  journal={Cryptology ePrint Archive},
  year={2024}
}

@inproceedings{ongaro2014search,
  title={In search of an understandable consensus algorithm},
  author={Ongaro, Diego and Ousterhout, John},
  booktitle={2014 USENIX annual technical conference (USENIX ATC 14)},
  pages={305--319},
  year={2014}
}

@inproceedings{oki1988viewstamped,
  title={Viewstamped replication: A new primary copy method to support highly-available distributed systems},
  author={Oki, Brian M and Liskov, Barbara H},
  booktitle={Proceedings of the seventh annual ACM Symposium on Principles of distributed computing},
  pages={8--17},
  year={1988}
}

@article{liskov2012viewstamped,
  title={Viewstamped replication revisited},
  author={Liskov, Barbara and Cowling, James},
  year={2012}
}

@inproceedings{junqueira2011zab,
  title={Zab: High-performance broadcast for primary-backup systems},
  author={Junqueira, Flavio P and Reed, Benjamin C and Serafini, Marco},
  booktitle={2011 IEEE/IFIP 41st International Conference on Dependable Systems \& Networks (DSN)},
  pages={245--256},
  year={2011},
  organization={IEEE}
}

@misc{microsoft2026throttling,
  author       = {Microsoft},
  title        = {Throttling Pattern},
  year         = {2026},
  howpublished = {\url{https://learn.microsoft.com/en-us/azure/architecture/patterns/throttling}},
  note         = {Accessed: 2026}
}

@misc{tendermint-timeouts,
  author       = {Tendermint Core},
  title        = {Configuration: Consensus Timeouts},
  year         = {2020},
  howpublished = {\url{https://docs.tendermint.com/v0.34/tendermint-core/configuration.html}},
  note         = {Accessed: 2026}
}

@misc{amazon-load-shedding,
  author       = {Amazon Builders' Library},
  title        = {Using Load Shedding to Avoid Overload},
  year         = {2026},
  howpublished = {\url{https://aws.amazon.com/builders-library/using-load-shedding-to-avoid-overload/}},
  note         = {Accessed: 2026}
}

@book{beyer2016site,
  title={Site reliability engineering: how Google runs production systems},
  author={Beyer, Betsy and Jones, Chris and Petoff, Jennifer and Murphy, Niall Richard},
  year={2016},
  publisher={" O'Reilly Media, Inc."}
}

@article{yang2026your,
  title={Your Loss is My Gain: Low Stake Attacks on Liquid Staking Pools},
  author={Yang, Sen and Yaish, Aviv and Gervais, Arthur and Zhang, Fan},
  journal={arXiv preprint arXiv:2605.01025},
  year={2026}
}

@article{zhang2025lifefin,
  title={Lifefin: Escaping Mempool Explosions in DAG-based BFT},
  author={Zhang, Jianting and Yang, Sen and Sonnino, Alberto and Loza, Sebasti{\'a}n and Kate, Aniket},
  journal={arXiv preprint arXiv:2511.15936},
  year={2025}
}

\ifvldb
\else
\newpage
\appendix

\section{Additional Related Work} \label{sec-additional-related-work}

\para{Studies of bounded models}
The bounded model has been studied in distributed protocol research.
Prior work~\cite{ricciardi1996impossibility, delporte2008finite, dolev2016possibility} proves the impossibility of solving sequential repeated reliable broadcast under the bounded model. Other work~\cite{delporte2008finite, dolev2010consensus} circumvents the impossibility by assuming a (weaker) transient fault model or a (stronger) failure detector~\cite{chandra1996unreliable}.
This paper shows that the bounded model also plays a role in the liveness guarantee of DAG-based BFT protocols due to the new features introduced by maintaining a DAG-based mempool. Prior theoretical analyses of distributed protocols often overlook such bounded-resource considerations, leaving substantial room to uncover similar vulnerabilities in future designs.

For each surveyed protocol, its specification allows dissemination to advance while commitment is stalled, uses later-round vertices for commitment, and retains undecided causal history without an independent bounded-resource committing path. These shared properties place every surveyed protocol within the scope defined in \S~\ref{sec-liveness-issue}.

\para{Prefix-based CFT protocols}
Prefix-based crash fault-tolerant (CFT) protocols, such as Zab~\cite{junqueira2011zab}, Raft~\cite{ongaro2014search}, and Viewstamped Replication~\cite{oki1988viewstamped, liskov2012viewstamped}, maintain a single totally ordered log and commit a monotonically growing prefix once the corresponding log entries (i.e., transactions) are replicated by a quorum of nodes. While they may also accumulate uncommitted log entries until they exhaust resources and introduce liveness issues during asynchrony, the root cause of the resource exhaustion is not inherent to their ordering logic: a (leader) node might keep receiving new requests from clients and attempt to add them to its log without waiting for the majority vote for the log prefix, leading to potential resource exhaustion.
In this case, these prefix-based CFT protocols can apply backpressure or flow-control mechanisms to avoid memory explosions and mitigate liveness issues: an explicit view-change mechanism selects a new leader to take over the consensus task with the existing logs.
However, as discussed in \apdix~\ref{sec-attempt}, these solutions do not resolve the circular dependency in the continuous-growth optimistic path studied here.

\section{Security Analysis of Sailfish-\sysname} \label{sec:detailed-security-analysis-sailfish}
In this section, we give a detailed security analysis of Sailfish-\sysname, presented in \S~\ref{sec:sailfish-sysname}.

Compared to the vanilla Sailfish protocol, Sailfish-\sysname introduces an ACS-based fallback mechanism to accomplish the ordering task with bounded working storage. To be more specific, nodes handle at most $n$ PoST blocks and retain only the fixed proposal state and current-iteration messages during the ACS instance, while leveraging the underlying committing rules to commit DAG vertices according to the ACS output.

The modification is that Sailfish-\sysname may select a new round $r^*$ leader vertex $\post_L^{r^*}$ to replace the predefined round $r^*$ leader vertex $v_L^{r^*}$ after each ACS instance. Thus, we adhere to the security analysis of Sailfish~\cite[Appendix B]{sailfish} while considering the subtle modification introduced by \sysname.
For each fallback view $v$, ACS agreement and the deterministic output construction ensure that all correct nodes obtain the same ACS output $V$.
Let $r_v$ be the highest PoST round selected in view $v$. The transition rule (that is, before signing $\mathsf{READY}$ in view $v+1$, a correct node creates its round-$r_v+2$ transition vertex), together with quorum intersection between the correct signers of $AC_{v+1}$ and the correct proposers represented in its ACS output, ensures that $r_{v+1}\geq r_v+2$. Thus, each round is selected by at most one fallback view, and the following arguments compose across fallback views.

Following Sailfish, we say that a leader vertex $v_k$ is directly committed by node $\node{i}$ if $\node{i}$ invokes the \Call{commit\_leader}{$v_k$} function (Figure~\ref{fig:sailfish-fallback-commit}, $\aln$~\ref{step:commit-leader-start}). In contrast, we say that a leader vertex $v_l$ is indirectly committed if $v_l$ is added to $\mathit{leaderStack}$ by $\node{i}$ (Figure~\ref{fig:sailfish-fallback-commit}, $\aln$~\ref{step:sailfish-indirect-commit}). Additionally, we say $\node{i}$ consecutively directly commits leader vertices $v_k$ and $v_k'$ if $\node{i}$ directly commits $v_k$ and $v_k'$ in rounds $r$ and $r'$, respectively, and does not directly commit any leader vertex between $r$ and $r'$.

\subsection{Safety}\label{sec:sf-detailed-safety-proof}
We first analyze the safety of Sailfish-\sysname. Safety indicates that all correct nodes commit transactions in the same order.

\begin{claim}\label{claim:no-equivocation}
    For any two correct nodes $\node{i}$ and $\node{j}$, at any given time $t$ and round $r$, if there are two DAG vertices $v_k$ and $v_k'$ s.t. $v_k\in DAG_i[r] \land v_k'\in DAG_j[r] \land v_k.\creator = v_k'.\creator$, then $v_k = v_k'$.
\end{claim}
\begin{myproof}{}
    For the sake of contradiction, if $v_k \neq v_k'$, then $v_k.\creator$ (called $\node{k}$ without loss of generality) must create two different vertices in round $r$. Note that in Sailfish-\sysname, a node $\node{i}$ adds a vertex $v_k$ to its local DAG $DAG_i[r]$ after delivering $v_k$ through RBC or ACS. However, both events require $2f+1$ nodes to acknowledge $v_k$. By the standard quorum intersection argument, at least one correct node must have equivocated if $v_k \neq v_k'$, which is a contradiction.
\end{myproof}

\begin{claim}\label{claim:at-most-f}
    Given any round number $r^*$ that is decided by an ACS instance, at most $2f$ vertices can be created in round $r^*+1$.
\end{claim}
\begin{myproof}{}
    Let $V$ be the ACS output defining $r^*=\max\{\post.\vertex.\round\mid\post\in V\}$. By ACS validity, $V$ contains PoST blocks from a set $\mathcal{P}$ of at least $f+1$ correct creators. Each creator in $\mathcal{P}$ constructs its PoST only after receiving the activation certificate and atomically stopping optimistic-path vertex creation; hence, its PoST contains its last optimistically created vertex.

    Suppose, for contradiction, that a set $\mathcal{Q}$ of $2f+1$ distinct creators produces round $r^*+1$ optimistic vertices. Since $|\mathcal{P}|\geq f+1$, $|\mathcal{Q}|=2f+1$, and $n=3f+1$, their creator sets intersect. Let $\node{k}\in\mathcal{P}\cap\mathcal{Q}$. Node $\node{k}$ is correct, so it created its round $r^*+1$ vertex before freezing and constructing $\post_k$. Therefore, $\post_k.\vertex.\round\geq r^*+1$, contradicting the definition of $r^*$. After ACS, the graceful transition starts correct nodes at round $r^*+2$, so they create no additional round $r^*+1$ vertices.
\end{myproof}

\para{Proof of \Cref{proposition-highest-leader-not-commit}} By \Cref{claim:at-most-f}, at most $2f$ vertices are created in round $r^*+1$. Therefore, before the fallback terminates, the predefined round $r^*$ leader vertex cannot collect $\geq2f+1$ votes and thus cannot be directly committed according to Sailfish's direct committing rule. Moreover, recall that the indirect committing rule requires a round $r'>r^*$ leader vertex to be directly committed; in that case, the fallback instance must terminate. This completes the proof.

\begin{claim} \label{claim:same-acs-leader}
    If a node $\node{i}$ directly or indirectly commits a leader vertex $v$ in round $r$, and a node $\node{j}$ directly or indirectly commits a leader vertex $v'$ in round $r$, then $v = v'$.
\end{claim}
\begin{myproof}{}
    If no fallback view selects round $r$, by Claim~\ref{claim:no-equivocation}, both $\node{i}$ and $\node{j}$ have the same predefined leader vertex. Otherwise, let $r^*=r$ be the round selected by its unique fallback view. If $v_L^r=\post_L^{r^*}$, we are trivially done. Otherwise, we show that no correct node directly or indirectly commits $v_L^r$ when $v_L^r\neq \post_L^{r^*}$.

    By Claim~\ref{claim:at-most-f}, we know that at most $2f$ vertices are created in round $r+1$. Therefore, $v_L^r$ cannot be directly committed upon receipt of $2f+1$ vertices in round $r+1$ (Figure~\ref{fig:sailfish-fallback-commit}, $\aln$~\ref{step:sailfish-direct-commit}). Additionally, $v_L^r$ cannot be directly committed through the ACS-based fallback mechanism because $v_L^r\neq \post_L^{r^*}$. Therefore, nodes will not directly commit $v_L^r$. Moreover, because nodes are forced to connect to blocks in the ACS output (which excludes $v_L^r$) after the ACS instance, there is no path from any subsequent leader vertex to $v_L^r$. By the code of the \Call{commit\_leader}{} function, $v_L^r$ will not be indirectly committed. This completes the proof.
\end{myproof}


\begin{claim}\label{claim:leader-before-post}
    If a correct node $\node{i}$ directly commits a leader vertex $v_k$ in round $r = r^*-1$ (where round $r^*$ is the round decided by an ACS instance), then all correct nodes must directly commit $v_k$ in round $r$.
\end{claim}
\begin{myproof}{}
    According to Figure~\ref{fig:sailfish-fallback-commit}, there are two cases in which a node $\node{i}$ can directly commit a round $r=r^*-1$ leader vertex. If $\node{i}$ directly commits $v_k$ through the ACS-based fallback mechanism (Figure~\ref{fig:sailfish-fallback-commit}, $\aln$~\ref{step:sailfish-fallback-direct-before-post}), then, by the agreement property of ACS, all correct nodes must perform \Call{finalize\_fallback}{} to directly commit $v_k$. If $\node{i}$ directly commits $v_k$ upon receiving $2f+1$ round $r+1$ vertices (Figure~\ref{fig:sailfish-fallback-commit}, $\aln$~\ref{step:sailfish-direct-commit}), then there are at least $f+1$ correct round $r+1$ vertices connecting to $v_k$. By ACS validity, the output $V$ contains at least $f+1$ correct PoST blocks; thus, by a quorum intersection argument, at least one correct PoST block in the ACS output $V$ must connect to $v_k$. In this case, all other correct nodes that do not commit $v_k$ upon receiving $2f+1$ round $r+1$ vertices must directly commit $v_k$ through the ACS-based fallback mechanism.
\end{myproof}

\begin{claim}\label{claim:leader-path}
    If a correct node $\node{i}$ directly commits a leader vertex $v_k$ in round $r$ and no fallback view selects round $r+1$, then for every leader vertex $v_l$ in round $r'$ such that $r'>r$, there exists a path from $v_l$ to $v_k$.
\end{claim}
\begin{myproof}{}
    There are two scenarios 1) and 2) depending on whether round $r$ is selected by a fallback view.

    1) If no fallback view selects $r$, $v_k$ is chosen in the optimistic path, and there exists a set $\mathcal{Q}$ of $2f+1$ round $r+1$ vertices that connect to $v_k$. Let $\mathcal{H} \subseteq \mathcal{Q}$ be the set of vertices created by correct nodes. Note that $|\mathcal{H}| \geq f+1$. We then complete the proof by showing that the statement holds for any $r'>r$.

    \textbf{Case} $r'=r+1$: Since no fallback view selects round $r+1$, $v_l$ must be the predefined leader vertex (i.e., the creation of $v_l$ satisfies the constraint described in \S~\ref{sec:sailfish-protocol-review}). If $v_l \in \mathcal{H}$, we are trivially done. Otherwise, the vertices in $\mathcal{H}$ are from correct non-leader nodes in round $r+1$. Since $|\mathcal{H}| \geq f+1$ and the nodes creating the vertices in $\mathcal{H}$ do not send a round $r$ $\NoVote$ message, by the quorum intersection argument, the round $r'$ leader node cannot form a $\NoVoteCert_{r}$ for $v_l$ such that $v_l$ does not connect to $v_k$. Therefore, if $v_l$ exists, it must connect to $v_k$, and there is a path from $v_l$ to $v_k$.

    \textbf{Case} $r'>r+1$: Since Sailfish adopts RBC to propagate vertices, all vertices in $\mathcal{H}$ ($|\mathcal{H}| \geq f+1$) will eventually be delivered by all correct nodes and added to $DAG[r+1]$. Furthermore, according to the construction of the DAG, a round $r+2$ vertex must connect to at least $2f+1$ round $r+1$ vertices. By a quorum intersection argument, each round $r+2$ vertex (including the round $r+2$ leader vertex) must have a path to $v_k$. Similarly, each round $r+3$ vertex must have a path to a round $r+2$ vertex and thus has a path to $v_k$. By transitivity, all vertices in rounds $r'' > r+3$ have a path to $v_k$. This implies that $v_l$ in round $r'>r+1$ must have a path to $v_k$.

    2) If a fallback view selects $r$, $v_k$ is included in its ACS output $V$. By the transition rule of Sailfish-\sysname, all first post-recovery vertices in round $r+2$ connect to every member of $V$ and thus have a path to $v_k$. If $r'=r+2$, then we are done. Otherwise, as in the second case under 1), we can prove that $v_l$ must have a path to $v_k$. Later fallback transitions preserve this path because their selected rounds strictly increase, and their transition vertices extend their ACS outputs, so the same argument applies inductively.
\end{myproof}

\begin{claim}\label{claim:leader-transitive}
    If a correct node $\node{i}$ directly commits a leader vertex $v_k$ in round $r$ and a correct node $\node{j}$ directly commits a leader vertex $v_l$ in round $r' \geq r$, then $\node{j}$ (directly or indirectly) commits $v_k$ in round $r$.
\end{claim}
\begin{myproof}{}
    If $r=r'$, by Claim~\ref{claim:same-acs-leader}, $v_k=v_l$ and we are trivially done. When $r' > r$, there are two cases.

    \textbf{Case} A fallback view selects round $r+1$: By Claim~\ref{claim:leader-before-post}, all correct nodes (including $\node{j}$) directly commit $v_k$ in round $r$.

    \textbf{Case} No fallback view selects round $r+1$: By Claim~\ref{claim:leader-path}, there exists a path from $v_l$ to $v_k$. By the code of the \Call{commit\_leader}{} function, after directly committing $v_l$, $\node{j}$ indirectly commits leader vertices $v_m$ in smaller rounds such that there exists a path from $v_l$ to $v_m$, until it reaches a round $r''<r'$ in which it previously directly committed a leader vertex. If $r''<r<r'$, $\node{j}$ will indirectly commit $v_k$ in round $r$. Otherwise, by an inductive argument, $\node{j}$ must have indirectly committed $v_k$ by directly committing the round $r''$ leader vertex.

\end{myproof}

\begin{claim}\label{claim:total-order}
    Let $v_k$ and $v_k'$ be two leader vertices consecutively directly committed by a party $\node{i}$ in rounds $r_i$ and $r_i' > r_i$, respectively. Let $v_l$ and $v_l'$ be two leader vertices consecutively directly committed by $\node{j}$ in rounds $r_j$ and $r_j'>r_j$, respectively. Then, $\node{i}$ and $\node{j}$ commit the same leader vertices between rounds $\max(r_i, r_j)$ and $\min(r_i', r_j')$ and in the same order.
\end{claim}
\begin{myproof}{}
    If $r_i'<r_j$ or $r_j'<r_i$, then there are no rounds between $\max(r_i, r_j)$ and $\min(r_i', r_j')$, and we are trivially done. Otherwise, assume without loss of generality that $r_i \leq r_j < r_i'$. By Claim~\ref{claim:same-acs-leader} and \ref{claim:leader-transitive}, both $\node{i}$ and $\node{j}$ will (directly or indirectly) commit the same leader vertex in the round $\min(r_i', r_j')$. Assume $\min(r_i', r_j') = r_i'$. Given that Sailfish-\sysname propagates vertices via RBC, both $DAG_i$ and $DAG_j$ will contain $v_k'$ and all vertices that have a path from $v_k'$.

    Recall that, in the implementation of the \Call{commit\_leader}{} function, after committing the leader vertex $v_k'$, nodes indirectly commit leader vertices in smaller rounds until they reach a round in which they previously directly committed a leader vertex. Consequently, both $\node{i}$ and $\node{j}$ will be able to indirectly commit all leader vertices from $r_i'$ to $r_j$ (i.e., from $\min(r_i', r_j')$ to $\max(r_i, r_j)$ under our general assumption). Furthermore, thanks to the deterministic code of \Call{commit\_leader}{}, both nodes will commit the same leader vertices from $\min(r_i', r_j')$ to $\max(r_i, r_j)$ in the same order.
\end{myproof}

By applying Claim~\ref{claim:total-order} inductively to any pair of correct nodes, we derive the following corollary.

\begin{corollary}\label{corollary:total-order}
    Correct nodes commit the same leader vertices in the same order.
\end{corollary}

\begin{lemma}[Safety]
    Sailfish-\sysname satisfies safety.
\end{lemma}
\begin{myproof}{}
    By Corollary~\ref{corollary:total-order}, nodes will commit the same leader vertices in the same order. With a deterministic ordering algorithm implemented in the \Call{order\_vertices}{} function, nodes consistently order the committed leader vertices as well as their causal histories. Given the same transaction payload in each DAG vertex, all correct nodes commit transactions in the same order. The safety property is guaranteed.
\end{myproof}

\subsection{Liveness}
We now analyze the liveness of Sailfish-\sysname. Liveness indicates that the system can continuously commit new transactions. We consider nodes to have limited resources for maintaining pending DAG vertices. A node enters a preparatory phase before exhausting its resources: it may stop creating DAG vertices but reserves bounded resources for $\mathsf{READY}$ messages, activation certificates, PoST blocks, commit witnesses/evidence, and the fixed proposal and current-iteration state of one ACS instance. Commit progress can cancel only this preparatory state; after activation, the node remains in fallback until the view's ACS decision. Views are completed in order: stale messages are ignored, while a valid future-view activation certificate triggers catch-up through all unfinished earlier views.

Node $\node{i}$ sends one signed $\mathsf{READY}(v_{fb})$ vote for its current fallback view when either of the following conditions holds:
{
\makeatletter
\def\@listi{\leftmargin10pt \labelwidth\z@ \labelsep5pt}    
\makeatother
\begin{itemize}
    \item \textbf{Condition 1:} $\node{i}$ approaches its configured resource limit;
    \item \textbf{Condition 2:} $\node{i}$ receives a valid $\mathsf{READY}(v_{fb})$ message and its committed round does not advance during a timeout $T_{st}$.
\end{itemize}
}

Any $2f+1$ distinct $\mathsf{READY}(v_{fb})$ signatures form an activation certificate $AC_{v_{fb}}$. A correct node constructs its PoST only after it receives $AC_{v_{fb}}$, atomically stops optimistic-path vertex creation, and takes a fresh snapshot of its last created vertex. The node then remains in fallback until ACS terminates.

\begin{claim}\label{claim-sailfish-receive-post}
    If commitment stops, then all correct nodes eventually receive an activation certificate for view $v_{fb}$.
\end{claim}
\begin{myproof}{}
    If a correct node has already completed view $v_{fb}$, then an activation certificate for that view must already have formed. Otherwise, after any preceding fallback, the common output supplies enough references for each correct node to create its required transition vertex without waiting for new-round messages. If commitment stops, every correct node then eventually reaches its configured limit or lets the no-commit timer started by a current-view $\mathsf{READY}$ expire and multicasts its $\mathsf{READY}(v_{fb})$ vote. Thus, the $2f+1$ correct signatures alone suffice to form $AC_{v_{fb}}$. Any correct node that forms or receives the certificate disseminates it, so all correct nodes eventually receive it.
\end{myproof}

\begin{claim}\label{claim-sailfish-handle-post}
    If one correct node activates fallback view $v_{fb}$, then every correct node eventually completes the same ACS instance.
\end{claim}
\begin{myproof}{}
    A correct node activates only after receiving $AC_{v_{fb}}$, which it disseminates. Hence, every correct node whose unfinished view is $v_{fb}$ eventually receives the certificate after GST; a node in an earlier view invokes catch-up, finalizes the intervening views' ACS outputs in order, and then processes $AC_{v_{fb}}$. If no correct node has yet decided this ACS instance, every correct node freezes optimistic vertex creation, constructs and disseminates a fresh PoST, and signs every valid fresh PoST for that view. Sailfish's certified DAG makes the vertices in correct PoST blocks available, so the correct nodes invoke $\ACSPropose$, satisfying ACS termination. If a correct node has already decided, ACS agreement instead guarantees that every other correct node eventually outputs the same decision. Thus, all correct nodes finalize the same ACS output for $v_{fb}$.
\end{myproof}

\begin{claim}\label{claim-sailfish-commit-dag}
    Sailfish-\sysname can continuously commit new DAG vertices even though correct nodes have limited resources for maintaining pending DAG vertices.
\end{claim}
\begin{myproof}{}
    By \textbf{Condition 1}, a correct node $\node{i}$ enters the preparatory phase before exhausting its working storage. The configured limit reserves sufficient resources for activation, PoST processing, one ACS instance, compact commit witnesses, and streaming synchronization. There are two cases:

    \textbf{Case} The optimistic path keeps committing: Some correct node eventually commits a later leader directly. If it uses the first-message path, it multicasts the corresponding $2f+1$-message witness after receiving $\mathsf{READY}(v_{fb})$; if it uses the RBC-delivery path, RBC totality eventually exposes the same evidence. Thus, $\node{i}$ obtains valid direct-commit evidence, commits that leader and any pending predecessors through the indirect rule, and returns to the optimistic path (\Cref{fig:sailfish-fallback-basic}, $\aln$~\ref{step:sf_return_op_start}-\ref{step:sf_return_op_end}). If a previously formed $AC_{v_{fb}}$ later arrives, \Cref{claim-sailfish-handle-post} ensures that all correct nodes instead complete the same ACS instance and then return to the optimistic path.

    \textbf{Case} The optimistic path stops committing: By \Cref{claim-sailfish-receive-post}, all correct nodes eventually receive $AC_{v_{fb}}$, and by \Cref{claim-sailfish-handle-post}, they all complete the same ACS instance. ACS outputs a common set $V$ with $|V|\geq n-f$, so every correct node obtains enough transition references to move to a new round. Finalization clears the completed view's temporary state, allowing the same bounded ACS workspace and argument to be reused for the next view. Thus, by induction, every activated fallback view completes, even if fallbacks occur repeatedly, and the underlying Sailfish protocol can again continuously generate and commit new DAG vertices between them.
\end{myproof}

\begin{lemma}[Liveness]
    Sailfish-\sysname satisfies liveness.
\end{lemma}
\begin{myproof}{}
    By \Cref{claim-sailfish-commit-dag}, Sailfish-\sysname can continuously generate and commit new DAG vertices. As the protocol allows nodes to create new-round DAG vertices only when there are at least $2f+1$ vertices for the last round, at least $f+1$ correct nodes can continuously generate and commit new DAG vertices. By fair inclusion, transactions sent to all correct nodes are eventually included and delivered, satisfying liveness.
\end{myproof}

\section{Security Analysis of Mysticeti-\sysname} \label{sec:detailed-security-analysis-mysticeti}
In this section, we give a detailed security analysis of Mysticeti-\sysname, presented in \S~\ref{sec:mysticeti-sysname}.

Recall that, to order DAG vertices, the Mysticeti protocol iteratively assigns \textit{all} predefined leader vertices either $\tocommit$ or $\toskip$ status. By applying the decision rules to interpret the DAG pattern of each leader vertex, Mysticeti allows nodes to order leader vertices consistently. When integrating \sysname into Mysticeti, the only modification to DAG vertex ordering is that Mysticeti-\sysname may select a new leader vertex $\post_L^{r^*}$ to replace the predefined leader vertex $v_L^{r^*}$ from the same round $r^*$ after each ACS instance. Thus, we follow the security proof of Mysticeti~\cite[Section V and Appendix C]{mysticeti} to complete our proof while considering the subtle modification introduced by \sysname.
For each fallback view $v$, ACS agreement and the deterministic output construction ensure that all correct nodes obtain the same ACS output $V$.
Let $r_v$ be the highest PoST round selected in view $v$. The transition rule (that is, before signing $\mathsf{READY}$ in view $v+1$, a correct node creates its round-$r_v+2$ transition vertex), together with quorum intersection between the correct signers of $AC_{v+1}$ and the correct proposers represented in its ACS output, ensures that $r_{v+1}\geq r_v+2$. Thus, each round is selected by at most one fallback view, and the following arguments compose across fallback views.

\subsection{Safety}\label{sec:my-safety-detailed-proof}
We first analyze the safety of Mysticeti-\sysname. Safety indicates that all correct nodes commit transactions in the same order.
\begin{claim}\label{claim:mt-no-skip}
    If a correct node commits a leader vertex $v$ in round $r$, then no correct node decides to directly skip $v$.
\end{claim}
\begin{myproof}{}
    We prove it by contradiction. Recall that a node $\node{i}$ decides to directly skip a leader vertex $v$ in round $r$ if $\node{i}$ observes a \textit{skip pattern} for $v$, that is, at least $2f+1$ round $r+1$ vertices do not reference $v$. If another correct node $\node{j}$ commits $v$, by the decision rules, there is at least one \textit{certificate pattern} for $v$, that is, at least $2f+1$ round $r+1$ vertices reference $v$. By a quorum intersection argument, at least one correct node must have proposed two vertices in round $r+1$ that, respectively, reference and do not reference $v$, which is a contradiction.
\end{myproof}

\begin{claim}\label{claim:mt-certificate-path}
    For any round $r$ leader vertex $v$, if $2f+1$ vertices in round $r'$ from distinct nodes certify $v$, then all leader vertices in round $r''>r'$ will have a path to a certificate for $v$.
\end{claim}
\begin{myproof}{}
    Recall that a round $r'$ vertex $v'$ is said to certify a round $r$ vertex $v$ (where $r'>r+1$) if $v'$ includes a \textit{certificate pattern} in its causal history. $v'$ is also called a certificate for $v$. We prove this statement by contradiction. Assume that there is a leader vertex $v''$ in round $r''>r'$ that does not reference a certificate for $v$. Since $v''$ must reference at least $2f+1$ vertices from the previous round, if $r''=r'+1$, by a quorum intersection argument, a correct node must have equivocated in round $r'$, which is a contradiction. Thus, all vertices in round $r'+1$ reference a certificate for $v$. By transitivity, all vertices (including leader vertices) in round $r''>r'$ have a path to a certificate for $v$.
\end{myproof}

\begin{claim}
    If a correct node $\node{i}$ directly commits a leader vertex $v$ in round $r$, then no correct node decides to skip $v$.
\end{claim}
\begin{myproof}{}
    We prove it by contradiction. Assume that a correct node $\node{i}$ directly commits a round $r$ leader vertex $v$ while another correct node $\node{j}$ decides to skip $v$. We consider all cases in which $\node{j}$ decides to skip $v$.

    \textbf{Case} $\node{j}$ decides to directly skip $v$: This is a contradiction of Claim~\ref{claim:mt-no-skip}.

    \textbf{Case} $\node{j}$ decides to indirectly skip $v$: According to the indirect decision rule, there is a committed leader vertex $v'$ in round $r'>r+2$ that neither references nor has a path to a certificate for $v$. However, since $\node{i}$ directly commits $v$, there are $2f+1$ certificates for $v$, contradicting Claim~\ref{claim:mt-certificate-path}.
\end{myproof}

\begin{claim}\label{claim:mt-at-most-2f}
    Given any round number $r^*$ that is decided by an ACS instance, at most $2f$ vertices from distinct nodes can be created in round $r^*+1$.
\end{claim}
\begin{myproof}{}
    Let $V$ be the ACS output defining $r^*=\max\{\post.\vertex.\round\mid\post\in V\}$. By ACS validity, $V$ contains PoST blocks from a set $\mathcal{P}$ of at least $f+1$ correct creators. Each creator in $\mathcal{P}$ constructs its PoST only after receiving the activation certificate and atomically stopping optimistic-path vertex creation; hence, its PoST contains its last optimistically created vertex.

    Suppose, for contradiction, that a set $\mathcal{Q}$ of $2f+1$ distinct creators produces round $r^*+1$ optimistic vertices. Since $|\mathcal{P}|\geq f+1$, $|\mathcal{Q}|=2f+1$, and $n=3f+1$, their creator sets intersect. Let $\node{k}\in\mathcal{P}\cap\mathcal{Q}$. Node $\node{k}$ is correct, so it created its round $r^*+1$ vertex before freezing and constructing $\post_k$. Therefore, $\post_k.\vertex.\round\geq r^*+1$, contradicting the definition of $r^*$. After ACS, the graceful transition starts correct nodes at round $r^*+2$, so they create no additional round $r^*+1$ vertices.
\end{myproof}

\para{Proof of \Cref{proposition-highest-leader-not-commit}} By \Cref{claim:mt-at-most-2f}, at most $2f$ vertices are created in round $r^*+1$. Therefore, before the fallback terminates, the predefined round $r^*$ leader vertex cannot collect $\geq2f+1$ votes and thus cannot be directly committed according to Mysticeti's direct committing rule. Moreover, recall that the indirect committing rule requires a round $r'>r^*$ leader vertex to be directly committed; in that case, the fallback instance must terminate. This completes the proof.

\begin{claim}\label{claim:mt-same-vertex}
    If a node $\node{i}$ marks a leader vertex $v$ in round $r$ as $\tocommit$ or $\toskip$, and a node $\node{j}$ marks a leader vertex $v'$ in round $r$ as $\tocommit$ or $\toskip$, then $v = v'$.
\end{claim}
\begin{myproof}{}
    If no fallback view selects round $r$, we are trivially done as both $v$ and $v'$ are predefined. Otherwise, let $r^*=r$ be the round selected by its unique fallback view. If $v_L=\post_L$ (i.e., the ACS instance does not change the leader vertex), we are also trivially done.

    We now consider the case where $r=r^*$ and $v_L\neq\post_L$. Without loss of generality, let $v=v_L$ and $v'=\post_L$. This indicates that $\node{i}$ has decided $v$ before terminating the ACS instance. By the decision rules, the earliest round in which a node can decide a leader vertex is round $r+1$ (i.e., the node directly skips it after observing that $2f+1$ round $r+1$ vertices do not reference it). However, by Claim~\ref{claim:mt-at-most-2f}, neither $v$ nor $v'$ can be decided in round $r+1$. $\node{i}$ must wait for $2f+1$ round $r+2$ vertices to decide $v$. However, upon receiving $f+1$ round $r+2$ vertices that reference the common ACS output, $\node{i}$ must be able to terminate the ACS instance. This is because these $f+1$ vertices contain at least one correct vertex, indicating that at least one correct node has terminated the ACS instance. $\node{i}$ then updates the round $r$ leader vertex to $\post_L$, leading to a contradiction.
\end{myproof}



\begin{claim}\label{claim:mt-same-order-leader}
    All correct nodes consistently decide the status of the leader vertex in each round, that is, if two correct nodes $\node{i}$ and $\node{j}$ decide the status of a leader vertex, then both either commit or skip the same leader vertex.
\end{claim}
\begin{myproof}{}
    Let $[v^i]_{i=0}^k$ and $[v^i]_{i=0}^l$ denote the statuses of the leader vertices for $\node{i}$ and $\node{j}$, where $k$ and $l$ are the indices of the highest committed rounds, respectively. By definition, $\node{i}$ directly commits the round $k$ leader vertex, and $\node{j}$ directly commits the round $l$ leader vertex. Assume without loss of generality that $k \leq l$. We now use induction to prove statement $P(r)$ for $0\leq r \leq k$: if $\node{i}$ and $\node{j}$ both decide the round $r$ leader vertex, then both either commit or skip the same leader vertex.

    Base case $r=k$: If $\node{i}$ directly commits the round $k$ leader vertex $v_k$, then, by Claims~\ref{claim:mt-no-skip} and \ref{claim:mt-same-vertex}, $\node{j}$ must commit $v_k$ if it decides the round $k$ leader vertex.

    Inductive hypothesis and step: Assume that $P(r)$ is true for $m+1 \leq r \leq k$. We now prove $P(m)$. As in the base case, if $\node{i}$ directly commits a round $m$ leader vertex $v_m$, then $\node{j}$ commits $v_m$ if it decides the round $m$ leader vertex. Additionally, if $\node{i}$ directly skips $v_m$, by the inverse of Claim~\ref{claim:mt-no-skip}, $\node{j}$ skips $v_m$. We now analyze the only remaining case, in which $\node{i}$ and $\node{j}$ indirectly decide $v_m$. Let $m'$ denote the first round with a round number higher than the decision round of $v_m$. There are rounds $m_i\geq m'$ and $m_j \geq m'$ such that $\node{i}$ commits a leader vertex $v_i$ in $m_i$ while skipping all leader vertices in $[m', m_i)$ and $\node{j}$ commits a leader vertex $v_j$ in $m_j$ while skipping all leader vertices in $[m', m_j)$. Since $m_i \leq k$, by the induction hypothesis, we have $m_i=m_j$ and $v_i=v_j=v$. This means that $\node{i}$ and $\node{j}$ use the same leader vertex to indirectly decide $v_m$. Thus, $\node{i}$ and $\node{j}$ have the same DAG pattern for $v_m$ with the same causal history of $v$ and consistently decide the round $m$ leader vertex.
\end{myproof}

\begin{claim}[Data availability]\label{claim:mt-data-availability}
    For any round $r$ leader vertex $v$ that is marked $\tocommit$, all correct nodes can derive the entire causal history of $v$.
\end{claim}
\begin{myproof}{}
    If round $r$ is not selected by any fallback view, we are trivially done, as \sysname does not affect the underlying committing rules (i.e., a node votes for a leader vertex only if it receives the entire causal history of the leader vertex). Otherwise, consider the unique fallback view selecting $r$. For a certified PoST block $\post_i$, let $\mathcal{P}$ be the creators of the $2f+1$ round $r'=\post_i.\vertex.\round-1$ parents referenced by $\post_i.\vertex$, and let $\mathcal{S}$ be the $2f+1$ signers certifying $\post_i$. The parent signatures carried by $\post_i$ prove that every identity in $\mathcal{P}$ signed its claimed round, creator identity, and parent digest. Since $|\mathcal{P}|=|\mathcal{S}|=2f+1$ and $n=3f+1$, $|\mathcal{P}\cap\mathcal{S}|\geq f+1$; hence, the intersection contains a correct node $\node{j}$. Because a correct $\node{j}\in\mathcal{P}$ created its round $r'$ parent, it is processing in a round at least $r'$. Therefore, $\post_i.\vertex.\round-\mathit{currentRound}_j\leq1$, and $\node{j}$ synchronizes the complete causal history of $\post_i$ if that history is missing and retains it before signing $\post_i$ (Figure~\ref{fig:mysticeti-fallback-basic}, $\aln$~\ref{step:mt-post-available-verify-0}-\ref{step:mt-post-available-verify-1}). Thus, every certified PoST selected by ACS has its causal history retained by at least one correct node, from which all correct nodes can derive it. The same argument applies to every member of the ACS output $V$, so histories represented by $V$ remain obtainable during the transition.
\end{myproof}

\begin{lemma}[Safety]
    Mysticeti-\sysname satisfies safety.
\end{lemma}
\begin{myproof}{}
    Mysticeti-\sysname requires nodes to decide a leader vertex in each round. Therefore, by Claim~\ref{claim:mt-same-order-leader}, all correct nodes will commit the same leader vertices in the same order. By \Cref{claim:mt-data-availability}, correct nodes can collect the causal histories of all committed leader vertices. With a deterministic ordering algorithm, all correct nodes order DAG vertices in the same order. Given the same transaction payload in each DAG vertex, all correct nodes commit transactions in the same order. The safety property is guaranteed.
\end{myproof}

\subsection{Liveness}\label{sec-my-liveness-proof}
We now analyze the liveness of Mysticeti-\sysname. Liveness indicates that the system can continuously commit new transactions. We consider nodes to have limited resources for maintaining pending DAG vertices. A node enters a preparatory phase before exhausting its resources: it may stop creating DAG vertices but reserves bounded resources for $\mathsf{READY}$ messages, activation certificates, PoST history synchronization, commit witnesses/evidence, and the fixed proposal and current-iteration state of one ACS instance. Commit progress can cancel only this preparatory state; after activation, the node remains in fallback until the view's ACS decision. Views are completed in order: stale messages are ignored, while a valid future-view activation certificate triggers catch-up through all unfinished earlier views.

Node $\node{i}$ sends one signed $\mathsf{READY}(v_{fb})$ vote for its current fallback view when either of the following conditions holds:
{
\makeatletter
\def\@listi{\leftmargin10pt \labelwidth\z@ \labelsep5pt}    
\makeatother
\begin{itemize}
    \item \textbf{Condition 1:} $\node{i}$ approaches its configured resource limit;
    \item \textbf{Condition 2:} $\node{i}$ receives a valid $\mathsf{READY}(v_{fb})$ message and its committed round does not advance during a timeout $T_{st}$.
\end{itemize}
}

Any $2f+1$ distinct $\mathsf{READY}(v_{fb})$ signatures form an activation certificate $AC_{v_{fb}}$. A correct node constructs its PoST only after it receives $AC_{v_{fb}}$, atomically stops optimistic-path vertex creation, and takes a fresh snapshot of its last created vertex. The node then remains in fallback until ACS terminates.

\begin{claim}\label{claim-mysticeti-receive-post}
    If commitment stops, then all correct nodes eventually receive an activation certificate for view $v_{fb}$.
\end{claim}
\begin{myproof}{}
    If a correct node has already completed view $v_{fb}$, then an activation certificate for that view must already have formed. Otherwise, after any preceding fallback, the common output supplies enough references for each correct node to create its required transition vertex without waiting for new-round messages. If commitment stops, every correct node then eventually reaches its configured limit or lets the no-commit timer started by a current-view $\mathsf{READY}$ expire and multicasts its $\mathsf{READY}(v_{fb})$ vote. Thus, the $2f+1$ correct signatures alone suffice to form $AC_{v_{fb}}$. Any correct node that forms or receives the certificate disseminates it, so all correct nodes eventually receive it.
\end{myproof}

\begin{claim}\label{claim-mysticeti-handle-post}
    If one correct node activates fallback view $v_{fb}$, then every correct node eventually completes the same ACS instance.
\end{claim}
\begin{myproof}{}
    A correct node activates only after receiving $AC_{v_{fb}}$, which it disseminates. Hence, every correct node whose unfinished view is $v_{fb}$ eventually receives the certificate after GST; a node in an earlier view invokes catch-up, finalizes the intervening views' ACS outputs in order, and then processes $AC_{v_{fb}}$. If no correct node has yet decided this ACS instance, every correct node freezes optimistic vertex creation, constructs and disseminates a fresh PoST, and verifies every valid fresh PoST for that view. For a correct PoST, the bounded Mysticeti history synchronization terminates and the correct nodes return signatures, invoke $\ACSPropose$, and satisfy ACS termination. If a correct node has already decided, ACS agreement instead guarantees that every other correct node eventually outputs the same decision. Thus, all correct nodes finalize the same ACS output for $v_{fb}$.
\end{myproof}

\begin{claim}\label{claim-mysticeti-commit-dag}
    Mysticeti-\sysname can continuously commit new DAG vertices even though correct nodes have limited resources for maintaining pending DAG vertices.
\end{claim}
\begin{myproof}{}
    By \textbf{Condition 1}, a correct node $\node{i}$ enters the preparatory phase before exhausting its working storage. The configured limit reserves sufficient resources for: (i) activation messages; (ii) synchronizing missing DAG vertices required to validate PoST blocks; (iii) managing up to $n$ PoST blocks and one ACS instance; and (iv) processing compact commit witnesses and streaming a committed prefix while retaining only a bounded live suffix. There are two cases:

    \textbf{Case} The optimistic path keeps committing: A correct node that observes a direct-commit pattern uses its next-round vertex as a carrier referencing the $2f+1$ certificate vertices. Broadcasts by correct nodes and bounded history synchronization ensure that $\node{i}$ eventually receives such a carrier and reconstructs the same direct-commit evidence. It therefore commits that leader and any pending predecessors through Mysticeti's decision rules, observes commit progress, and returns to the optimistic path (\Cref{fig:mysticeti-fallback-basic}, $\aln$~\ref{step:mt_return_op_start}-\ref{step:mt_return_op_end}). If a previously formed $AC_{v_{fb}}$ later arrives, \Cref{claim-mysticeti-handle-post} ensures that all correct nodes instead complete the same ACS instance and then return to the optimistic path.

    \textbf{Case} The optimistic path stops committing: By \Cref{claim-mysticeti-receive-post}, all correct nodes eventually receive $AC_{v_{fb}}$, and by \Cref{claim-mysticeti-handle-post}, they all complete the same ACS instance. ACS outputs a common set $V$ with $|V|\geq n-f$ and marks a newly selected leader vertex $\post_L$ as $\tocommit$. By \Cref{claim:mt-data-availability}, all correct nodes can derive its causal history and commit pending DAG vertices. Finalization supplies enough transition references to move to a new round and clears the completed view's temporary state, allowing the same bounded ACS workspace and argument to be reused for the next view. Thus, by induction, every activated fallback view completes, even if fallbacks occur repeatedly, and Mysticeti can again continuously generate and commit new DAG vertices between them.
\end{myproof}

\begin{lemma}[Liveness]
    Mysticeti-\sysname satisfies liveness.
\end{lemma}
\begin{myproof}{}
    By \Cref{claim-mysticeti-commit-dag}, Mysticeti-\sysname can continuously generate and commit new DAG vertices. As the protocol allows nodes to create new-round DAG vertices only when there are at least $2f+1$ vertices for the last round, at least $f+1$ correct nodes can continuously generate and commit new DAG vertices. By fair inclusion, transactions sent to all correct nodes are eventually included and delivered, satisfying liveness.
\end{myproof}

\section{The Core Pseudocode of Mysticeti-\sysname} \label{sec-pseudo-my-sysname}

\begin{figure}[t]
    \scriptsize
	\begin{boxedminipage}[t]{\linewidth}
		\textbf{Variables:}
		\begin{itemize}[nosep]
                \item[] \gtext{$DAG_i[] - $ An array of sets of vertices (indexed by rounds)}
                \item[] $v_{fb} \gets 0$ \Comment{the monotonically increasing fallback-view number}
                \item[] $\mymode \gets OP$ \Comment{$OP$: optimistic; $PP$: preparatory; $FP$: activated fallback}
                \item[] struct PoST block $\post$ $\{ \cdots, \post.\mathit{parentSigs} \}$   \Comment{add signed parent digests}
                \item[] $\mathit{postBuffer} \gets \{\}$ \Comment{PoST blocks missing history data}
		\end{itemize}

			\begin{algorithmic}[1]
				\Procedure{create\_new\_vertex}{$r$}
                \If{$\mymode \neq OP$}
                    \State \Return $\bot$
                \EndIf
                \State \gtext{$\cdots$} \Comment{\gtext{return a round $r$ DAG vertex if processing in the optimistic path}}
            \EndProcedure
            \vspace{-0.2em}

            \Event{a local fallback condition after creating any required transition vertex} \label{step:mt-ready-start}
                \If{$\mymode = OP$}
                    \State $\mymode \gets PP$ \Comment{preserve resources without invoking ACS}
                    \State if not already sent in this view, multicast $\sig{\mathsf{READY}(v_{fb})}_i$
                \EndIf
            \EndEvent \label{step:mt-ready-end}
            \vspace{-0.2em}

            \Event{receiving valid $\sig{\mathsf{READY}(v_{fb})}_j$}
                \State start $T_{st}$ for view $v_{fb}$ if no such timer is running
            \EndEvent
            \vspace{-0.2em}

            \Event{receiving $2f+1$ distinct $\sig{\mathsf{READY}(v_{fb})}_*$} \label{step:mt-activation-cert-start}
                \State $AC_{v_{fb}} \gets$ a certificate formed by the $2f+1$ signatures
            \EndEvent \label{step:mt-activation-cert-end}
            \vspace{-0.2em}

            \Event{forming or receiving a valid $AC_u$} \label{step:mt-switch_fallback}
                \If{$u<v_{fb}$} \State \Return \Comment{ignore a stale view} \EndIf
                \If{$u>v_{fb}$} \State \Call{catch\_up}{$AC_u$}; \Return \Comment{complete earlier views in order, then $AC_u$} \EndIf
                \State multicast $AC_u$ if not already disseminated
                \If{$\mymode \neq FP$}
                    \State $\mymode \gets FP$ \Comment{irrevocably activate the fallback view}
                    \State $r \gets \max \{ r \mid \exists B \in DAG_i[r] \land B.\creator = \node{i} \}$
                    \State \Call{create\_new\_post}{$r, AC_u$} \Comment{snapshot after freezing}
                \EndIf
            \EndEvent
            \vspace{-0.2em}

            \Event{committing transactions while $\mymode=PP$} \label{step:mt_return_op_start}
            \State reset $T_{st}$
            \State $\mymode\gets OP$ 
            \EndEvent   \label{step:mt_return_op_end}
            \vspace{-0.2em}

            \Procedure{create\_new\_post}{$r, AC_{v_{fb}}$} \label{step:mt-create_post_start}  \Comment{PoST \texttt{Propose}}
                \State $\post_{i}.\view \gets v_{fb}$   \Comment{bind the fresh state to the activated view}
                \State $\post_{i}.\vertex \gets$ $v$ s.t $v \in DAG_i[r] \land v.\creator = \node{i}$\strut
                \State $\post_{i}.\creator \gets \node{i}$
                \State $\post_i.\mathit{activation} \gets AC_{v_{fb}}$
                \State $\post_i.\mathit{parentSigs} \gets$ creators' signatures for all parent digests of $\post_i.\vertex$
                \State multicast $\post_{i}$
            \EndProcedure   \label{step:mt-create_post_end}
            \vspace{-0.2em}

            \Event{receiving $\post_{j}$}   \label{step:mt-vote_post_start}
                \State process $\post_j.\mathit{activation}$ as a received activation certificate
                \If{$\post_j.\view=v_{fb} \land \post_j.\mathit{activation}$ is a valid $AC_{v_{fb}}$}
                    \If{\Call{is\_post\_valid}{$\post_{j}$}} \Comment{creator uniqueness, parent signatures, and vertex validity}
                        \If{ ($\post_{j}.\vertex.\round{-}\mathit{currentRound} \leq$ 1)${\land}$\Call{miss\_history}{$\post_{j}$} }  \label{step:mt-post-available-verify-0}
                            \State  $\mathit{postBuffer} \gets \mathit{postBuffer} \cup \{\post_{j}\}$
                            \State \Call{synchronize\_post}{$\post_{j}$} \label{step:mt-post-available-verify-1} \Comment{sync missing blocks and then call back}
                        \Else
                            \State send $\sig{\post_{j}}_i$ to $\node{j}$ \Comment{sign $\post_j$ and send it back}
                        \EndIf
                    \EndIf
                \EndIf
            \EndEvent   \label{step:mt-vote_post_end}
            \vspace{-0.2em}

            \Event{receiving $2f+1$ $\sig{\post_{i}}_*$ for $\post_i$ while $\mymode=FP$ and $\post_i.\view=v_{fb}$} \label{step:mt-acs_propose_start}
                \State $\post_{i}.\cert \gets$ a certificate formed from $2f+1$ $\sig{\post_{i}}_*$
                \State multicast $\post_i$; $\ACSPropose_i(v_{fb},\post_{i})$
            \EndEvent   \label{step:mt-acs_propose_end}
                \vspace{-0.2em}
        \algstore{bkbreak2}
		\end{algorithmic}
    \end{boxedminipage}
    \caption{Mysticeti-\sysname: vertex creation for node $\node{i}$; the code shown in \gtext{gray} comes from Mysticeti~\cite{mysticeti}}
    \label{fig:mysticeti-fallback-basic}
\end{figure}

\begin{figure}[t]
    \scriptsize
	\begin{boxedminipage}[t]{\linewidth}
		\textbf{Variables:}
            {
                \makeatletter
                \def\@listi{\leftmargin17pt \labelwidth\z@ \labelsep5pt}    
                \makeatother
		\begin{itemize}[nosep]
			\item[] \gtext{$\mathit{currentRound} \gets 1; \mathit{committedRound} \gets 0;$}
                \item[] \gtext{$\mathit{lastReferences} - $ The parent or transition references for a new block}
                \item[] $\post_L$ - the leader block selected by the ACS instance
		\end{itemize}
		}
		\begin{algorithmic}[1]
                \algrestore{bkbreak2}
                \begin{graytext}

    		\Procedure{try\_decide}{$ $} \label{step:mt-try-decide-0}
                \State $\mathit{sequence} \gets []$
                \For{$r \in  [\mathit{currentRound} \text{ down to } \mathit{committedRound}+1]$ }
                \If{$\exists v \in DAG_i[r]: v.\creator=L_r$}
                    \State $\mathit{status} \gets$ \Call{try\_direct\_decide}{$v$} \Comment{direct decide} \label{step:mt-dir-decide}
                    \If{$\neg\mathit{status}$.\Call{is\_decided}{$ $}}
                    \State $\mathit{status} \gets$ \Call{try\_indirect\_decide}{$v, \mathit{sequence}$} \Comment{indirect decide} \label{step:mt-indir-decide}
                    \EndIf
                    \State $\mathit{sequence} \gets \mathit{status} || \mathit{sequence}$
                \EndIf
                \EndFor
                \State $\mathit{decided} \gets []$
                \For{$\mathit{status} \in \mathit{sequence}$}
                \If{$\neg\mathit{status}$.\Call{is\_decided}{$ $}} break
                \EndIf
                \State $\mathit{decided} \gets \mathit{decided} || \mathit{status}$
                \EndFor
                \State \Call{finalize\_decided\_vertices}{$\mathit{decided}$}
                \EndProcedure   \label{step:mt-try-decide-1}
                \vspace{-0.2em}
                \end{graytext}

                \Event{$\ACSDecide_i(v_{fb},V)$} \label{step:mt-acs-decide-start}
                \If{$\forall\post \in V, v_{fb} = \post.\view$}
                \State \Call{finalize\_fallback}{$V$}
                \EndIf
                \EndEvent   \label{step:mt-acs-decide-end}
                \vspace{-0.2em}

                \Procedure{finalize\_fallback}{$V$} \label{step:mt-finalize-fallback-start}
                \State $r_{fb} \gets \max \{\post.\vertex.\round \mid \post \in V\}$ \label{step:mt-set-acs-round}
                \If{$\exists \post' \in V:$ $\post'.\vertex.\creator = L_{r_{fb}} \land \post'.\vertex.\round = r_{fb}$}  \label{step:mt-finalize-fallback-predefine-0}
                \State $\post_L \gets \post'$   \label{step:mt-finalize-fallback-predefine-1}
                \Else   \label{step:mt-finalize-fallback-newdefine-0}
                \State $\post_L \gets$ deterministically select a $\post'\in V$ s.t. $\post'.\vertex.\round = r_{fb}$
                \EndIf
                \State $\mathit{L_{r_{fb}}} \gets \post_{L}.\creator$
                \State mark $\post_L$ as $\tocommit$ 
                \State $\mathit{lastReferences \gets V}$ \label{step:mt-fallback-commit-enter-op}
                \State $\mathit{currentRound \gets r_{fb}+2}$ \Comment{graceful transition} \label{step:mt-fallback-commit-graceful-transition}
                \State \Call{try\_decide}{$ $}
                \State clear temporary state and timers for view $v_{fb}$
                \State $v_{fb} \gets v_{fb}+1$ \Comment{complete this fallback view}
                \State $\mymode \gets OP$ \Comment{switch to the optimistic path}
                \EndProcedure   \label{step:mt-finalize-fallback-end}
                \vspace{-0.2em}
		\end{algorithmic}
    \end{boxedminipage}
    \caption{Mysticeti-\sysname: vertex commitment for $\node{i}$; the code shown in \gtext{gray} comes from Mysticeti~\cite[Algorithm 3]{mysticeti}}
    \label{fig:mysticeti-fallback-commit}
\end{figure}

\Cref{fig:mysticeti-fallback-basic} and \Cref{fig:mysticeti-fallback-commit} give the algorithms for vertex creation and vertex commitment in Mysticeti-\sysname, respectively. These algorithms are described in detail in \S~\ref{sec:mysticeti-sysname}.

\section{Extended Discussion of PoST} \label{sec-post-discussion}

The ACS protocol does not inherently ensure the validity of proposals: malicious nodes can $\ACSPropose$ invalid vertices that contain nonexistent causal histories or come from stale rounds.
If an invalid leader vertex is selected by the ACS-based fallback mechanism, nodes may fail to retrieve the uncommitted history of vertices or risk committing vertices inconsistently.

To elaborate, in uncertified DAG-based BFT protocols~\cite{mysticeti, cordial-miners, jovanovic2024mahi}, DAG vertices are not guaranteed to be available. If Byzantine nodes reference ancestors that are never shared with others, correct nodes will not be able to retrieve the missing ancestors of the ACS output, introducing new liveness issues.

Moreover, in both certified and uncertified DAG-based BFT protocols, if Byzantine nodes use stale-round DAG vertices as their ACS proposals, the ACS outputs will contain a subset of old-round DAG vertices. As a result, some correct nodes may have already committed previous leader vertices via the underlying DAG protocol, while others commit a newly selected leader vertex (from the ACS output) that indirectly skips them. To understand why this safety issue exists, consider Bullshark~\cite{bullshark} (which has been deployed in production), where a round $r$ leader vertex $v_L^r$ is directly committed when $f+1$ DAG vertices in round $r+1$ connect to it. Assume that these $f+1$ DAG vertices in round $r+1$ are created by a set of nodes $\mathcal{N}_1$ consisting of $f$ Byzantine nodes and one correct node $\node{c}$. The remaining $2f$ correct nodes (denoted by a set $\mathcal{N}_2$) have round $r+1$ DAG vertices that do not connect to $v_L^r$. In this case, if Byzantine nodes from $\mathcal{N}_1$ use their round $r$ (not the latest round $r+1$) DAG vertices as ACS proposals and correct nodes from $\mathcal{N}_2$ use their round $r+1$ DAG vertices as ACS proposals, then the ACS output $V$ might consist of $2f+1$ DAG vertices that do not connect to $v_L^r$ (i.e., $\node{c}$'s round $r+1$ DAG vertex is excluded from $V$). As a result, correct nodes from $\mathcal{N}_2$ will skip committing $v_L^r$, leading to inconsistency.

\section{Details of Inflation Attacks on Sailfish and Mysticeti}\label{sec-concrete-attack-example}
In this section, we describe a specific implementation of inflation attacks on two state-of-the-art protocols: Sailfish and Mysticeti.

\begin{figure}[ht]
    \centering
    \includegraphics[width=0.95\linewidth]{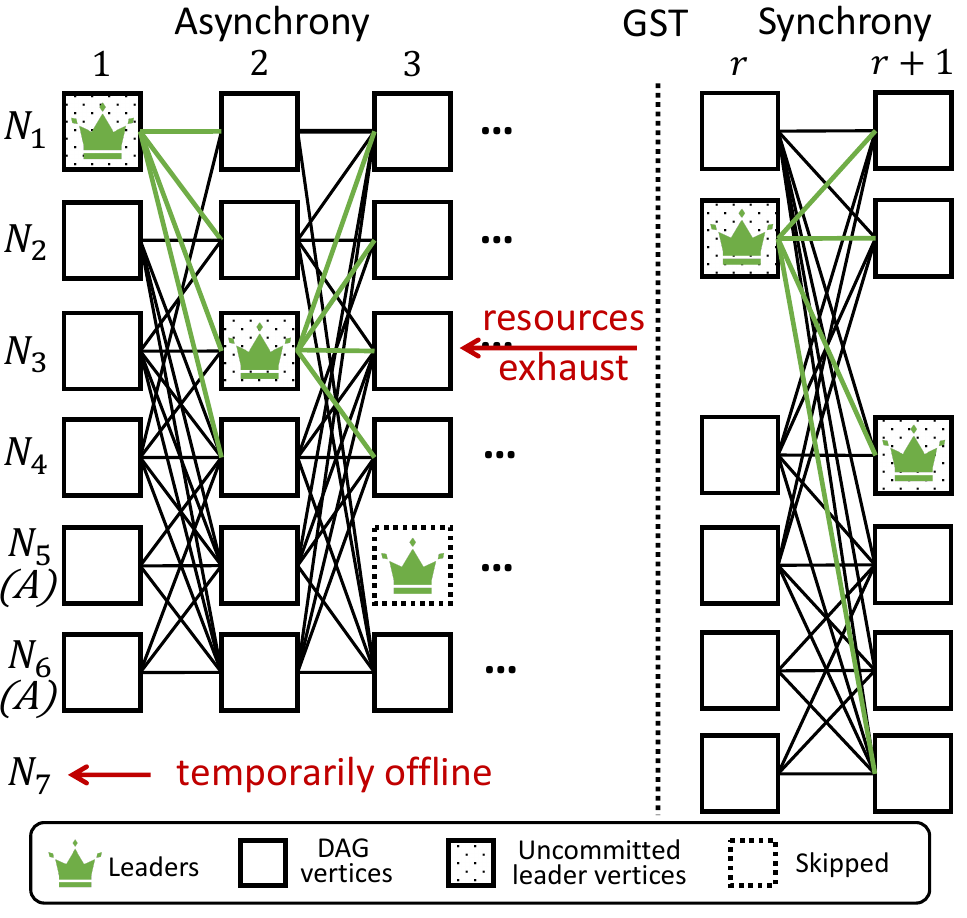}
    \caption{Overview of the inflation attack in Sailfish~\cite{sailfish} and Mysticeti~\cite{mysticeti}, where $\node{5}$ and $\node{6}$ are attackers and $\node{7}$ is temporarily offline: attackers skip proposing their leader vertices and do not vote for leader vertices. By doing this, attackers exhaust $\node{3}$'s resources before the network recovers to synchrony}
    \label{fig:attack-overview}
\end{figure}

As illustrated in~\Cref{fig:attack-overview}, the adversary controls $f$ nodes out of $3f+1$ nodes. For simplicity, we use $f=2$ as an example, and therefore the dissemination quorum $q_d=5$ and the committing quorum $q_c=5$.
Recall that in Sailfish and Mysticeti, a leader vertex can only be committed if it receives the required connections from subsequent vertices (\S~\ref{sec:sailfish-protocol-review} and \S~\ref{sec:mysticeti-protocol-review}). A leader that lacks the committing quorum remains uncommitted, and its retained history continues to consume resources. Our inflation attack deliberately creates this condition, eventually causing mempool explosions.

More specifically, the adversary first makes a correct node temporarily unavailable by launching a DDoS attack or causing a misconfiguration error on it. Then, the adversary uses the following approach to prevent any leader vertex from being committed.

In a round where a correct node is selected to propose the leader vertex, Byzantine nodes (i.e., $\node{5}$ and $\node{6}$) create vertices in the next round that deliberately do not connect to the leader vertex. Given that the adversary controls $f$ nodes and at least one correct node is temporarily offline due to the DDoS attack, the correct leader vertex can only be connected by at most $2f$ vertices. Because of the requirement $q_c = 2f+1$ for Mysticeti and Sailfish, the correct leader vertex cannot be committed.

In a round where a Byzantine node is selected to propose the leader vertex, the adversary deliberately skips this task, resulting in no leader vertex being created for that round. Consequently, the vertices proposed in previous rounds still remain uncommitted.

Note that during the above process, the data dissemination task is always successful as there are $3f>q_d$ nodes generating new vertices.
As a result, correct nodes must maintain an ever-growing uncommitted DAG in their mempools and eventually exhaust their resources.


If a correct node reaches its resource limit before GST (e.g., $\node{3}$ in \Cref{fig:attack-overview}), enforcing that limit prevents it from creating further vertices even after the network recovers to synchrony. The adversary can then adopt one of two strategies, depending on how the implementation handles the mempool explosion. If the node recovers by adding resources, the adversary can repeatedly trigger the same condition; no further DDoS is needed while the stopped node cannot contribute to the $q_c=2f+1$ committing quorum. Alternatively, the adversary can stop creating vertices, leaving only $2f<q_d$ nodes able to advance rounds. In either case, the unmodified protocol has no recovery action that both respects the resource limit and restores progress.

\section{Mempool Explosions in Non-DAG Decoupled BFT}\label{sec-mem-dis}
In the main body, we show how a mempool explosion can compromise the liveness of partially synchronous DAG BFT protocols. In this section, we discuss whether the adversary can exploit a mempool explosion in non-DAG decoupled BFT protocols to compromise their liveness.

\subsection{Mempool explosions in linear-chain BFT protocols}\label{sec-mem-dis-linear}
The adversary can exploit a mempool explosion in most implementations of linear-chain protocols~\cite{hotstuff-code,jolteon-code,diem-code}, including the HotStuff family of protocols~\cite{hotstuff,hotstuff-2,hotstuff-1,jolteon,streamlet,fasthotstuff,librabft}.
However, the vulnerability depends on the specific implementation of their pacemaker module~\cite{hotstuff}.

In these implementations, the following scenario can occur: during a prolonged period of asynchrony, the protocol may experience multiple rounds that time out. In each round, the leader successfully propagates a proposal to the entire committee but fails to collect enough votes to form a certificate (due to the asynchrony or the exploited attacks). As a result, validators retain uncertified blocks while waiting to determine which one will be used for progress. This approach, however, is unsustainable under prolonged periods of asynchrony, as it requires unbounded resources.

To withstand this vulnerability, these implementations could adopt a strategy in which validators discard uncertified blocks~\cite{jolteon} after a fixed number of timeout rounds~\cite{hotstuff,jolteon}. However, this solution presents a challenge, as validators typically commit to persisting and serving any block they sign~\cite{librabft,hotstuff-code}. A promising direction for future work is to explore whether validators could retain only the latest two or three rounds of blocks (depending on the protocol~\cite{hotstuff,fasthotstuff,jolteon}). This would ensure that each validator's storage requirements grow only linearly with the size of the committee.

We finally note that all implementations of these linear-chain BFT protocols that separate data dissemination from consensus~\cite{hotstuff-code,jolteon-code} are inherently sensitive to mempool-explosion attacks. Validators continuously generate batches of transactions intended for inclusion in future blocks. This forces committee members to hold onto these uncommitted batches, and, to the best of our knowledge, no literature currently explores a cleanup strategy for such batches. Once again, these protocols could mitigate the issue by retaining only a fixed number of batches per round and producing empty batches upon detecting a prolonged loss of liveness. However, this approach is delicate, as it complicates formal liveness arguments and may introduce unforeseen performance degradation during temporary and brief losses of network synchrony.

\subsection{Mempool explosions in multi-chain BFT protocols}\label{sec-mem-dis-multi}
Recent decoupled BFT protocols (such as Autobahn~\cite{autobahn}, Star~\cite{duan2024dashing}, and Arete~\cite{arete}) adopt a multi-chain structure to separate data dissemination from consensus. In these protocols, nodes can propagate multiple blocks of transactions in parallel and append the blocks to multiple chains. The consensus task is performed separately by an existing linear-chain BFT protocol (such as Hotstuff~\cite{hotstuff} or PBFT~\cite{pbft}), in which nodes take the hash digests of disseminated blocks as input and agree on a global order across multiple chains.

Our inflation attacks can be extended to these multi-chain BFT protocols. Specifically, the adversary proactively participates in the data dissemination task by continuously proposing new blocks, but it prevents the protocol from agreeing on the global order by making an arbitrary correct node unavailable for a while. The adversary can achieve this by launching a DDoS attack or causing the correct node to be misconfigured. As a result, correct nodes will experience a mempool explosion because of the ever-growing set of uncommitted blocks, compromising the protocol's liveness.

To mitigate this vulnerability, these protocols could force nodes to stop producing new blocks once their resources are nearly exhausted and to preserve sufficient resources only for the ordering task (i.e., for a BFT instance in that task). However, this solution also relies on the specific implementation of the employed BFT protocol and faces the same issues discussed in \Cref{sec-mem-dis-linear}. To the best of our knowledge, none of these multi-chain BFT protocols consider the management of limited resources for their liveness guarantees in either theoretical analyses or implementations~\cite{autobahn-code, arete-code}. It is worth emphasizing that traditional garbage collection mechanisms cannot circumvent this issue, as they rely on consensus to determine which data can be discarded, but consensus is never reached in the above cases. An independent and urgent follow-up topic for both academia and industry is to explore similar liveness vulnerabilities and design efficient mitigations for these BFT protocols.

\fi

\end{document}